\documentclass[twocolumn, twocolappendix]{aastex631}

\begin{document}

\title{Near-Infrared Spectroscopy of the Sun and Solar Analog Star HD\,76151: \\ Compiling an Extensive Line List in $Y$-, $J$-, $H$-, and $K$-Bands}

\correspondingauthor{T\. {ı}mur~\c{S}ah\.{ı}n}
\email{timursahin@akdeniz.edu.tr}

\author[0000-0003-3863-6525]{Sena~Aleyna~\c{S}ent\"{u}rk}

\author[0000-0002-0296-233X]{T\. {ı}mur~\c{S}ah\.{ı}n}
\affiliation{Akdeniz University, Faculty of Science, Department of Space Sciences and Technologies \\
07058, Antalya, Turkey}

\author[0000-0003-3884-974X]{Ferhat~G\"{u}ney}
\affiliation{Akdeniz University, Faculty of Science, Department of Space Sciences and Technologies \\
07058, Antalya, Turkey}

\author[0000-0003-3510-1509]{Sel\c cuk B\.Il\.Ir}
\affiliation{Istanbul University, Faculty of Science, Department of Astronomy and Space Sciences, 34119, Beyaz\i t, Istanbul, Turkey}

\author[0000-0002-9397-2778]{Mahmut~Mari\c{s}mak}
\affiliation{Akdeniz University, Faculty of Science, Department of Space Sciences and Technologies \\
07058, Antalya, Turkey}




\begin{abstract}

Determining the physical nature of a star requires precise knowledge of its stellar atmospheric parameters, including effective temperature, surface gravity, and metallicity. This study presents a new atomic line list covering a broad spectral range (1--2.5 {\textmu}m; $YJHK$-bands) for iron (Fe) and $\alpha$-elements (Ca, Mg, Ti, Si) to improve stellar parameter determination using near-infrared (NIR) spectroscopy. We highlight the limitations of existing line lists, stemming primarily from inconsistencies in oscillator strengths for ionized iron lines within the APOGEE DR17. The line list was validated using the high-resolution and high-quality disc-center NIR spectra of the Sun and its solar analog HD\,76151. As a result of the spectroscopic analyses, the effective temperature of HD\,76151 was calculated as 5790$\pm$170 K, surface gravity as 4.35$\pm$0.18 cgs, metal abundance as 0.24$\pm$0.09 dex, and microturbulence velocity of 0.30$^{\rm +0.5}_{\rm -0.3}$ km~s$^{-1}$ by combining the optical and NIR line lists. A comparison of the model atmospheric parameters calculated for HD\,76151 with the {\sc PARSEC} isochrones resulted in a stellar mass of $1.053_{-0.068}^{+0.056} M_{\odot}$, radius $1.125_{-0.011}^{+0.035} R_{\odot}$, and an age of 5.5$^{\rm +2.5}_{\rm -2.1}$ Gyr. For the first time, kinematic and dynamical orbital analyses of HD\,76151 using a combination of {\it Gaia} astrometric and spectroscopic data showed that the star was born in a metal-rich region within the Solar circle and is a member of the thin disc population. Thus, the slightly metal-rich nature of the star, as reflected in its spectroscopic analysis, was confirmed by dynamical orbital analysis. 

\end{abstract}



\section{Introduction} \label{sec:intro}
In the infrared (IR) region, the atmosphere of the Earth is not fully transparent, and observation from the ground is only possible for certain wavelengths. Many pioneering ground-based spectroscopic surveys, including the Radial Velocity Experiment (RAVE; \citealp[]{Steinmetz2006}), Sloan Extension for Galactic Understanding and Exploration \citep[SEGUE-1 and SEGUE-2;][]{Yanny2009, Eisenstein2011}, the Apache Point Observatory Galactic Evolution Experiment (APOGEE; \citealp[] {Majewski2017}; \citealp[]{Prieto2008}), the Galactic ArchaeoLogy with HERMES project (GALAH; \citealp[]{DeSilva2015}), {\it Gaia}-RVS \citep{Cropper2018}, the Large sky Area Multi-Object Fibre Spectroscopic Telescope (LAMOST; \citealp[]{Luo2015}), the {\it Gaia}-ESO spectroscopic survey \citep{Gilmore2012}, and projects involving next-generation optical and near-infrared spectrographs such as the 4-m Multi-Object Spectroscopic Telescope (4MOST; \citealp[]{deJong2012, deJong2019}), the WHT Enhanced Area Velocity Explorer (WEAVE; \citealp[]{Dalton2012}), the Multi-Object Optical and Near-Infrared Spectrograph (MOONS) on ESO’s Very Large Telescope (\citealp[]{Taylor2018}), the Prime Focus Spectrograph (PFS; \citealp[]{Takada2014}), the Maunakea Spectroscopic Explorer (MSE; \citealp[]{McConnachie2016}) and finally the 4-m East Anatolian Telescope\footnote{It is expected to see its first light in 2024.} (DAG; \citealp{DAG2022}) will extend the six-dimensional phase space into a multidimensional information space, incorporating chemical abundance measurements of a wide range of tracers for the chemical evolution of several thousand stars, not only in the optical region but also in the NIR region. 

In limited contexts, abundance measurements can aid the investigation of exoplanets and theories related to planetary formation. Accurately determining the mass of a planet requires the knowledge of the mass of the host star. Similarly, determining the radius of a planet depends on precise determination of the radius of the accompanying star. In a broader context, these abundances can also be utilized to construct a complete chemodynamic map of the Milky Way and play a vital role in investigating the origins and evolution of stellar populations. This requires high resolution spectra for accurate determination of atmospheric parameters and metallicity. It is also necessary to have a good grid of model atmospheres. Most importantly, it requires a complete line list with precise estimates of atomic data such as $\log gf$-values (\citealp[]{Barklem2002}, \citealp[]{Caffau2013}). The number of line lists available, particularly in the NIR region, is remarkably limited compared with that in the optical region. This remains an urgent concern because of the increasing number of spectrometers that will operate in the near future, particularly in the NIR region. 

  \begin{figure*}
    \centering
    \includegraphics[width=1.00\textwidth]{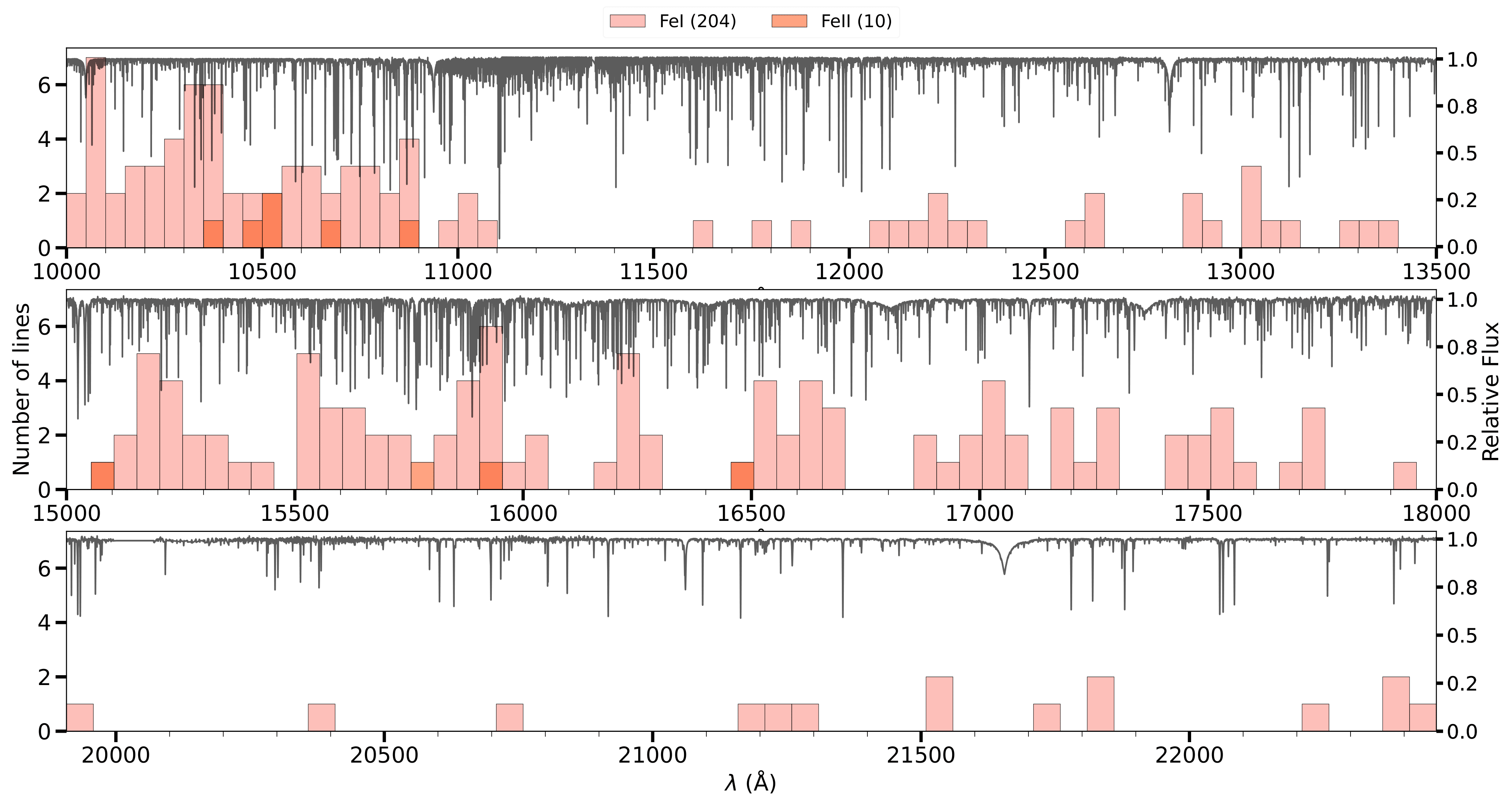}
    \caption{The KPNO-NIR solar spectrum and the number of identified lines in the respective wavelength regions (each bar indicates 50 Å region of the spectrum). }
\label{fig:fig1}
\end{figure*}

Compared to the spectrum obtained in the optical region, a major advantage of working in the NIR region is that the spectrum obtained in this region is less susceptible to blending lines and there are fewer lines that tend to pose challenges in accurately correcting the continuum level. However, most published NIR line lists contain several lines that are not properly identified or lack well-calibrated oscillator strengths, and some are theoretical (\citealp[]{Ryde2009}). Therefore, this issue for line lists with theoretical $\log gf$ estimates of questionable accuracy should be addressed in future studies. 

The solar spectrum serves as a common reference point for spectroscopic studies of F-G-K spectral type stars in both the optical and NIR regions because the Sun is an excellent benchmark owing to its well-characterized atmosphere and extensive observational data in the infrared region. This is particularly useful in assessing the observability and sensitivity of atomic data for theoretically predicted spectral lines, as well as the accuracy of the atomic data. The high-resolution solar spectrum significantly improves both the line identification process and the measurement of equivalent widths (EW).

In this study, with wavelength coverage from 1 {\textmu}m to 2.5 {\textmu}m, we provide a new line list with an increased number of atomic transitions for several species, in particular, Fe\,{\sc i}, Fe\,{\sc ii}, and $\alpha$-elements (Ca, Mg, Ti, and Si). These elements have been subjected to spectroscopic analysis in previous studies restricted to certain wavelength ranges (\citealp[]{Nave1994}; \citealp[]{Melendez1999}; \citealp[]{Borrero2003}; \citealp[]{Smith2013}; \citealp[]{Jofre2014}; \citealp[]{Andreasen2016}; \citealp[]{Kondo2019}; \citealp[]{DR17}\footnote{\url{https://data.sdss.org/sas/dr17/apogee/spectro/speclib/linelists/synspec/}}). Among these preliminary studies listed above, the number of studies in which lines were subjected to EW analysis was limited. 

The use of the new line list in spectroscopic analyses of main-sequence stars of the F, G, and K spectral types in the vicinity of the Sun will contribute significantly to Galactic archaeological research, since these stars are long-lived and have a relatively thin convective layer \citep{Soubiran2022, Lopez-Valdivia2024}. An extensive line list, such as that of iron, would also be extremely useful for several high-resolution spectral libraries developed in the near-infrared region and obtained using ground-based instruments. The APOGEE spectroscopic survey (\citealp[]{Zamora2015}; \citealp{Majewski2017}), as one of the first examples of fiber-fed multi-object spectrographs, provided spectra for hundreds of thousands of stars in the $H$-band (1.5 – 1.7 {\textmu}m, $R=\lambda/\Delta\lambda=$ 22\,500). CRIRES-POP \citep{Dorn2014} provides spectra at 1 – 5 {\textmu}m. Particularly in certain parts of this wavelength range (e.g., beyond 2.3 {\textmu}m), the telluric sky and telescope appear to glow because of their inherent temperatures, making observations in these windows challenging. Similarly, NIRSPEC \citep{McLean1998} on the Keck II telescope provided spectra at ($R \approx$ 25\,000) 1.5-1.8 {\textmu}m for bulge stars. To validate the line list and ensure the accuracy of the derived stellar parameters, we employed a high resolution ($R \approx$ 45\,000) spectrum of the solar analog star (\citealp[]{Datson2015}, \citealp[]{Martos2023}), HD\,76151, acquired using an Immersion GRating INfrared Spectrometer (IGRINS; \citealp[]{Park2014}) from the IGRINS spectral library\footnote{The IGRINS library (\citealp[]{Park2018}) provides spectra in the $H$ (1.49 – 1.80 {\textmu}m) and $K$ (1.96 – 2.46 {\textmu}m) bands.}. This spectrum offers superior resolution compared to other commonly used instruments, such as NIRSPEC, allowing for more detailed line analysis. For the nine Fe\,{\sc ii} transitions included in our analysis, astrophysical loggf values were calculated. While a range of loggf values exists for the remaining lines, we carefully selected those that best reproduced the observed solar abundance for each transition. This manual adjustment process was crucial, especially considering the limitations of automated procedures like those used in the APOGEE line list.

The remainder of this paper is organized as follows. Section 2 focuses on the development of a new line list. It presents an evaluation of atomic line libraries and catalogs, along with line identification and selection processes. Subsequently, Section 2 discusses the star selection process and inconsistencies identified within the ionized Fe line data listed in APOGEE DR17 \citep{DR17}. The spectra used in this study are described in Section 3. Section 4 details the spectral analysis method and model atmosphere analysis employed in this study. Section 5 utilizes both optical and NIR spectra of HD\,76151 to test the line list and derive its key astrophysical parameters. To provide a comprehensive characterization of HD\,76151, Section 5 also determines fundamental parameters and its space velocity components, the Galactic orbital parameters, birthplace, and current location within the Galaxy. Finally, Section 6 presents concluding remarks, summarizing the key findings of this study.

\section{Construction of a new line list in the infrared} \label{sec:linelist}

We present a new line list in the NIR for Fe and $\alpha$ elements (Ca, Mg, Si, and Ti) based on high-resolution spectra for the Sun and a solar analog star, HD\,76151. Initially covering the APOGEE \citep{Majewski2017} project's wavelength range, the list was subsequently extended to include $Y$-, $J$-, $H$- and $K$-bands (details in Figure~\ref{fig:fig1}). Notably, the $H$-band showed a high abundance of Si, Ca, and Ti transitions.

Line measurements and identification of the echelle spectra used in this study were performed using {\sc LIME} code \citep{Sahin2017} in a classical spectroscopic framework (details in Section 4).  

\begin{figure*}
    \centering
    \includegraphics[width=0.99\textwidth]{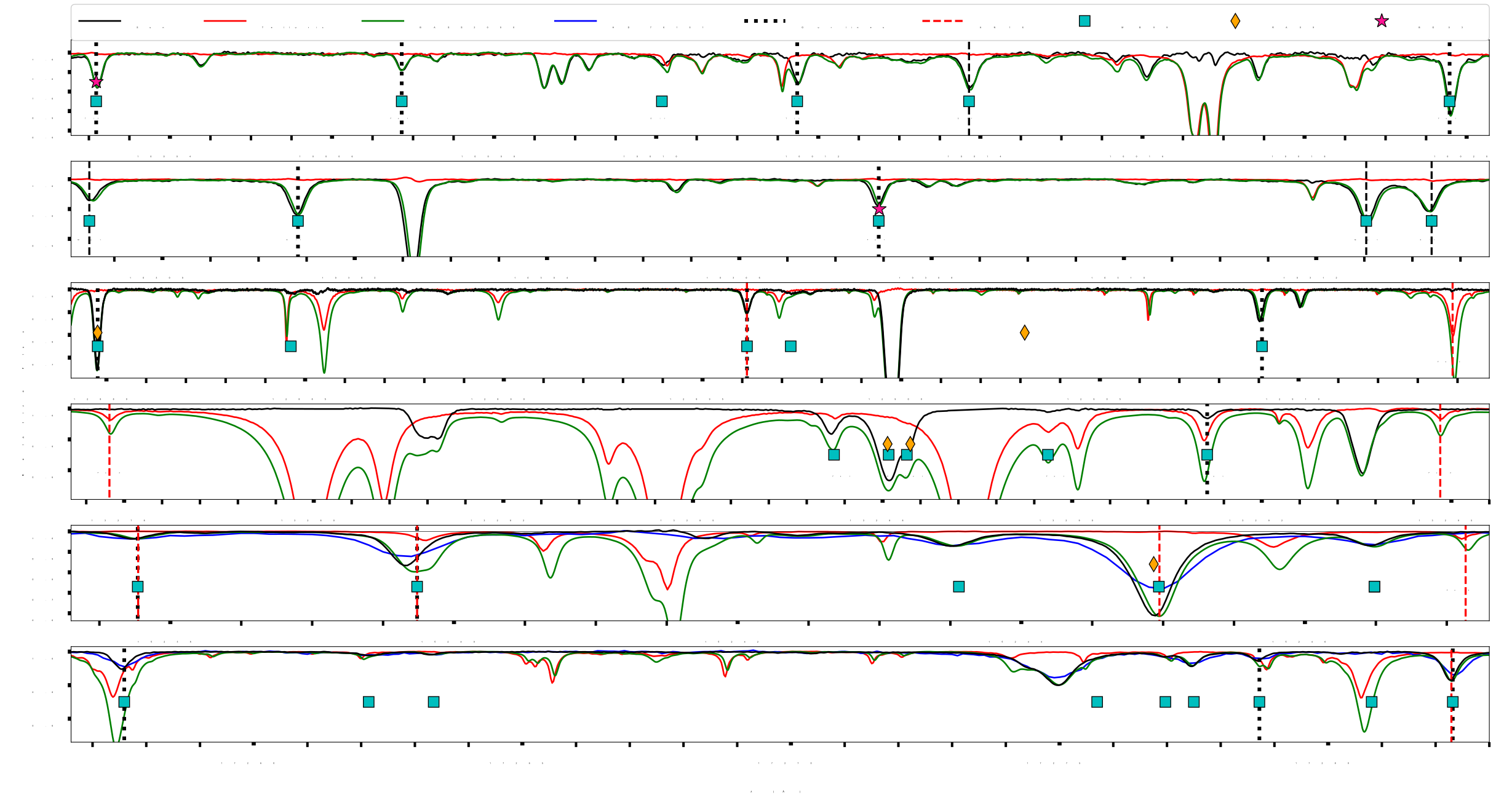}
    \caption{NIR solar spectrum (black) and the NIR spectrum of HD\,76151 (turquoise). The BASS2000 solar spectrum used by \citet{Andreasen2016}[AN16] (green) and the NIR telluric Kitt Peak Natiaonal Observatory (KPNO) spectrum (red) are also included. The neutral and ionised Fe lines reported by \citet{Andreasen2016}, \citet{Melendez1999}[MB99], \citet{Rayner2009}[RA09], and \citet{Borrero2003}[BO03] are included as coloured dots. The black and red dotted vertical lines indicate the lines used in this study and included in the study of \citet{Andreasen2016}, respectively.}
    \label{fig:fig2}
\end{figure*}

Atomic data for the new line list, were compiled from several different sources and 
compared with atomic data obtained from known atomic databases in the literature, such as Kurucz's 
database (\citealp[]{Kurucz1995}), the National Institute of Standards and Technology (NIST) Atomic 
Spectra Database\footnote{NIST Atomic Spectra Database can be reached at 
\url{http://physics.nist.gov/PhysRefData/ASD}. The NIST and VALD databases were solely employed to 
verify the accuracy of atomic data, specifically the $\log gf$ values.}, and the Vienna Atomic Line 
Database\footnote{\url{http://vald.astro.uu.se}} (VALD3; \citealp[]{Ryabchikova2015}).
 
All the line lists from previous studies were inspected against the observed spectra (Section 3). We found that most line center wavelengths and transition probabilities ($\log gf$ values) were theoretically calculated. Rigorous tests were conducted on these line lists during the initial evaluations.

Our initial approach involves replicating the line list and analysis presented by \citet{Andreasen2016}. However, upon careful examination, we discovered significant contamination by atmospheric lines in the original line list. This compromised the accuracy of the equivalent width measurements and necessitated the creation of a new and more reliable line list. Because the target star in the original study was identified as a binary system, we focused our analysis on a single star to mitigate potential complexities. The aforementioned challenges related to the original line list and the binary nature of the target star necessitate more in-depth examination of the original study's methodology to substantiate the reliability of our results.

\begin{table*}
\setlength{\tabcolsep}{5.0pt}
\centering
\caption{Line statistics for reviewed studies in the infrared region.}
\label{tab:1}
\begin{tabular}{l|c|c|c|c|c|c|c|c|c|c}
\hline
Reference &Wav. Range (\AA)& Total Lines & Fe\,{\sc i} & Fe\,{\sc ii} & Mg\,{\sc i} & Si\,{\sc i} & Ca\,{\sc i} & Ca\,{\sc ii} & Ti\,{\sc i} & Ti\,{\sc ii} \\
\hline
\citet{Melendez1999} & 10\,000 - 17\,980 & 2\,218 & 1\,163 & 13 & 79 & 181 & 52 & 7 & 82 & 1  \\
\citet{Borrero2003} &~~9\,860 - 15\,650& 83 & 42 & - & - & 17 & 4 & - & 6 & -  \\
\citet{Rayner2009} &~~8\,180 - 38\,658& 608 & 299 & - & 48 & 114 & 19 & 10 & 16 & -  \\
\citet{Ryde2010} &15\,320 - 15\,700& 104 & 63 & - & - & 19 & - & - & 7 & -  \\
\citet{Smith2013} &15\,159 - 16\,828& 53 & 9 & - & 7 & 9 & 4 & - & 5 & -  \\
\citet{Afsar2016, Afsar2018} &15\,020 - 23\,379& 120 & 27 & - & 11 & 12 & 11 & - & 28 & 7  \\
\citet{Andreasen2016} &10\,070 - 24\,648& 330 & 317 & 13 & - & - & - & - & - & -  \\
\citet{DR17} &15\,000 - 17\,000& 51\,227 & 4\,740 & 2\,979 & 256 & 332 & 346 & 72 & 1\,899 & 181  \\
\citet{Kondo2019} &~~9\,117 - 13\,291& 171 & 171 & - & - & - & - & - & - & - \\
Kurucz$^{*}$ &~~9\,780 - 24\,999& 72\,964 & 8\,295 & 2\,600 & 2\,118 & 1\,184 & 822 & 159 & 1\,480 & 160 \\
\hline
\hline
Inspected lines &~~8\,180 - 38\,658& 127\,878 & 15\,126 & 5\,605 & 2\,519 & 1\,868 & 1\,258 & 248 & 3\,523 & 349  \\
\hline
This study &10\,003 - 23\,308& 267 & 204 & 10 & 2 & 28 & 11 & 1 & 10 & 1  \\
\hline
\hline
\end{tabular}
    \vspace{1ex}
\footnotesize{($^{*}$)\url{http://kurucz.harvard.edu/linelists/gfnew/gfallwn08oct17.dat}\\ An inspection of the Kurucz line list was conducted within the $Y$-, $J$-, $H$-, and $K$-band wavelength range.} 
\end{table*}

\citet{Andreasen2016} created a list of lines for 313 neutral Fe (Fe\,{\sc i}) and 13 single-ionized Fe (Fe\,{\sc ii}) from 1 to 2.5 $\mu$m and reported astrophysically calculated transition probabilities. They tested the line list on the BASS2000 spectrum of Sun and HD\,20010, a metal-poor ([Fe/H] = -0.23$\pm$0.14 dex) sub-giant star. Surprisingly, despite our initial expectations, this star was identified as a double system in the SIMBAD database. Therefore, it is not feasible to conduct similar tests on the HD\,20010 spectrum. Thanks to the IGRINS team, the IGRINS ($R$=45\,000) library contains several telluric-free spectra of various spectral types (e.g., nine F-type stars and 21 G type stars) and offers a spectrum in the range 1.45 – 2.45 $\mu$m range (\citealp[]{Oh2010}; \citealp[]{Park2014}). Among the 21 G stars, we used the IGRINS spectrum of HD\,76151, a solar analog thin disc star with high proper motion. HD\,76151 has frequently been used in exoplanet studies. No evidence of planets was found in the NASA Exoplanet Archive on HD\,76151. Additionally, a study on spectroscopic parameter determination of 451 stars in the HARPS GTO planet search program by \citet{Sousa2008} reported no planets for this particular star. The reported renormalised unit weight error (RUWE) value of $\approx$1.0 for HD\,76151 indicates a good data quality from {\it Gaia} Data Release 3 \citep[{\it Gaia} DR3,][]{Gaia2023}, and provides no evidence of the presence of a secondary object\footnote{\url{https://gea.esac.esa.int/archive/documentation/GDR2/Gaia_archive/chap_datamodel/sec_dm_main_tables/ssec_dm_ruwe.html}}. An RUWE value higher than 1.4 is often considered indicative of binarity \citep{Belokurov2020}. The proper motion anomaly was extremely low, excluding a companion with a mass larger than the Jupiter mass of up to 30 AU\footnote{Pierre Kervella; private communication}. 

The model atmospheric parameters for the BASS2000 solar spectrum reported by \citet{Andreasen2016} using an ATLAS9 model atmosphere are as follows: $T_{\rm eff} =$ 5777 K, $\log g = 4.438$ cgs and $\xi=$ 1.0 km s$^{\rm -1}$. They adopted a logarithmic abundance of 7.47 dex as the solar metallicity ([Fe/H]) from \citet{Gonzalez-Laws2000}. By applying our analysis procedures to Andreasen's line list, we identified a model providing excitation and ionization equilibrium, namely $T_{\rm eff} =$ 5790$_{-125}^{+125}$ K, $\log g =$ 4.40$_{-0.23}^{+0.16}$ cgs, $\xi = 0.68_{-0.50}^{+0.50}$ km s$^{\rm -1}$. To repeat the analyses carried out with their published line list, we obtained the NIR portion of solar spectrum\footnote{It was acquired from the Kitt Peak National Observatory's McMath Solar Telescope with its accompanying Fourier Transform Spectrometer at a precision of 0.004 $\mu$m at 1 $\mu$m to 0.1 $\mu$m at 5 $\mu$m. The spectrum had a signal-to-noise ratio ranging from 2000 at 5 $\mu$m to 5000 at 1 $\mu$m.} (i.e., the BASS200 solar spectrum) provided by Delgado Mena\footnote{private communication}. Our investigation of their line list showed that the BASS2000 solar spectrum employed in their analysis was contaminated by telluric lines. Furthermore, certain lines did not meet the criteria for measurable and unblended lines. A similar telluric contribution was detected in additional neutral (e.g., 15\,237.77 \AA, 15\,239.74 \AA, and 15\,244.97 \AA) and ionized (15\,247.13 \AA) Fe lines. The Fe\,{\sc i} lines at 15\,230.32 \AA\,and 15\,251.01 \AA\, are also among these lines that are contributed by telluric lines. To address the fact that the BASS2000 solar spectrum employed in the original study by \citet{Andreasen2016} was not corrected for atmospheric absorption, potentially introducing systematic errors into the equivalent width measurements, we adopted the NSF/NOAA-produced, telluric-free NSO/Kitt Peak FTS solar spectrum\footnote{To evaluate the impact of telluric contributions on the KPNO-NIR spectrum, we compared it to Baker's telluric-free solar spectrum \citep{Baker2020}.} (i.e., the KPNO-NIR solar spectrum, $R\approx$700\,000 \citep{wallace2003atlas}, as opposed to the BASS2000 solar spectrum.  

To enhance the completeness and reliability of the initial line list, we employed a multiplet analysis technique (Rowland multiplet number, RMT). This approach relies on the systematic identification of related groups of lines within a multiplet \citep{Moore1954}, has proven to be instrumental in expanding the line list and identifying potential errors in previously published data. By applying the multiplet analysis technique to the line list of \citet{Andreasen2016}, using the RMT information from \citet{Nave1994}, we identified 39 additional Fe\,{\sc i} transitions. Of these, five were deemed unsuitable for EW analysis for various reasons, and three were excluded because of blending issues. Ultimately, 18 of these newly identified Fe\,{\sc i} lines were incorporated into our final line list and are marked in blue in Tables \ref{tab:A3}, \ref{tab:A4}, \ref{tab:A5}, and \ref{tab:A6}. The remaining 13 Fe\,{\sc i} lines exhibited significant line-to-line scatter in the calculated solar iron abundances and were subsequently removed.

In addition to the line list by \citet{Andreasen2016}, we conducted an extensive literature search encompassing the studies of \citet{Melendez1999},  \citet{Borrero2003}, \citet{Rayner2009}, \citet{Ryde2009}, \citet{Smith2013}, \citet{Afsar2016, Afsar2018}, \citet{Kondo2019}, and APOGEE DR17 \citep{DR17} to construct a robust and comprehensive line list in the infrared region. Table \ref{tab:1} provides a detailed overview of the line statistics derived from these sources.

A critical component of this study involves the application of multiplet analysis to each of the aforementioned line lists. This allowed for the identification of additional spectral transitions and detection of potential errors in previously published data. Appendix A provides detailed information on the individual studies (details in Appendices A1-A7), comparative analyses of the atomic data used in each study, and the test results for APOGEE DR17 (Appendix A8).

To validate the resulting line list, the echelle spectra of both the Sun and a solar analog star (HD\,76151) were employed. Details of the spectra are provided in the following section.

\begin{table*}
\setlength{\tabcolsep}{11.5pt}
\centering
\caption{Model atmosphere parameters from {\sc HARPS} and {\sc IGRINS} spectra of HD\,76151. Model parameters for optical solar (KPNO-OPTICAL) and NIR solar (KPNO-NIR) spectrum are also presented.}
\label{tab:2}
\centering
\begin{tabular}{lcccccc}
\hline
\hline
Star	&Spectrum	& Wav. Range& $T_{\rm eff}$	&	$\log g$ 	&	[Fe/H] &  $\xi$ \\
    &  &(\AA)  & (K) & (cgs) & (dex) & (km s$^{\rm -1}$) \\
    \hline
    \hline
HD\,76151    &OPTICAL (HARPS)  & 3\,780-6\,900 &	5780$\pm$88&	4.35$\pm$0.16	&	0.14$\pm$0.08   & 0.69$\pm$0.50    \\ 
HD\,76151               &NIR (IGRINS) & 14\,500-24\,500  &	5780$\pm$178	    &	4.31$\pm$0.25       	&	0.19$\pm$0.17	& 2.10$\pm$0.50    \\
HD\,76151              &OPTICAL+NIR & 3\,780-24\,500  &	5790$\pm$170	    &	4.35$\pm$0.18       	&	0.24$\pm$0.09	& 0.30$\pm$0.50    \\
\hline
Sun          &KPNO-OPTICAL   & 4\,000-7\,000 &	5790$\pm$45	    &	4.40$\pm$0.09		&	0.00$\pm$0.04   & 0.66$\pm$0.50    \\ 
Sun           &KPNO-NIR    & 11\,100-54\,110  &	5780$\pm$55	    &	4.40$\pm$0.22	&	0.00$\pm$0.03& 1.08$\pm$0.50    \\ 
Sun           &OPTICAL+NIR    & 4\,000-54\,110  &	5790$\pm$78	    &	4.44$\pm$0.14	&	0.00$\pm$0.05   & 0.90$\pm$0.50 \\
\hline
\end{tabular}
\end{table*}

\section{Observations \label{sec:observations}}

This study utilizes high-resolution, high-signal-to-noise spectra from both the Sun and HD\,76151 for spectral analysis. The spectra of stars cover a wide wavelength range, encompassing both the optical and NIR regions.

The decision to utilize both the optical and near-infrared spectra for both the Sun and HD 76151 was driven by the need for a comprehensive and robust analysis. By covering a wide spectral range, we aimed to: (1) maximize the effectiveness of the newly constructed infrared line list, allowing for direct testing and refinement in the target spectral region; (2) compare model atmospheric parameters and elemental abundances derived from different spectral regions, assess the consistency of results, and identify potential region-specific discrepancies; and (3) enhance the overall reliability of the abundance analysis by combining information from multiple spectral regions to reduce uncertainties associated with any single spectral region.

The high-resolution ($R\approx$700\,000) NIR solar spectrum (hereafter, KPNO-NIR) was obtained using a Fourier Transform Spectrometer (FTS) at KPNO, which is free of telluric lines. The characteristics of the KPNO-NIR solar spectrum are shown in Figure~\ref{fig:fig1}. The high-resolution solar spectrum used for spectroscopic analysis in the optical region was obtained from \cite{kurucz1984}. The HARPS \citep{Mayor2003} echelle spectrum (MJD: 55305.01205; Exp: 900 s; $R$=115\,000; $S/N$=356.7)\footnote{The star was observed by S. Udry (P.I.) in the frame work of a study titled ``a census of the super-earth and neptune-mass planet population around solar type stars''.} of HD\,76151 was obtained from the ESO Science Archive\footnote{\url{https://archive.eso.org/wdb/wdb/adp/phase3_main/form}}, which covers the wavelength range 3780--6910 \AA. The radial velocity for the HARPS spectrum obtained using the cross-correlation method via NARVAL spectrum of the Sun\footnote{It is from the {\it Gaia} Benchmark stars library pipeline, \url{https://www.blancocuaresma.com/s/benchmarkstars}. } for HD\,76151 is $V_{\rm Rad}=32.05\pm0.05$ km s$^{\rm -1}$. The IGRINS spectrum from the IGRINS library was corrected for the radial velocity. A further test on the IGRINS spectrum using the cross-correlation method along with the NIR solar spectrum provided the following shifts in the $V_{\rm Rad}$=0.39 km s$^{\rm -1}$ for the $H$-band and 0.31 km s$^{\rm -1}$ for the $K$-Band IGRINS spectrum. The {\it Gaia} DR3 \citep{Gaia2023} radial velocity of the star (Gaia DR3 5760701787150565888, $V_{\rm Rad}$= 31.99$\pm$0.12  km s$^{\rm -1}$) is in agreement with the radial velocity of the star from its HARPS (i.e. $V_{\rm Rad}$=32.05$\pm$0.05 km s$^{\rm -1}$) and IGRINS (i.e. $V_{\rm Rad}$=32.08 km s$^{\rm -1}$) spectra within error limits. 

Most of the identified lines were suitable for the EW analysis. We employed the {\sc LIME} code (\citealp[]{Sahin2017}) for continuum normalization, line measurement and verification. This code facilitates an interactive identification process similar to that described by \citet{Sahin2011}. It operates on a continuum-normalized spectrum and provides access to atomic data (RMT, $\log gf$, and lower-level excitation potential-LEP) from its library for the candidate line.

\begin{figure}
    \centering
    \includegraphics[width=0.47\textwidth]{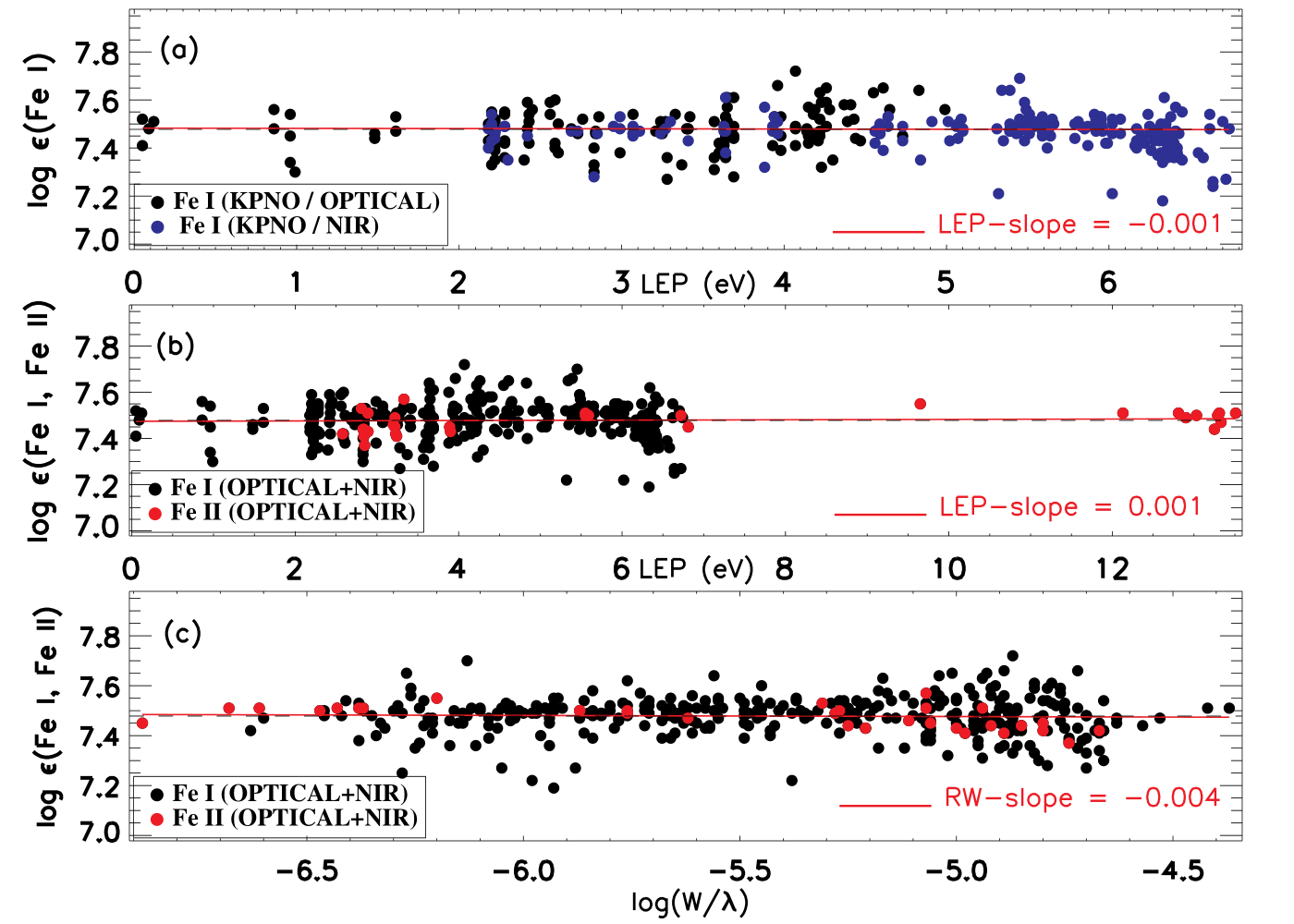}
    \includegraphics[width=0.47\textwidth]{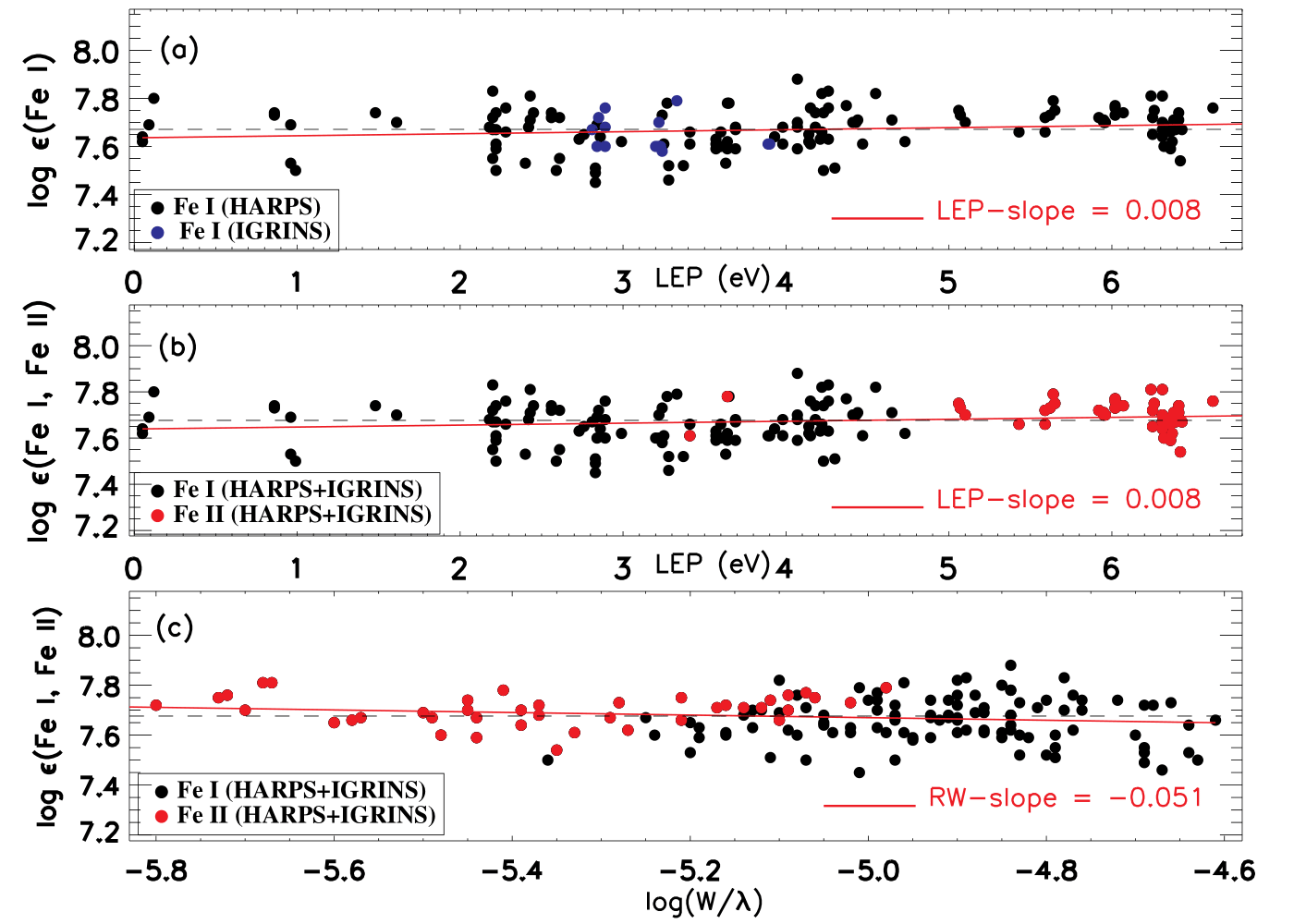} 
    \caption{An example of the solar analysis performed with the MOOG LTE line analysis code in the optical and NIR is shown. In the upper panel, the top and middle panels, respectively, show the determination of the photospheric model atmosphere parameters for the Sun (KPNO-OPTICAL and KPNO-NIR) as a function of the LEP and the reduced EW using the MOOG stellar atmosphere code. The lower panel is for the HARPS and IGRINS spectra of HD\,76151. The solid dashed lines in all panels represent least squares fits to the data. The bottom panels show the wavelength distribution of the abundances calculated for the Fe lines in both panels.}
    \label{fig:3}
\end{figure}

For the EW measurements, a first-degree polynomial was used to fit the local continuum. Subsequently, Gaussian profiles were employed to measure the EWs. For strong lines, direct integration is preferred over the Gaussian approximation.
 
Figure~\ref{fig:fig2} shows the NIR telluric spectrum used for the telluric line positions, together with the IGRINS echelle spectrum of HD\,76151 and BASS2000 solar spectrum\footnote{They used the solar spectrum: \url{https://bass2000.obspm.fr/solar_spect.php}} contaminated by telluric lines. The telluric-free KPNO-NIR solar spectrum across various wavelengths is shown in the same figure. This spectrum was used for solar analysis in NIR. The BASS2000 solar spectrum in Figure~\ref{fig:fig2} was also included for comparison purposes and was contaminated by telluric lines. The IGRINS spectrum of HD\,76151 is also free of telluric lines. The lines (12\,913.876 \AA, 13\,277.306 \AA, 13\,294.853 \AA, and 15\,247.133 \AA) listed as ionized iron transitions and used for elemental abundance analysis in the BASS2000 solar spectrum by \citet{Andreasen2016} were found to be telluric lines (Figure~\ref{fig:fig2}). In contrast to BASS2000 solar spectrum, it can be seen that the neutral iron lines at 13\,291.78 \AA\, and 17\,257.59 \AA\, are free of telluric contributions in the KPNO-NIR solar spectrum.  

\section{Spectroscopic analyses \label{sec:spectroscopy}}

In this study, we carried out an abundance analysis utilizing {\sc ATLAS9} model atmospheres computed in local thermodynamic equilibrium (LTE) using the ODFNEW method \citep{Castelli2004}.
The line analysis code MOOG (\citealp[]{Sneden1973})\footnote{The MOOG code, written in the FORTRAN programming language, allows element abundances to be obtained under the condition of local thermodynamic equilibrium (LTE) and is available at \url{http://www.as.utexas.edu/~chris/moog.html}} was used to obtain the model parameters ($T_{\rm eff}$, $\log g$, $v_{\rm t}$, [Fe/H]) for stars. For details of the abundance analysis of stars, see \citet{Sahin2009}, \citet{Sahin2011,Sahin2016}, \citet{Sahin2020}, \citet{Sahin2023}\footnote{The line list used for the spectral analyses in the optical region in this study is presented in Tables 1 and 2 of the same study.}, and \citet{Marismak2024}. 

\begin{figure}
    \centering
    \includegraphics[width=0.48\textwidth]{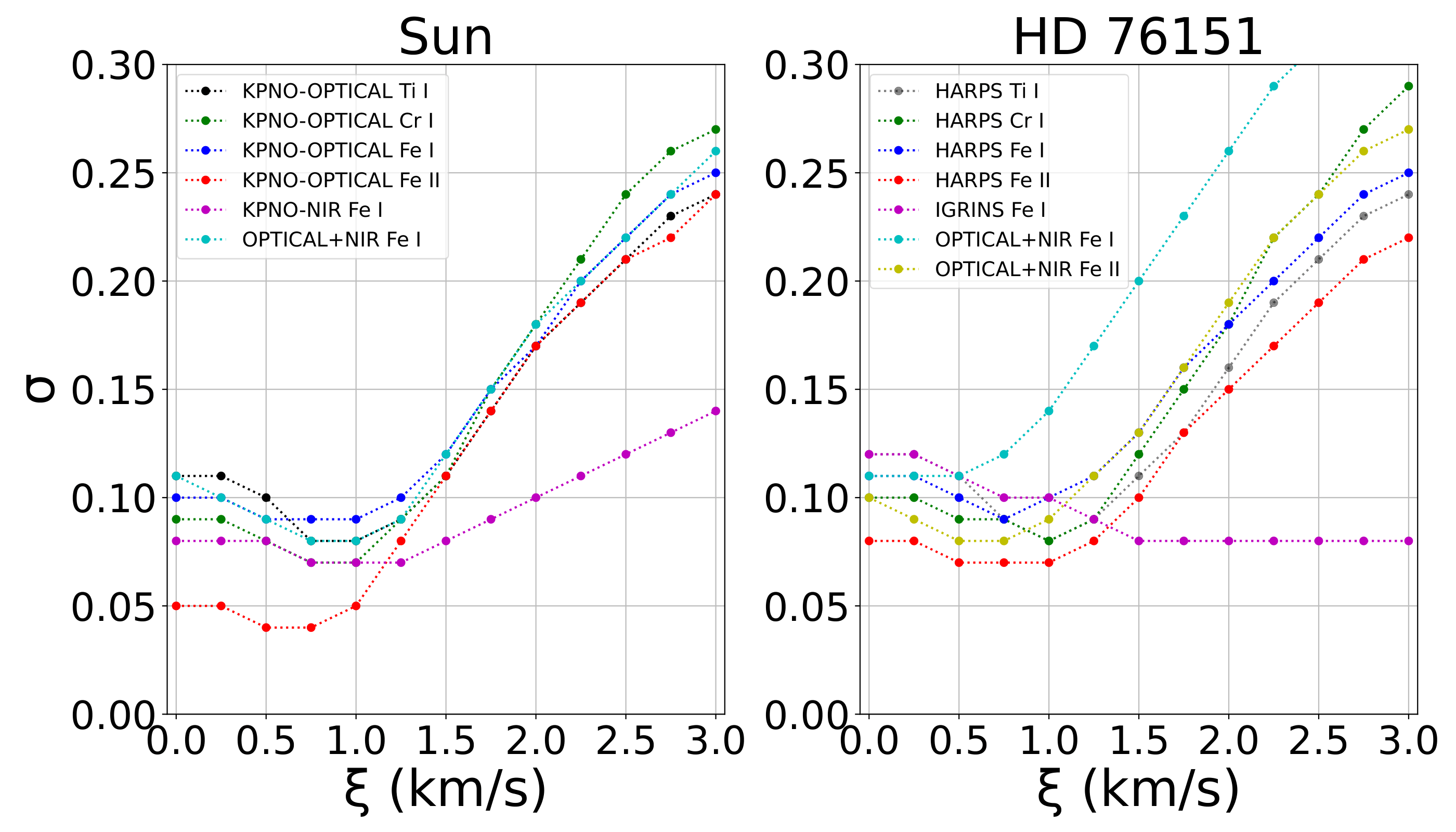}
    \caption{The standard deviation of the Ti, Cr, and Fe abundances from the suite of Ti\,{\sc i}, Cr\,{\sc i}, Fe\,{\sc i}, and Fe\,{\sc ii} lines as a function of $\xi$.}
\label{fig:micro}
\end{figure}

The list of lines that can be used for $\alpha$ element abundance calculations for Mg\,{\sc i}, Si\,{\sc i}, Ca\,{\sc i}, and Ti\,{\sc i/ii} is given in Tables \ref{tab:A2} and \ref{tab:A3}. Neutral and ionized Fe lines were used to determine model parameters in the NIR region (Tables \ref{tab:A3}-\ref{tab:A7}). Sample figures for the method used to determine the model parameters for HD\,76151 and the solar spectrum are shown in Figure~\ref{fig:3}. The upper panel of both figures depicts the relationship between the logarithmic abundance\footnote{We follow the standard definitions for elemental abundances and ratios. For element $X$, the logarithmic absolute abundance is calculated as the number of atoms of X per 10$^{\rm 12}$ hydrogen atoms, $\log \epsilon(X) = \log_{\rm 10}(N_{\rm X}/N_{\rm H}) + 12.0$. For elements $X$ and $Y$, the log abundance ratio relative to the solar ratio is defined as $[X/Y] = \log_{\rm 10}(N_{\rm X}/N_{\rm Y}) - \log_{\rm 10}(N_{\rm X}/N_{\rm Y})_{\rm \odot}.$} ($\log \epsilon({\rm Fe})$) values and LEP. Accordingly, the effective temperature value for the star was obtained under the condition that the calculated abundance values for Fe\,{\sc i} were independent of the LEP values (i.e., spectroscopic excitation technique\footnote{It is sensitive to neutral spectral lines with a broad range of excitation potentials.}). To determine the microturbulence parameter, the elemental abundance values measured for neutral and single ionized atoms (i.e., Fe\,{\sc i} and Fe\,{\sc ii}) of the selected elements should be independent of their EW values (i.e., the LTE principle). For our sample of Fe\,{\sc i} lines, these two conditions were imposed simultaneously (Figure~\ref{fig:3}a,b). Microturbulence can also be determined separately from Ti\,{\sc i}, Cr\,{\sc i}, Fe\,{\sc i}, and Fe\,{\sc ii}. For a given model, we computed the dispersion in the Ti, Cr, and Fe abundances over a range in $\xi$ from $\approx$0 to 3.0 km s$^{\rm -1}$ for HD\,76151 and the Sun. Figure \ref{fig:micro} shows the dispersion ($\sigma$) of Ti\,{\sc i}, Cr\,{\sc i}, Fe\,{\sc i}, and Fe\,{\sc ii}. The dispersion test applied to HD\,76151 (right panel in Figure~\ref{fig:micro}) using Fe\,{\sc i} lines yielded a micro-turbulent velocity of $\approx$0.8 km s$^{\rm -1}$. This method also provided a value of 1.0 km s$^{\rm -1}$ for both the Ti\,{\sc i} and Cr\,{\sc i} lines. The Fe\,{\sc ii} yielded a $\xi$ ranging from 0.5 to 1.0 km s$^{\rm -1}$. When both methods are evaluated together, the measurement uncertainty for HD\,76151 in the HARPS spectrum was estimated to be $\approx$0.5 km s$^{\rm -1}$. In particular, the dependence of $\xi$ on the neutral iron lines in the IGRINS spectrum weakens beyond 1.0 km s$^{\rm -1}$ and remains constant after 1.5 km s$^{\rm -1}$. The $\xi$ value that minimizes the slope via the LTE principle method in the IGRINS spectrum is 2.1 km s$^{\rm -1}$, with an estimated error margin of 0.5 km s$^{\rm -1}$. The $\xi$ derived via optical+NIR line list for HD\,76151 ranges from 0.0 to 0.5 km s$^{\rm -1}$ for neutral Fe lines and 0.5 to 0.8 km s$^{\rm -1}$ for ionized Fe lines (right panel in Figure~\ref{fig:micro}). The micro-turbulent velocity that minimized the slope in the LTE principle method from the optical+NIR line list was 0.3 km s$^{\rm -1}$. Thus, the microturbulent velocity obtained using the optical+IR list for the star was 0.3$\pm$0.5 km s$^{\rm -1}$. 

To determine the surface gravity ($\log g$), we analyzed the iron abundances calculated using MOOG. This analysis ensured the presence of ionization equilibrium, a state where Fe\,{\sc i} and Fe\,{\sc ii} lines yield identical iron abundances. Subsequently, the metallicity ([Fe/H]) was iteratively refined until the derived iron abundance converged with the abundance initially assumed for constructing the model atmosphere. 

The uncertainty in the surface temperature derived originates from the error associated with the slope of the relationship between Fe\,{\sc i} abundance and LEPs of the lines. In the case of the HARPS spectrum of HD\,76151, a significant change in this slope was observed with a temperature variation of $\pm$88 K in the assumed model atmosphere (Figure~\ref{fig:3}a). Additionally, a 1$\sigma$ difference in the abundance ratio [X/H] between the neutral and ionized iron lines translates into an approximate change of $\approx$0.2 dex in the $\log g$ of the star from its HARPS spectrum. For the IGRINS spectrum of the star, the same method indicates the errors reported in Table~\ref{tab:2} where the model parameters obtained by the line analysis for the HARPS and IGRINS spectra of HD\,76151 are listed. 

\begin{figure}
    \centering
    \includegraphics[width=0.49\textwidth]{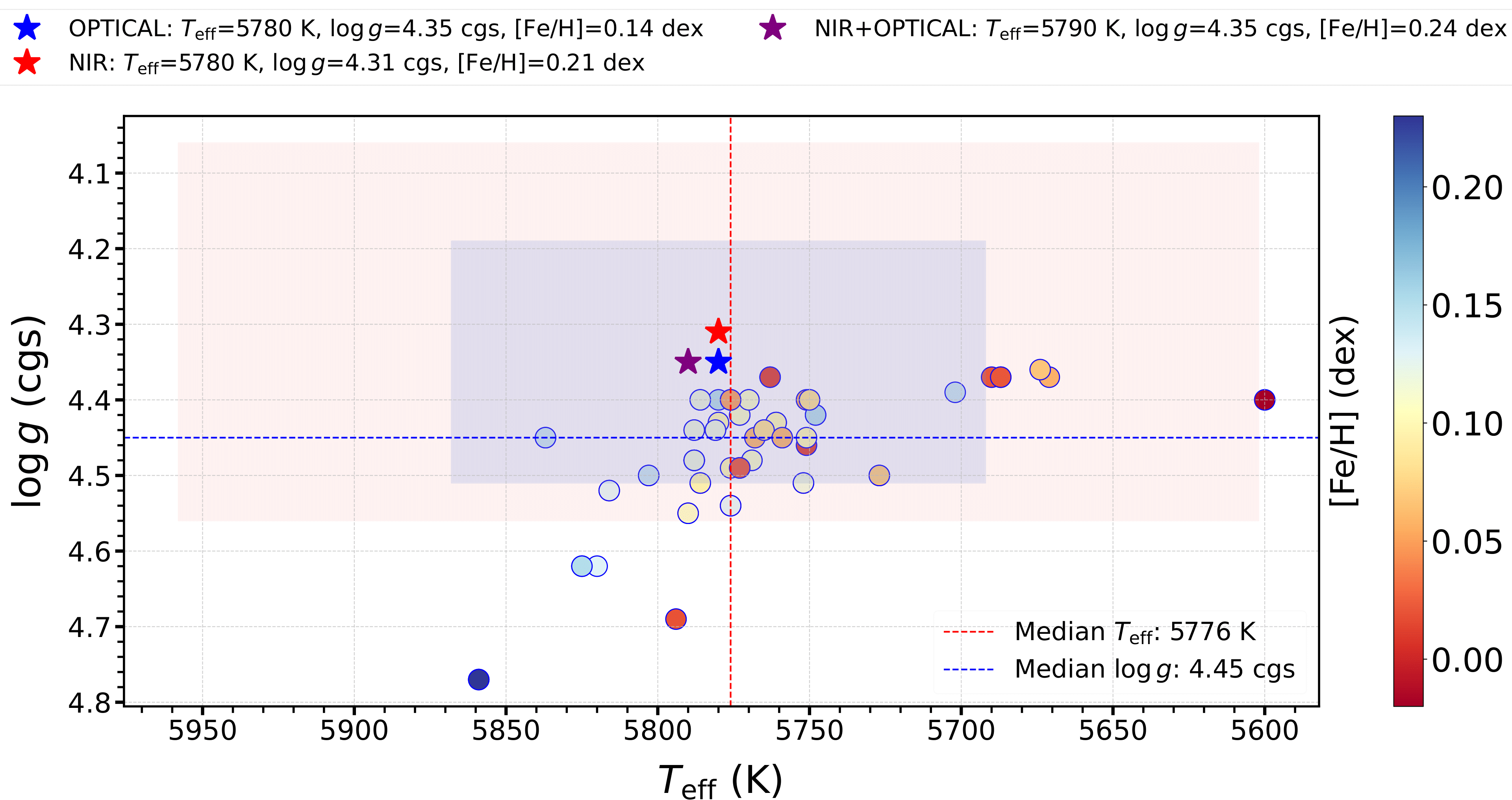}
    \caption{The faint blue area in the image represents the errors in the model parameters ($\Delta T_{\rm eff}=$ 88 K, $\Delta \log g=$ 0.16 cgs) obtained from the HARPS spectrum of HD\,76151 in the optical region (blue star symbol), while the faint red area represents the errors in the model parameters ($\Delta T_{\rm eff}=$ 178 K, $\Delta \log g=$ 0.25 cgs) obtained from the IGRINS spectrum of HD\,76151 in the infrared (red star symbol). The model parameters from both optical and infrared are presented with a star symbol in purple.}
    \label{fig:model_param}
\end{figure}

The model parameters obtained in this study for the KPNO-NIR and KPNO-OPTICAL solar spectra are presented in Table \ref{tab:2}. The stellar parameters reported in the literature for the star exhibits large variations (Figure \ref{fig:model_param}). The model parameter analyses were performed in the optical and NIR regions, which encompass all the identified lines from the HARPS and IGRINS spectra of HD\,76151, indicate stellar atmospheric parameters that are in good agreement with those obtained by spectroscopic analyses in this study. The same is true for Sun (labeled OPTICAL+NIR in Table \ref{tab:2}). Figure \ref{fig:model_param} displays the graphical representations of the data presented in Table \ref{tab:A8} which depicts $T_{\rm eff}$, $\log~g$, [Fe/H], $\alpha$-element abundances (i.e., [Mg/Fe], [Si/Fe], [Ca/Fe], and [Ti/Fe]), radius ($R$), mass ($M$), and age ($t$) reported in the literature, and in this study for HD\,76151 with their uncertainties.

Errors in the effective temperatures from the spectroscopic excitation technique may be due to systematic errors in oscillator strength as a function of the excitation potential. The same applies to errors in the EWs. To decrease systematic effects, we used a differential analysis technique for a star (the Sun) with known parameters (\citealp[]{Fulbright2006,Koch2008}). 

Accurate atomic data and stellar model atmosphere are essential for success (\citealp[]{Barklem2002}). The selection of damping parameters can also significantly impact the derived atmospheric parameters, particularly for elements with intricate line profiles. Future investigations may benefit from the development of more sophisticated damping parameter models. In this study, we searched for damping constants (i.e., van der Waals) for 273 lines. 237 lines of them have listed the damping constants in the Kurucz database. An alternative search with the VALD database provided damping constants for 20 lines. For the remaining 16 lines, the damping constants were obtained using the unsold approximation.

When determining model atmospheric parameters, it is often difficult to avoid degeneracies and covariance, particularly between $T_{\rm eff}$ and $\log g$ (\citealp[]{McWilliam1995}). To check the probable  degeneracy between the model atmospheric parameters for spectral analyses in the $H$- and $K$-bands, we employed the line depth ratio method. The line pairs reported by \citet{Fukue2015} were used for temperature determination. In the current study, nine pairs of lines were identified for temperature determination using line depth ratios in both the KPNO-NIR solar spectrum and IGRINS $H$-band spectra of HD\,76151. Following the testing of line pairs\footnote{Line pairs are reported in Table 4 of \citet{Fukue2015}.} of 2, 3, 6, and 9, the effective temperature obtained for the IR spectrum of the Sun was found to be $T_{\rm eff}=5685\pm99$ K. This value is consistent with the spectroscopic temperature obtained for the Sun using the KPNO-NIR solar spectrum in this study, $T_{\rm eff}=5780\pm55$ K. The same pair of lines on the IGRINS spectrum of HD\,76151 provided $T_{\rm eff}=5690\pm 38$ K. The spectroscopic temperature derived from the IGRINS spectrum of the star is $T_{\rm eff}=5780\pm 178$ K. Among the lines used for the line ratios, the new line list contains the Fe\,{\sc i} transitions at 15\,194.50 \AA\, and 15\,207.54 \AA. The latter transition was added with the RMT analysis applied to Andreasen's line list (see Section 2 for details of the RMT analysis). As can be seen, the effective temperature values obtained by the line-depth ratio method are consistent with the spectroscopic results obtained from the infrared spectra for both the Sun and HD\,76151. 

The abundance results may be influenced by various systematic uncertainties, including those related to the treatment of convection and potential non-LTE effects on the formation of atomic lines. To explore the potential systematic uncertainties affecting our abundance results, we included detailed information on convection and non-LTE corrections in Appendix B. This section provides in-depth discussions and analyses of these factors, including their potential impact on the derived elemental abundances. 

\begin{table}
\tiny
\centering
\setlength{\tabcolsep}{2.95pt}
\caption{The abundances of the observed species for HD\,76151 (HARPS spectrum) are presented. Simultaneously, the solar abundances obtained from the KPNO-OPTICAL solar spectrum in this study (TS) and those obtained by \citet{Asplund2009}[ASP] are provided. Bold values indicate elemental abundances measured in the KPNO-NIR solar spectrum and IGRINS spectrum of HD\,76151. The abundances determined using a combined optical and NIR line list are indicated in blue.}  
\label{tab:3}
    \centering
    \begin{tabular}{l|c|c|c|c|c|c|c}
    \hline
        & \multicolumn{3}{c|}{\bf HD\,76151} & \multicolumn{4}{c}{\bf Sun}\\
    \cline{2-8}  
    \bf Species   &   [X/Fe] & $\sigma$ &  $N$  &  $\log \epsilon_{\odot}$(X$_{\rm TS}$) &  $N$     & $\log \epsilon_{\odot}$(X$_{\rm ASP}$)  & $\Delta \log \epsilon_{\rm \odot}(X)$ \\
    \cline{2-2}  
    \cline{5-5}
    \cline{7-7}
     &  (dex)  &   &  & (dex) &   & (dex)   & \\
\cline{1-8}
Na\,{\sc i}      & 0.04 & 0.11 & 1 & 6.16 $\pm$ 0.07 & 2 & 6.24 $\pm$ 0.04 & -0.08 \\ 

Mg\,{\sc i}*      & 0.00 & 0.08 & 1 & 7.58 $\pm$ 0.00 & 2 & 7.60 $\pm$ 0.04 & -0.02 \\ 
\bf Mg\,{\sc i}*  & \bf - & \bf - & \bf 0 & \bf 7.49 $\pm$ 0.01 & \bf 2 &  7.60 $\pm$ 0.04 &  -0.11 \\ 
 Mg\,{\sc i}*  & - & - & 0 & 7.52 $\pm$ 0.06 & 4 & 7.60 $\pm$ 0.04 & -0.08 \\

Si\,{\sc i}      & 0.03 & 0.13 & 10 & 7.50 $\pm$ 0.07 & 12 & 7.51 $\pm$ 0.03 & -0.01 \\ 
 \bf Si\,{\sc i} & \bf -0.01 & \bf 0.20 & \bf 1 & \bf 7.49 $\pm$ 0.11 & \bf 28 &  7.51 $\pm$ 0.03 &  -0.02 \\ 
 Si\,{\sc i} &  -0.07 &  0.16 &  11 &  7.50 $\pm$ 0.01 &  40 &  7.51 $\pm$ 0.03 & -0.01 \\ 
 
Ca\,{\sc i}      & 0.00 & 0.14 & 16 & 6.34 $\pm$ 0.08 & 18 & 6.34 $\pm$ 0.04 & 0.00 \\ 
 \bf Ca\,{\sc i} & \bf - & \bf - & \bf 0 & \bf 6.29 $\pm$ 0.11 & \bf 11 &  6.34 $\pm$ 0.04 & \bf -0.05 \\
Ca\,{\sc i} &  - &  - &  0 &  6.29 $\pm$ 0.09 &  29 &  6.34 $\pm$ 0.04 &  -0.05 \\

 \bf Ca\,{\sc ii} & \bf -0.05 & \bf 0.17 & \bf 1 & \bf 6.22 $\pm$ 0.00 & \bf 1 &  6.34 $\pm$ 0.04 &  -0.12 \\ 
Ca\,{\sc ii} &  -0.11 &  0.10 &  1 &  6.23 $\pm$ 0.00 &  1 &  6.34 $\pm$ 0.04 &  -0.11 \\ 
 
Sc\,{\sc ii} & 0.03 & 0.17 & 3 & 3.23 $\pm$ 0.08 & 7 & 3.15 $\pm$ 0.04 & 0.08 \\ 

Ti\,{\sc i} & -0.04 & 0.16 & 36 & 4.96 $\pm$ 0.09 & 43 & 4.95 $\pm$ 0.05 & 0.01 \\ 
 \bf Ti\,{\sc i} & \bf 0.12 & \bf 0.21 & \bf 1 & \bf 4.91 $\pm$ 0.12 & \bf 10 &  4.95 $\pm$ 0.05 &  -0.04 \\ 
Ti\,{\sc i} &  -0.05 &  0.18 &  37 &  4.91 $\pm$ 0.09 &  53 &  4.95 $\pm$ 0.05 &  -0.04 \\ 

Ti\,{\sc ii} & -0.02 & 0.14 & 9 & 4.99 $\pm$ 0.08 & 12 & 4.95 $\pm$ 0.05 & 0.04 \\ 
 \bf Ti\,{\sc ii}* & \bf - & \bf - & \bf 0 & \bf 4.82 $\pm$ 0.00 & \bf 1 &   4.95 $\pm$ 0.05 &  -0.13 \\
Ti\,{\sc ii} &  - &  - &  0 &  4.95 $\pm$ 0.08 &  13 &   4.95 $\pm$ 0.05 &  0.00 \\
 
V\,{\sc i}* & -0.03 & 0.09 & 4 & 3.90 $\pm$ 0.03 & 5 & 3.93 $\pm$ 0.08 & -0.03 \\ 
Cr\,{\sc i} & 0.00 & 0.14 & 16 & 5.71 $\pm$ 0.07 & 19 & 5.64 $\pm$ 0.04 & 0.07 \\ 
Cr\,{\sc ii} & 0.03 & 0.17 & 2 & 5.64 $\pm$ 0.14 & 3 & 5.64 $\pm$ 0.04 & 0.00 \\ 
Mn\,{\sc i} & -0.01 & 0.20 & 9 & 5.62 $\pm$ 0.13 & 13 & 5.43 $\pm$ 0.05 & 0.19 \\ 

Fe\,{\sc i} & -0.02 & 0.15 & 90 & 7.54 $\pm$ 0.09 & 132 & 7.50 $\pm$ 0.04 & 0.04 \\ 
 \bf Fe\,{\sc i} & \bf 0.06 & \bf 0.20 & \bf 74 & \bf 7.46 $\pm$ 0.07 & \bf 204 &  7.50 $\pm$ 0.04 &  -0.04 \\ 
Fe\,{\sc i} &  0.02 &  0.17 &  164 &  7.48 $\pm$ 0.08 & 336 &  7.50 $\pm$ 0.04 &  -0.02 \\
 
Fe\,{\sc ii} & 0.00 & 0.11 & 14 & 7.51 $\pm$ 0.04 & 17 & 7.50 $\pm$ 0.04 & 0.01 \\ 
 \bf Fe\,{\sc ii} & \bf 0.00 & \bf 0.24 & \bf 2 & \bf 7.50 $\pm$ 0.02 & \bf 10 & 7.50 $\pm$ 0.04 &  0.00 \\
Fe\,{\sc ii} & 0.00 &  0.15 &  16 &  7.47 $\pm$ 0.05 & 27 &  7.50 $\pm$ 0.04 & -0.03 \\
 
Co\,{\sc i}* & 0.04 & 0.21 & 5 & 4.90 $\pm$ 0.14 & 6 & 4.99 $\pm$ 0.07 & -0.09 \\ 
Ni\,{\sc i} & -0.01 & 0.15 & 45 & 6.28 $\pm$ 0.09 & 54 & 6.22 $\pm$ 0.04 & 0.06 \\ 
Zn\,{\sc i}*& 0.17 & 0.08 & 1 & 4.60 $\pm$ 0.00 & 2 & 4.56 $\pm$ 0.05 & 0.04 \\ 
Sr\,{\sc i}* & 0.10 & 0.08 & 1 & 2.86 $\pm$ 0.00 & 1 & 2.87 $\pm$ 0.07 & -0.01 \\ 
Y\,{\sc ii}* & 0.06 & 0.13 & 2 & 2.20 $\pm$ 0.04 & 2 & 2.21 $\pm$ 0.05 & -0.01 \\ 
Zr\,{\sc ii}* & -0.01 & 0.08 & 1 & 2.70 $\pm$ 0.00 & 1 & 2.58 $\pm$ 0.04 & 0.12 \\ 
Ba\,{\sc ii}* & -0.19 & 0.21 & 3 & 2.52 $\pm$ 0.19 & 4 & 2.18 $\pm$ 0.09 & 0.34 \\ 
Ce\,{\sc ii}* & 0.04 & 0.08 & 1 & 1.58 $\pm$ 0.02 & 2 & 1.58 $\pm$ 0.04 & 0.00 \\ 
Nd\,{\sc ii}* & -0.13 & 0.12 & 1 & 1.33 $\pm$ 0.09 & 3 & 1.42 $\pm$ 0.04 & -0.09 \\ \hline
    \end{tabular}
\vspace{1ex}
     \footnotesize{$\Delta \log \epsilon_{\odot}(X)=\log \epsilon_{\odot}(X_{\rm TS}) - \log \epsilon_{\odot}(X_{\rm ASP})$}\\  
     (*) The abundance was determined using the spectrum synthesis method.
 \end{table}
 
\begin{table}
\tiny
\centering
\setlength{\tabcolsep}{1.40pt}
\caption{Sensitivity of the derived abundances to uncertainties in the model atmospheric parameters from the KPNO-OPTICAL solar spectrum and the HARPS spectrum of HD\,76151 (upper section). Results from the KPNO-NIR solar spectrum and IGRINS spectrum of HD\,76151 (middle section). The abundance analysis results derived from atomic transitions in both optical and NIR wavelength regions (bottom section).}
\label{tab:4}
\centering
\begin{tabular}{l|c|c|c|c|cc|c|c|c}
\hline
\multicolumn{10}{c}{$\Delta \log \epsilon$ (OPTICAL)} \\
\cline{1-10}
\multicolumn{5}{c|}{Sun} & \multicolumn{5}{c}{HD\,76151} \\
\cline{1-10}
\multicolumn{5}{c|}{~~~~~~5790~~~~~~4.40~~~~~~0.00~~~~~~~0.66} & \multicolumn{5}{c}{~5780~~~~~~4.35~~~~~~0.14~~~~~~~0.69} \\
\cline{1-10}
Species & $\Delta T_{\rm eff}$	&	$\Delta \log g$ 	&	$\Delta$[Fe/H] &  $\Delta \xi$ & & $\Delta T_{\rm eff}$	&	$\Delta \log g$ 	&	$\Delta$[Fe/H] &  $\Delta \xi$ \\
\cline{2-10}
 ~    & (+45) & (+0.09) & (+0.04) & (+0.50)&   &(+88) & (+0.16) & (+0.08) & (+0.50) \\
  \hline
Na\,{\sc i}   & +0.03  & -0.02  & +0.00  & -0.06  & ~ & +0.05   & -0.06  & +0.02 & -0.04 \\
Mg\,{\sc i}*  & +0.05  & +0.04  & +0.00  & +0.02  & ~ &  +0.18  & +0.03  & +0.00 & -0.07 \\
Si\,{\sc i}   & +0.01  & +0.01  & +0.01  & -0.03  & ~ &  +0.02  & +0.00  & +0.01 & -0.05  \\
Ca\,{\sc i}   & +0.03  & -0.02  & +0.00  & -0.11  & ~ &  +0.07  & -0.03  & +0.01 & -0.10  \\
Sc\,{\sc ii}  & +0.01  & +0.03  & +0.02  & -0.07  & ~ &  +0.00  & +0.05  & +0.03 & -0.15 \\
Ti\,{\sc i}   & +0.05  & +0.00  & +0.00  & -0.12  & ~ & +0.09   & -0.01  & +0.00 & -0.14 \\
 Ti\,{\sc ii} & +0.01  & +0.03  & +0.02  & -0.06  & ~ & +0.01   & +0.06  & +0.03 & -0.10 \\
 V\,{\sc i}*  & +0.09  & +0.05  & +0.00  & +0.02  & ~ &  +0.10  & +0.02  & +0.00 &  0.00 \\
 Cr\,{\sc i}  & +0.04  & -0.01  & +0.00  & -0.16  & ~ &  +0.07  & -0.03  & +0.00 & -0.17 \\
 Cr\,{\sc ii} & -0.01  & +0.02  & +0.01  & -0.10  & ~ &  -0.01  & +0.04  & +0.02 & -0.17 \\
 Mn\,{\sc i}  & +0.04  & -0.01  & +0.01  & -0.15  & ~ &  +0.08  & -0.01  & +0.01 & -0.18 \\
 Fe\,{\sc i}  & +0.04  & -0.01  & +0.01  & -0.14  & ~ & +0.07   & -0.02  & +0.00 & -0.16 \\ 
 Fe\,{\sc ii} & +0.00  & +0.03  & +0.02  & -0.12  & ~ & -0.02   & +0.05  & +0.03 & -0.16 \\
 Co\,{\sc i}* & +0.12  & +0.05  & +0.00  & -0.02  & ~ & +0.02   & -0.08  & +0.01 & -0.07 \\
 Ni\,{\sc i}  & +0.02  & -0.01  & +0.00  & -0.09  & ~ &  +0.05  & -0.01  & +0.01 & -0.11 \\
 Zn\,{\sc i}* & +0.02  & +0.03  & +0.01  & -0.12  & ~ & +0.05   & +0.05  & +0.03 & -0.20 \\
 Sr\,{\sc i}* & +0.05  & -0.02  & -0.01  & -0.17  & ~ &  +0.06  & -0.05  & +0.00 & -0.23 \\
 Y\,{\sc ii}* & +0.06  & +0.03  & +0.01  & -0.17  & ~ & +0.04   & +0.06  & +0.03 & -0.14 \\
 Zr\,{\sc ii}* & +0.02 & +0.05  & +0.01  & -0.18  & ~ & +0.05   & -0.02  & +0.03 & -0.30 \\
 Ba\,{\sc ii}* & +0.07 & +0.06  & +0.03  & -0.11  & ~ &  +0.03  & -0.01  & +0.05 & -0.11 \\
 Ce\,{\sc ii}* & +0.01 & +0.03  & +0.01  & -0.06  & ~ &  +0.00  & +0.05  & +0.03 & -0.10 \\
 Nd\,{\sc ii}* & +0.00 & +0.07  & +0.02  & +0.01  & ~ & +0.01   & +0.06  & +0.03 & -0.02\\ 
\hline
\multicolumn{10}{c}{$\Delta \log \epsilon$ (NIR)} \\
\cline{1-10}
\multicolumn{5}{c|}{Sun} & \multicolumn{5}{c}{HD\,76151} \\
\cline{1-5}
\cline{6-10}
\multicolumn{5}{c|}{~~~~~~5780~~~~~~4.40~~~~~~0.00~~~~~~~1.08} & \multicolumn{5}{c}{~5780~~~~4.31~~~~~~0.21~~~~~~~2.10} \\
\cline{1-10}
	Species& $\Delta T_{\rm eff}$	&	$\Delta \log g$ 	&	$\Delta$[Fe/H] &  $\Delta \xi$ & & $\Delta T_{\rm eff}$	&	$\Delta \log g$ 	&	$\Delta$[Fe/H] &  $\Delta \xi$ \\
\cline{2-10}
     & (+55) & (+0.22) & (+0.03) & (+0.50)&   &(+178) & (+0.25) & (+0.09) & (+0.50) \\
  \hline
 Si\,{\sc i} & +0.01  & +0.00  & +0.01  & -0.02  & ~ & +0.05  & +0.03  & +0.01 & -0.00 \\ 
 Ca\,{\sc i} & +0.02  & -0.01  & +0.00  & -0.03  & ~ &  ..  & ..  & .. & .. \\
 Ti\,{\sc i} &  +0.05 & +0.00  & +0.00  & -0.01  & ~ & +0.16  & +0.00  & +0.00 & -0.01 \\
 Ti\,{\sc ii} & +0.01  & +0.10  & +0.01  & -0.01  & ~ & ..  & ..  & .. & .. \\ 
 Fe\,{\sc i} & +0.03  & +0.00  & +0.00  & -0.02  & ~ &  +0.08  & +0.00  & +0.01 & -0.02 \\ 
 Fe\,{\sc ii} & -0.02  & +0.08  & +0.00  & -0.01  & ~ &  -0.05  & +0.10  & -0.02 & +0.08 \\
 
\hline
\multicolumn{10}{c}{$\Delta \log \epsilon$ (OPTICAL+NIR)} \\
\cline{1-10}
\multicolumn{5}{c|}{Sun} & \multicolumn{5}{c}{HD\,76151} \\
\cline{1-10}
\multicolumn{5}{c|}{~~~~~~5790~~~~~~4.44~~~~~~0.00~~~~~~~0.90} & \multicolumn{5}{c}{~5790~~~~4.35~~~~~~0.24~~~~~~~0.30} \\
\cline{1-10}
	Species& $\Delta T_{\rm eff}$	&	$\Delta \log g$ 	&	$\Delta$[Fe/H] &  $\Delta \xi$ & & $\Delta T_{\rm eff}$	&	$\Delta \log g$ 	&	$\Delta$[Fe/H] &  $\Delta \xi$ \\
\cline{2-10}
     & (+78) & (+0.14) & (+0.08) & (+0.50)&   &(+170) & (+0.18) & (+0.12) & (+0.50) \\
  \hline
Na\,{\sc i} & +0.05  & -0.04  & +0.01  & -0.04  & ~ & +0.09  & -0.09  & +0.03 & -0.03 \\

Mg\,{\sc i}* & -0.01  & +0.00  & +0.00  & -0.04  & ~ & +0.15  & -0.03  & +0.01  & +0.03 \\

Si\,{\sc i} & +0.01  & +0.00  & +0.01  & -0.03  & ~ &  +0.03   & +0.00   & +0.01  & -0.04  \\

Ca\,{\sc i} & +0.05  & -0.02  & +0.01  & -0.07  & ~ &  +0.11  & -0.06  & +0.01  & -0.09  \\

Sc\,{\sc ii} & +0.00  & +0.05  & +0.02  & -0.11  & ~ &  -0.01   & +0.08   & +0.03  & -0.13 \\

Ti\,{\sc i} & +0.07  & -0.01  & -0.01  & -0.10  & ~ &  +0.16   & -0.03   & +0.01  & -0.12 \\
 
Ti\,{\sc ii} & +0.00  & +0.05  & +0.02  & -0.10  & ~ &  -0.01   & +0.08  & +0.03 & -0.10 \\
 
V\,{\sc i}* & +0.04  & +0.02  & -0.01  & +0.03  & ~ &  +0.07  &  0.00  & +0.01 &  0.02 \\
 
 Cr\,{\sc i} & +0.07  & -0.03  & +0.00  & -0.17  & ~ &  +0.14  & -0.05  & +0.02 & -0.15 \\
 
Cr\,{\sc ii} & -0.02  & +0.05  & +0.02  & -0.12  & ~ & -0.05  & +0.06  & +0.03 & -0.15 \\
 
 Mn\,{\sc i} & +0.06  & -0.03  & +0.00  & -0.17  & ~ &  +0.13  & -0.04  & +0.02 & -0.16 \\
 
 Fe\,{\sc i} & +0.04  & -0.02  & +0.00  & -0.07  & ~ &  +0.10  & -0.03  & +0.02 & -0.09 \\ 
 
 Fe\,{\sc ii} &-0.02  & +0.06  & +0.02  & -0.09  & ~ &  -0.06  & +0.07  & +0.03 & -0.12 \\
 
Co\,{\sc i}* & +0.03  & +0.03  & +0.01  &  0.00  & ~ &  +0.09  & -0.02  & +0.02 & +0.03 \\
 
Ni\,{\sc i} & +0.05  & -0.01  & +0.01  & -0.08  & ~ &  +0.09  & -0.02 & +0.02 & -0.09 \\
 
 Zn\,{\sc i}* & +0.01  & 0.00  & +0.02  & -0.10  & ~ &  -0.06  & -0.08 & +0.04 & -0.07 \\
 
 Sr\,{\sc i}* & 0.00  & -0.01  & -0.01  & -0.08  & ~ &  +0.06  & +0.01  & +0.01 & -0.06 \\
 
 Y\,{\sc ii}* & +0.03  & +0.05  & +0.02  & -0.11  & ~ &  +0.06  & +0.08  & +0.03 &  0.00 \\
 
 Zr\,{\sc ii}* & -0.02  & -0.03  & +0.02  & -0.08  & ~ &  +0.04  & +0.06  & +0.03 & +0.05 \\
 
 Ba\,{\sc ii}* & +0.00  & +0.02  & +0.04  & -0.10  & ~ &  +0.03  & +0.00  & +0.05 & -0.07 \\
 
 Ce\,{\sc ii}* & -0.01  & +0.05  & +0.03  & -0.04  & ~ &  +0.02  & -0.02  & +0.04 & -0.08 \\
 
 Nd\,{\sc ii}* & 0.00  & +0.06  & +0.02  & +0.02  & ~ &  +0.05  & +0.07  & +0.04 &  +0.03\\ 
 
\hline
\end{tabular}
\vspace{1ex}
     (*) The error in abundance was determined using the spectrum synthesis method.
\end{table}

The abundances obtained for HD\,76151 using the HARPS and IGRINS spectra and for the solar photosphere using the KPNO-OPTICAL and KPNO-NIR solar spectra are listed in Table \ref{tab:3}. The Fe and $\alpha$-element abundances obtained for the star by independent model atmosphere analyses in the optical and NIR regions in this study were surprisingly consistent. The same was true for the KPNO-OPTICAL and KPNO-NIR spectra. The solar abundances from \citet{Asplund2009} are also included for comparison. In Table \ref{tab:3}, $\log \epsilon$ is the logarithm of abundance. The number of lines used in this analysis has also been provided. The errors reported in log$\epsilon$ abundances are represented by the 1$\sigma$ line-to-line scatter in abundance. [X/Fe] is the logarithmic abundance considering the Fe\,{\sc ii} abundance. The error in [X/Fe] is the square root of the sum of the squares of the errors in [X/H]\footnote{[X/H] is the logarithmic abundance ratio of hydrogen to the corresponding solar value.} and [Fe/H]. The formal errors in abundances resulting from uncertainties in the atmospheric parameters $T_{\rm eff}$, $\log g$ and $v_{\rm t}$ are summarized in Table \ref{tab:4} for changes relative to the model.

\section{Results and Discussion \label{sec:info}}

The subsequent section presents an evaluation of atomic line libraries and catalogs. This evaluation serves as a guide for the atomic line identification and selection criteria employed to create our new line list. The lines marked as doubtful in APOGEE DR17 \citep{DR17} are also under scrutiny. These include both neutral and ionized iron lines, as well as newly identified ionized iron lines in the $H$-band (Table \ref{tab:A1}). In addition, this section presents the key astrophysical parameters of HD\,76151 (Table \ref{tab:summary}). These values were compared with those reported in the literature (Table \ref{tab:A8}). Furthermore, this section provides a comprehensive characterization of HD\,76151 by determining its space velocity components, Galactic orbital parameters, birthplace, and current position within the Galaxy.

\subsection{Evolutionary status of HD\,76151}

To reveal the evolutionary situation, this study considered four sets of stellar parameters determined by the spectral method and analyzed the state of the star in the Kiel diagram using the PARSEC isochrone. Based on this, the age of the star was calculated. The isochrone fitting method also allowed us to obtain the fundamental astrophysical parameters of the star. For this purpose, the fundamental astrophysical parameters calculated for the stellar parameters obtained from the spectral analyses performed using the HARPS and IGRINS spectra of the star, as well as the stellar parameters obtained from the spectral analysis performed by combining the optical and NIR line lists are presented in Table \ref{tab:summary}. 

The metallicity ([Fe/H]) of the star was transformed into the mass fraction $Z$ to derive the age of the star using the equation given for {\sc PARSEC} isochrones \citep{Bressan2012} by Bovy\footnote{\url{https://github.com/jobovy/isodist/blob/master/isodist/Isochrone.py}}. 

\begin{equation}
\label{equ:1}
z = \frac{(z_{\rm x} - 0.2485 \times z_{\rm x})}{(2.78 \times z_{\rm x} + 1)}
\end{equation}

\begin{equation}
\label{equ:2}
z_{\rm x}=10^{\large \left [{\rm[ Fe/H]}+\log\large\left(\frac{z_{\odot}}{1-0.248-2.78\times z_{\odot}}\large \right)\large \right]}%
\end{equation}
Here, $z$ and $z_{\rm x}$ refer to elements heavier than helium and the intermediate operational functions, respectively. $z_{\rm \odot}$ is the solar metallicity which was adopted as 0.0154 \citep{Bressan2012}. We calculated $z$ for the four sets of stellar parameters for HD\,76151 (Table \ref{tab:summary}). A comparison of the model atmospheric parameters calculated for HD\,76151 with the {\sc PARSEC} isochrones resulted in a stellar mass of $1.053_{-0.068}^{+0.056} M_{\odot}$, radius $1.125_{-0.011}^{+0.035} R_{\odot}$, and an age of 5.5$^{\rm +2.5}_{\rm -2.1}$ Gyr for the stellar parameters obtained by combining the optical and NIR line lists. Moreover, the mass, radius, luminosity, and age for HD\,76151 calculated from the HARPS optical spectra are $M=1.015_{-0.039}^{+0.034}M_{\odot}$, $R=1.125_{-0.007}^{+0.008}R_{\odot}$, $\log L=0.105_{-0.055}^{+0.019}L_{\odot}$, and $t=6.8^{\rm +1.5}_{\rm -1.7}$ Gyr, respectively, while the fundamental astrophysical parameters for HD\,76151 from IGRINS infrared spectra are $M=1.046_{-0.076}^{+0.062}M_{\odot}$, $R=1.190_{-0.014}^{+0.009}R_{\odot}$, $\log L=0.154_{-0.109}^{+0.053}L_{\odot}$, and $t=6.8^{\rm +2.8}_{\rm -2.4}$ Gyr, respectively. Optical and NIR spectral analyses yielded an age of 6.8 Gyr. The stellar ages calculated using different sets of spectroscopic parameters were consistent within uncertainties. 

\begin{table*}[htp!]
\small
\setlength{\tabcolsep}{5pt}
\renewcommand{\arraystretch}{0.95}
\centering
\caption{Fundamental astrophysical parameters and kinematic and dynamic orbital parameters of HD\,76151 calculated from photometric, astrometric and spectroscopic data.}
\label{tab:summary}
\begin{tabular}{l|c|c|c|c}
\hline
Parameter                       & \multicolumn{4}{c}{Value} \\
\hline
($\alpha,~\delta)_{\rm J2000}$ (Sexagesimal) & \multicolumn{4}{c}{08:54:17.95, -05:26:04.05} \\
$(l, b)_{\rm J2000}$ (Decimal)               & \multicolumn{4}{c}{233.205785, +24.164218} \\	
($\mu_{\rm \alpha}\cos\delta$, $\mu_{\rm \delta}$) (mas yr$^{-1}$)& \multicolumn{4}{c}{-413.648$\pm$0.036, 30.619$\pm$0.026} \\
$\varpi$ (mas)	                & \multicolumn{4}{c}{59.3595$\pm$0.0408} \\
$d$ (pc)	                    & \multicolumn{4}{c}{16.85$\pm$0.01} \\
Spectral type	                & \multicolumn{4}{c}{G2V} \\
$V$ (mag)	                    & \multicolumn{4}{c}{6.00} \\
$U-B$ (mag)	                    & \multicolumn{4}{c}{0.22} \\
$B-V$ (mag)	                    & \multicolumn{4}{c}{0.67} \\
$E(B-V)$ (mag)                  & \multicolumn{4}{c}{0.00} \\
$G$ (mag)	                    & \multicolumn{4}{c}{5.847$\pm$0.003} \\
$G_{\rm BP}$-G$_{\rm RP}$ (mag)	& \multicolumn{4}{c}{0.825$\pm$0.005} \\
$M_{\rm G}$ (mag)               & \multicolumn{4}{c}{4.714$\pm$0.003} \\
\hline
                                & HARPS                      & IGRINS                       & HARPS+IGRINS                   & {\it Gaia} DR3\\
\hline
$T_{\rm eff}$ (K)               & 5780$\pm$88	             &	5780$\pm$178	            &  5790$\pm$170	                 & 5825$\pm$10    \\
$\log g$ (cgs)	                & 4.35$\pm$0.16              &	4.31$\pm$0.25	            &  4.35$\pm$0.18	             & 4.44$\pm$0.04  \\
{\rm [Fe/H]} (dex)	            & 0.14$\pm$0.08	             &	0.21$\pm$0.13	            &  0.24$\pm$0.09	             & -0.05$\pm$0.05 \\
$Z$	                            & 0.02054	                 &	0.02382       	            &  0.02536	                     & 0.01363	      \\
$\xi$ (km~s$^{-1}$) 	        & 0.69$\pm$0.50	             &	2.10$\pm$0.50	            &  0.30$\pm$0.50	             &	---	          \\
$V_{\rm R}$ (km~s$^{-1}$)       & 32.05$\pm$0.05	                     &	32.08	                    &   ---	                         & 31.99$\pm$0.12  \\
$M$ ($M_{\rm \odot}$)           & $1.015_{-0.039}^{+0.034}$	 &	$1.046_{-0.076}^{+0.062}$	& $1.053_{-0.068}^{+0.056}$	     & $0.968_{-0.015}^{+0.002}$   \\
$R$ ($R_{\rm \odot}$)           & $1.125_{-0.007}^{+0.008}$	 &	$1.190_{-0.014}^{+0.009}$	& $1.125_{-0.011}^{+0.035}$	     & $0.977_{-0.002}^{+0.029}$   \\
$\log L$ $(L_{\rm \odot}$)      & $0.105_{-0.055}^{+0.019}$	 &	$0.154_{-0.109}^{+0.053}$	& $0.105_{-0.057}^{+0.075}$	     & $-0.006_{-0.004}^{+0.032}$  \\
 $t$ (Gyr)                      &6.8$^{\rm +1.5}_{\rm -1.7}$	&  6.8$^{\rm +2.8}_{\rm -2.4}$ & 5.5$^{\rm +2.5}_{\rm -2.1}$    & 5.0$^{\rm +1.0}_{\rm -1.0}$ \\
$R_{\rm Birth}$ (kpc)           & 7.15$\pm$0.13              &	7.15$\pm$0.13               &  7.42$\pm$0.14                 & 7.93$\pm$0.13               \\
\hline
$(U,V,W)_{\rm LSR}$ (km~s$^{-1}$)  &  \multicolumn{4}{c}{ -30.84$\pm$0.25, -5.80$\pm$0.35, -4.40$\pm$0.22}\\
$S_{\rm LSR}$ (km~s$^{-1}$)  &  \multicolumn{4}{c}{ 31.69$\pm$0.49}\\
$R_{\rm a}$ (pc) &  \multicolumn{4}{c}{8580$\pm$10} \\
$R_{\rm p}$ (pc) &  \multicolumn{4}{c}{6969$\pm$6} \\
$Z_{\rm max}$ (pc) &  \multicolumn{4}{c}{59$\pm$1} \\
$e$ &  \multicolumn{4}{c}{0.104$\pm$0.001} \\
$P$ (Myr) &  \multicolumn{4}{c}{217$\pm$1} \\
\hline
\end{tabular}
\end{table*}

Photometric and astrometric data of HD\,76151 from {\it Gaia} DR3  \citep{Gaia2023} were also used to study stellar evolution. The distance relation was used to calculate the absolute magnitude ($M_{\rm G}$) of the star in the $G$-band.
\begin{equation}
    M_{\rm G}=G-5\times\log(1000/\varpi)+5-A_{\rm G}
\end{equation}
where $G$ is the apparent magnitude of the star, $\varpi$ is the trigonometric parallax in mas, and $A_{\rm G}$ is the interstellar extinction in the $G$ band. 

\begin{figure}
    \centering
    \includegraphics[width=0.48\textwidth]{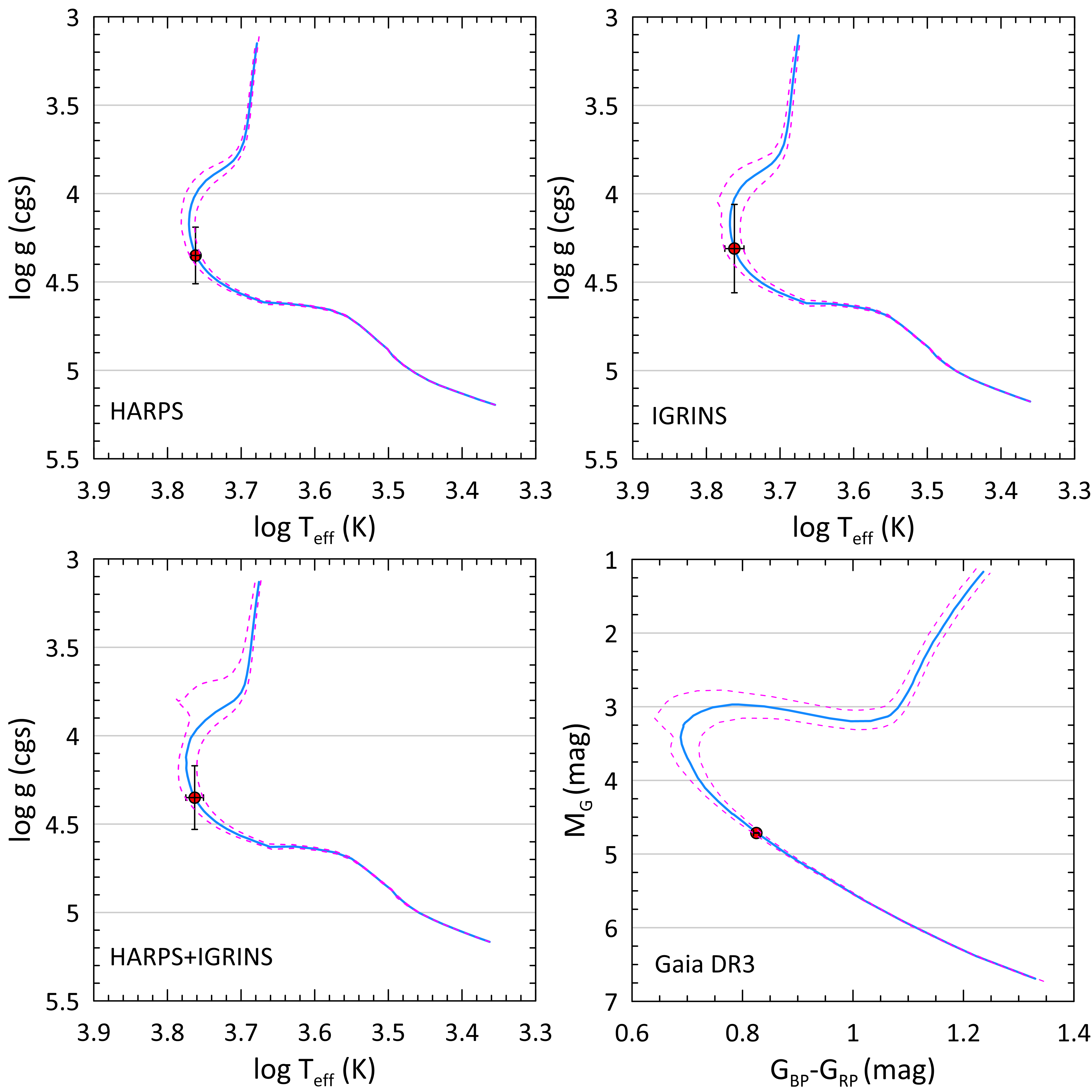}
    \caption{The positions of HD\,76151 on the kiel and color-magnitude diagrams. The blue curves show the {\sc PARSEC} isochrone representing the star, and the magenta dashed curves represent the uncertainties taken into account in the age determination.}
\label{fig:8}
\end{figure}

The foreground interstellar extinction, determined using the online {\sc Stilism} tool\footnote{\url{https://stilism.obspm.fr/}} by \citet{Capitanio2017}, was $E(B-V)=0$ mag. This conclusion seems correct, because the star is close to the Sun. Therefore, in this study, $G$-band absorption was assumed to be $A_{\rm G}=0$ mag. Using photometric and astrometric parameters listed in Table \ref{tab:summary}, the absolute magnitude of the star was calculated as $M_{\rm G}=4.714\pm0.003$ mag. The age of HD\,76151 was determined to be $t=5.0^{\rm +1.0}_{\rm -1.0}$ Gyr by plotting the star on a color-magnitude diagram (Figure \ref{fig:8}) and fitting the most appropriate {\sc PARSEC} isochrones. The {\sc PARSEC} isochrone that best fits the color-magnitude diagram of HD\,76151 shows that the star has a mass of $M=0.968_{-0.015}^{+0.002}M_{\odot}$, radius $R=0.977_{-0.002}^{+0.029}R_{\odot}$, and metal abundance of [Fe/H]=-0.05$\pm$0.05 dex (Table \ref{tab:summary}).

As can be seen from Table \ref{tab:A8}, when the literature over the last four decades is analysed, it is seen that the mass, radius and age values take values in the following intervals $0.88\leq M/M_{\odot} \leq 1.24$, $0.97\leq R/R_{\odot} \leq 1.12$, and $0.29 < t {\rm (Gyr)} \leq 6.8$, respectively. In this context, the age value obtained in this study is the largest age value determined to date. However, the mass and luminosity of the star are compatible with those reported for these parameters in the literature. Despite the agreement in mass and luminosity, this difference in age obtained for HD\,76151 is likely due to the updated isochrone curves. In this study, the fact that the four different methods gave ages in agreement with each other underlines the accuracy of the analyses. Considering the spectroscopic studies in the literature, it was observed that the elemental abundances reported for HD\,76151, especially on the basis of $\alpha$-elements, were close to solar or slightly metal-poor (Table \ref{tab:A8}). This result emphasises that HD\,76151 belongs to the thin disc in terms of its $\alpha$-element composition \citep{Plevne2020}.

\subsection{Kinematic and Dynamic Orbital Parameters}

In this study, we focused on the analysis of HD\,76151, particularly, the determination of its space velocity components and Galactic orbital parameters. The proper motion components, trigonometric parallax, and radial velocity of the star were obtained from {\it Gaia} DR3 \citep{Gaia2023}, and the astrometric and spectroscopic data are listed in Table \ref{tab:summary}. The space velocity components for HD\,76151 were calculated as $(U, V, W)=(-40.00\pm0.07, -20.01\pm0.09, -10.97\pm0.05)$ km s$^{-1}$ using the \texttt{galpy} code developed by \citet{Bovy2015}. The uncertainties associated with the space velocity components were evaluated using the algorithm proposed by \citet{Johnson1987}. Because the star is too close to the Sun, no differential rotation correction \citep{Mihalas1981} was applied to the space velocity components. However, the space velocity components are corrected for the peculiar velocity of the Sun, which is $(U, V, W)_\odot=(8.83\pm0.24, 14.19\pm0.34, 6.57\pm0.21$) km s$^{-1}$ given by \citet{Coskunoglu2011}. Thus, local standard of rest (LSR) space velocity components $(U, V, W)_{\rm LSR}$ of HD\,76151 are $-30.84\pm0.25$, $-5.80\pm0.35$, $-4.40\pm0.22$ km s$^{-1}$. The total space velocity of HD\,76151 was calculated as $S_{\rm LSR} = 31.39\pm0.49$ km s$^{-1}$, using the relation $S_{\rm LSR}=\sqrt{U_{\rm LSR}^2+V_{\rm LSR}^2+W_{\rm LSR}^2}$. The total space velocity of the star indicates that HD\,76151 is a young, thin-disc star \citep{Leggett1992}. The kinematic results are listed in Table \ref{tab:summary}. 

The \texttt{MWPotential2014} potential in the \texttt{galpy} library was used to obtain the Galactic orbital parameters of HD\,76151. The \texttt{MWPotential2014} model is a simplified representation of the Milky Way, assuming axis-symmetry and time-independence of the potential. It consists of a spherical bulge, dark matter halo, and Miyamoto-Nagai \citep{Miyamoto1975} disk potential. The spherical bulge represents the mass distribution of the Milky Way, and is defined as a spherical power-law density profile as described by \citet{Bovy2015}. In this analysis, the galactocentric distance and orbital velocity were assumed to be $R_{\rm GC}=8$ kpc and $V_{\rm rot}=220$ km s$^{-1}$, respectively \citep{Bovy2015, Bovy2012}. In addition, the distance of the Sun from the Galactic plane was adopted as $27\pm4$ pc as presented by \citet{Chen2000}.

\begin{figure}[h!]
\centering\includegraphics[width=1.0\linewidth]{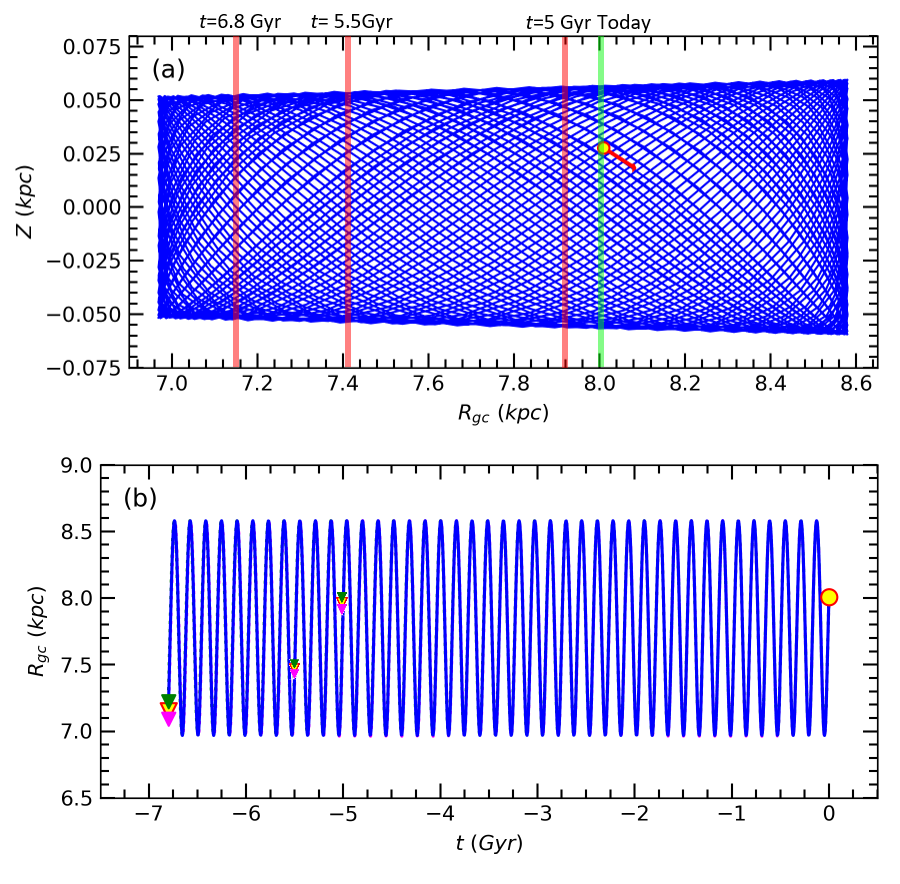}
\caption{The Galactic orbit and birth radii of HD\,76151 for three different age value in the $Z \times R_{\rm gc}$ (a) and $R_{\rm gc} \times t$ (b) diagrams. The filled yellow circles and triangles show the today and birth positions, respectively. The red arrow is the motion vector of HD\,76151 in today. The green and pink dotted lines show the orbit when parameter errors of the input values are considered, while the green and pink filled triangles represent the birth places of the HD\,76151 based on the lower and upper error estimates.}
\label{fig:galactic_orbits}
\end{figure} 

The Galactic orbit of HD\,76151 was integrated backward in time with 5 Myr steps up to an age of 7 Gyr from the star's present positions in the Galaxy. The output parameters that were estimated from the orbit analyses are listed in Table \ref{tab:summary}, where $R_{\rm a}$ and $R_{\rm p}$ are the apogalactic and perigalactic distances, respectively, $e$ is the eccentricity, $Z_{\rm max}$ is the maximum vertical distance from the Galactic plane and $P$ is the orbital period. The errors in the Galactic orbital parameters are calculated by considering the uncertainties in the proper motion components, trigonometric parallax, and radial velocity of HD\,76151. The positions of the star according to their distances from the Galactic center ($R_{\rm gc}$) and perpendicular to the Galactic plane ($z$) at different times are shown in Figure \ref{fig:galactic_orbits}a. The results show that HD\,75151 has a slightly flattened Galactic orbit. Moreover, the fact that the star is approximately $z=7$ pc from the Galactic plane indicates that the star belongs to the thin-disc population of the Galaxy \citep{Guctekin2019}.

The orbit of HD\,76151 on different planes, as a result of the dynamic orbit analyses, is shown in Figure \ref{fig:galactic_orbits}. The upper panel of Figure \ref{fig:galactic_orbits} presents side views of the HD\,76151 motions as functions of distance from the Galactic center and the Galactic plane \citep[e.g.,][]{Tasdemir2023, Yontan2023} while the lower panel shows the positions of the star at birth and today are represented by yellow filled triangles and circles, respectively. The eccentricity of the orbit of HD\,76151 does not exceed 0.11. The distance from the Galactic plane reaches a maximum at $z_{\rm max}=59\pm1$ pc for the star. However, the fact that HD\,76151 is poor in $\alpha$-element abundance, as indicated by the spectral analyses performed in this study, supports the results of the kinematic and dynamical orbital analyses of the star. These results indicate that HD\,76151 belongs to the young thin-disc population of the Milky Way \citep{Plevne2015}. 

To obtain the birth radius of HD\,76151 in the Galaxy, its orbit around the Galactic center was projected backwards in time, taking into account its age as determined by the {\sc PARSEC} isochrones \citep{Bressan2012}. Dynamical orbital analyses of HD\,76151 for ages of 5, 5.5 and 6.8 Gyr show that the star is born at distances of 7.15$\pm$0.13, 7.42$\pm$0.14 and 7.93$\pm$0.13 kpc from the Galactic center, respectively (see Figure \ref{fig:galactic_orbits}). This result supports the two-infall model of the Galactic disc \citep{Chiappini1997} and it is important evidence that HD\,76151 was born in a metal-rich environment. 
\vspace{1.5cm}

\section{Concluding remarks}

One of the advantages of this study is that all transitions included in the new line list are suitable for EW analysis, and the atomic data were confirmed by spectroscopic analysis of the optical and infrared solar spectra. In addition, during this confirmation phase, the atomic data ($\log gf$) reported in different line lists published in the literature over the past 40 years were independently tested and included in the line list, which provided the most compatible element abundance values for solar abundances from the optical and NIR solar spectra. This study provides two Mg\,{\sc i}, 28 Si\,{\sc i}, 11 Ca\,{\sc i}, one Ca\,{\sc ii}, 10 Ti\,{\sc i}, one Ti\,{\sc ii}, 204 Fe\,{\sc i}, and 10 Fe\,{\sc ii} transitions in the $Y$-, $J$-, $H$-, and $K$-band wavelength ranges, allowing the determination of model atmospheric parameters by providing ionization equilibrium in the NIR region, as is common in spectral analyses performed in the optical region, not only for Fe but also for Ti and Cr (see Table \ref{tab:2}).

Based on its spectral similarity to the Sun (\citealp[]{Lewin2020}; see their Table 2), HD\,76151 has been identified as a potential solar analog star. Solar analogs are stars with characteristics similar to those of the Sun, including mass, radius, and metallicity. Age is often excluded from the definition owing to difficulties in estimation. Alternatively, some researchers have defined solar analogs based on specific stellar parameters. A solar analog star is defined to have an effective temperature of 5772$\pm$300 K, a surface gravity of 4.44$\pm$0.30 cgs, and a metallicity of 0.0$\pm$0.3 dex (\citealp[]{Prvsa2016,Berke2023}). In this study, the stellar parameters from spectroscopic analysis with combined line list (OPTICAL + NIR) provided an effective temperature of 5790$\pm$170 K, a surface gravity of 4.35$\pm$0.18 cgs, and a metal abundance of 0.24$\pm$0.09 dex. Kinematic and dynamic orbital analyses indicated a thin disk population membership for HD\,76151. These results show that HD\,76151 is a solar analog star with stellar parameters determined in this study likely falling within the above range but is a slightly older ($t=5.5\pm2.3$ Gyr) star than the Sun.

By employing the new line list for the spectroscopic analysis of HD\,76151, along with both photometric and astrometric data from {\it Gaia} DR3, we were able to obtain the key astrophysical parameters of the star, such as its absolute bolometric magnitude for several different photometric bands, luminosity, and radius. Additionally, the astrometric data allowed us to determine the space velocity components, Galactic orbital parameters, birthplace, and current position within the Galaxy.

The primary focus of this study was the creation of a comprehensive line list for the analysis of stellar spectra in the infrared region. For F- and G-type stars, this line list will enable precise determination of model atmospheric parameters through NIR analysis. This capability will facilitate comparisons with the parameters obtained from astroseismological methods. Furthermore, the line list can be applied to both Sun-like stars and evolved stars of similar spectral types, allowing for a comprehensive comparison of the parameters derived from astroseismological and spectral methodologies. This line list will contribute to the confirmation of astrophysical parameters based on model atmospheric parameters, particularly in the context of studies conducted in the optical range.

\section*{Acknowledgments}
We thank the anonymous referees for their insightful and constructive suggestions, which significantly improved the paper. This study was supported by the Scientific and Technological Research Council of Turkey (TUBITAK) under Grant Number 121F265. The authors thank TUBITAK for their support. We thank Prof. Dr. Sunetra Giridhar for her suggestions and comments on improving this paper. We would also like to thank Remziye Canbay for her contribution to kinematic and dynamic orbital analyses. This research used NASA's (National Aeronautics and Space Administration) Astrophysics Data System and the SIMBAD Astronomical Database, operated at CDS, Strasbourg, France, and the NASA/IPAC Infrared Science Archive, which is operated by the Jet Propulsion Laboratory, California Institute of Technology, under contract with the National Aeronautics and Space Administration. 

\software{LIME (\citealp{Sahin2017}), SPECTRE (\citealp{Sneden1973}), MOOG (\citealp{Sneden1973}), and INSPECT program v1.0 (\citealp{Lind2012}), galpy \citep{Bovy2015}, MWPotential2014 \citep{Bovy2015}.}

\bibliography{sample631}{}
\bibliographystyle{aasjournal}

\begin{appendix}

\section{Expanded Literature Review for construction of the new line list}

\subsection{\citet{Melendez1999}}
After analyzing the 2\,218 lines reported by \citet{Melendez1999} in the wavelength range 1.0 - 1.79 \micron, a total of 185 common Fe\,{\sc i} and six common Fe\,{\sc ii} were detected. The $\log gf$ difference between the common Fe\,{\sc i} lines in both studies was 0.016$\pm$0.084 dex. The $\log gf$ difference for the common  Fe\,{\sc ii} lines was 0.000$\pm$0.074 dex. The reported oscillator strengths of the two common Mg\,{\sc i} lines were found to be identical. The $\log gf$ difference over the 25 common Si\,{\sc i} lines was 0.041$\pm$0.090 dex. The mean $\log gf$ difference between the eight common Ti\,{\sc i} lines in both studies was 0.066$\pm$0.124 dex. Compared to \citet{Melendez1999}, we report 15 additional transitions (four Fe\,{\sc i}, nine Fe\,{\sc ii}, one Ti\,{\sc ii}, and one Si\,{\sc i}) in the wavelength range of 1.0 to 1.79 \micron.  Furthermore, the blend lines reported by \citet{Melendez1999} are not included in this study. For example, the Fe\,{\sc i} line at 15\,246.49\footnote{The multiplet number for the line reported by \citet{Nave1994} was 2\,440.} \AA\,with a LEP of 6.31 eV was blended with the Fe\,{\sc i} line at 15\,246.49\footnote{\citet{Nave1994} did not report a multiplet number for line.} \AA\, (4.14 eV).

\subsection{\cite{Borrero2003}}
The line list published by \cite{Borrero2003} contains 83 lines in the wavelength range 9862 \AA\, of 15\,563 \AA. The average $\log gf$ difference calculated for the 27 Fe\,{\sc i} lines common to both studies was -0.079$\pm$0.060 dex. The mean $\log gf$ differences calculated for the five common Si\,{\sc i} lines, three common Ca\,{\sc i} lines, and four common Ti\,{\sc i} lines were less than 0.1 dex. 

\subsection{\citet{Rayner2009}}
Compared with \citet{Rayner2009}, we have 15 common lines in the present study. These 15 lines are listed as strong metal lines in the solar spectrum (see Table 6 in \citealp[]{Rayner2009}). They did not report atomic data. 

\subsection{\citet{Ryde2010}}
\citet{Ryde2010} reported 104 lines for 12 species in the wavelength range of 15\,324--15\,699 \AA. The average $\log gf$ difference, calculated over 11 common Fe\,{\sc i} lines was -0.034 ± 0.062 dex. In this study, one common Si\,{\sc i} line and two common Ti\,{\sc i} lines showed differences of 0.04 dex and -0.185$\pm$0.145 dex, respectively. The current study contains seven additional transitions, four Fe\,{\sc i} and three Fe\,{\sc ii}, in the wavelength range of 15\,323.55 - 15\,698.979 \AA\, which are within the wavelength limits determined by \citet{Ryde2010}.

\subsection{\citet{Smith2013}}
This study reveals a total of six common transitions in the wavelength range of 15\,159--16\,828 Å, which are one Si\,{\sc i}, two Ti\,{\sc i}, and three Fe\,{\sc i}, compared to the study conducted by \citet{Smith2013}. The latter study reported 53 lines for 13 species within the same wavelength range. The mean $\log gf$ difference for the three common iron lines was -0.063$\pm$0.049 dex, whereas the mean $\log gf$ difference for the common silicon and titanium lines was -0.155 dex and -0.178$\pm$0.181 dex, respectively. Furthermore, this study reported an additional 65 Fe\,{\sc i} lines and seven Fe\,{\sc ii} lines in the same wavelength range, whereas \citet{Smith2013} did not report any ionized iron, calcium, or titanium lines.

\subsection{\citet{Afsar2016,Afsar2018}}
Model atmospheric parameters from the optical spectrum were used to determine element abundance in the $H$- and $K$-bands. Six transitions at 22\,808.03 \AA, 17\,108.63 \AA, 15\,765.84 \AA, 15\,748.99 \AA, 15\,740.71\AA, and 15\,047.71 \AA\, are listed for Mg in the IGRINS spectral range. For the element Si, they listed 13 transitions  at 21\,354.26 \AA, 19\,928.92 \AA, 19\,722.55 \AA, 16\,680.77 \AA, 16\,434.93 \AA, 16\,381.55 \AA, 16\,241.85 \AA, 16\,215.69 \AA, 16\,163.71 \AA, 16\,094.8 \AA, 16\,060.02 \AA, 15\,960.08 \AA, 15\,888.44 \AA. A comparison of our line list with the lines used by \citet{Afsar2016} in their abundance analysis showed no common transitions in either study. However, we have four additional transitions at 15\,127.91 \AA, 15\,361.16 \AA, 20\,296.36 \AA, and  20\,343.84 \AA\, that are suitable for abundance analysis using the EW analysis technique in the same spectral region including the $H$- and $K$-bands where \citet{Afsar2016} reported no Ti transitions. In the same band range, we list one neutral Ti\,{\sc i} line at 15\,543.78 \AA\, and one ionized Ti\,{\sc ii} line at 15\,873.84 \AA. The Fe transitions listed by \citet{Afsar2016} were located in the optical region of their spectra.

\citet{Afsar2016, Afsar2018} used model atmospheric parameters from the optical spectrum to report abundances in the $H$- and $K$-bands. They reported 27 neutral transitions of Fe in the IGRINS spectral region, 17 of which are common to both studies. The mean $\log gf$ difference over the 17 common Fe\,{\sc i} lines was -0.015$\pm$0.074 dex. The spectrum synthesis method was used to determine the abundance of Fe lines. In this study, we used four common Ti lines (four Ti\,{\sc i} lines and one Ti\,{\sc ii} line). The mean $\log gf$ difference for the common Ti\,{\sc i} lines is -0.050$\pm$0.104 dex. For the common Ti\,{\sc ii} line, it was -0.05 dex. The new line list contains an additional Ca\,{\sc i} transition at 15\,118.20 \AA\,and a Ca\,{\sc ii} transition at 16\,649.88 \AA. 

In a recent study by \citet{Afsar2018}, 248 transitions in the wavelength range 4\,546 - 23\,379 \AA\, were reported. A comparison of these transitions with the present study revealed one common Si\,{\sc i}, one common Ca\,{\sc i}, four common Ti\,{\sc i}, one common Ti\,{\sc ii}, and 17 common neutral Fe lines. The mean $\log gf$ difference calculated for the common Fe\,{\sc i} lines was -0.015$\pm$0.074 dex. It is -0.270 dex for Si\,{\sc i} at 20\,343.84 \AA, -0.050$\pm$0.104 dex for Ti\,{\sc i}, 0.16 dex for Ca\,{\sc i} line at 19\,917.19 \AA\,, and -0.05 dex for Ti\,{\sc ii} line at 15\,873.84 \AA. The present study contributes to the literature by providing 249 lines in the wavelength range of $\approx$ 10\,000 \AA\, to 23\,379.136 \AA\, which is the upper wavelength limit in \citet{Afsar2018}. These included 188 Fe\,{\sc i}, 15 Fe\,{\sc ii}, two Mg\,{\sc i}, 27 Si\,{\sc i}, 10 Ca\,{\sc i}, one Ca\,{\sc ii}, and six Ti\,{\sc i}.

\subsection{\citet{Kondo2019}}
The line list published by \citet{Kondo2019} in the wavelength range 0.91-1.33 \micron\, consists of selected lines from the line lists reported by VALD and MB99. However, they tested their line list in the spectra of two red giant stars (Arcturus and $\mu$ Leo). The list contains only Fe\,{\sc i} lines. Of the 171 Fe\,{\sc i} lines, 97 were selected from the VALD and 75 from the MB99. In this study, ionized iron lines were not observed. The difference in the mean $\log gf$ calculated for the 56 common Fe\,{\sc i} lines compared with the line list generated in the present study was -0.163$\pm$0.211 dex. We report 76 additional lines (6 Ti\,{\sc i}, 29 Fe\,{\sc i}, 24 Si\,{\sc i}, 9 Ca\,{\sc i}, 2 Mg\,{\sc i}, and 6 Fe\,{\sc ii}) in the 0.91-1.33 \micron\,wavelength range, and 141 additional lines beyond the upper wavelength limit included in \citet{Kondo2019}. 

\subsection{Comparison of the line list to DR17: ionized Fe lines}

APOGEE DR17 \citep{DR17} lists 4\,740 Fe\,{\sc i} lines and 2\,979 Fe\,{\sc ii} lines among a total of 10\,944 lines in the wavelength range of 15\,000 to 17\,007 \AA. The observability of all the iron lines in APOGEE DR17 in the NIR solar spectrum in the $H$-band was visually inspected and the common lines are listed in Table \ref{tab:A1}. Among the visually inspected lines, those suitable for EW measurements in the solar spectrum were subjected to the RMT analysis. Regrettably, the RMT test was only performed on Fe\,{\sc i} lines, and to the best of our knowledge, there is no reference in the literature for the RMT analysis of Fe\,{\sc ii} lines, particularly, in the NIR region. For the Fe\,{\sc i} lines, we used RMT tables reported by \citet{Nave1994}. This test provided additional iron lines. Table \ref{tab:A1} designates (as ``new'') the two additional Fe\,{\sc i} lines listed in the APOGEE $H$-band wavelength range and also lists three Fe\,{\sc ii} lines at 15\,078.21 \AA, 15\,792.03 \AA, and 16\,468.51 \AA\, which fall within the wavelength range of APOGEE DR17. These lines are listed in APOGEE DR17 but not in the other examined line catalogs. In this study, trio was tested in the observed spectra for abundance analysis for the first time. 

Table \ref{tab:A1} also shows that the APOGEE $H$-band terminates at 17\,005 \AA\, and it does not extend the upper wavelength limit of 18\,000 \AA\, for the $H$-band. It is worth noting that the present study shifted the APOGEE $H$-band upper limit to 18\,000 \AA\, for Fe lines, which is beneficial for IGRINS $H$-band spectroscopy. In Table \ref{tab:A1}, the additional transitions detected in the present study within the range $17\,007-18\,000$ \AA\, are also labelled as ``new''. The table further provides the differences between the $\log gf$ data employed in this study (TS) and the oscillator strengths listed in APOGEE DR17. 

Compared to APOGEE DR17 \citep{DR17}, the line list in the present study contained 74 common Fe\,{\sc i} lines for which the mean $\log gf$ difference was -0.068$\pm$0.347 dex. In contrast, the difference between the nine common Fe\,{\sc ii} transitions detected in the first test run is 7.096$\pm$2.042 dex. Later in the text, we explore the issue of oscillator strength values provided for ionized iron lines in APOGEE DR17.

\setcounter{figure}{0}
\renewcommand{\thefigure}{A\arabic{figure}}

\begin{figure}
    \centering
    \includegraphics[width=0.47\textwidth]{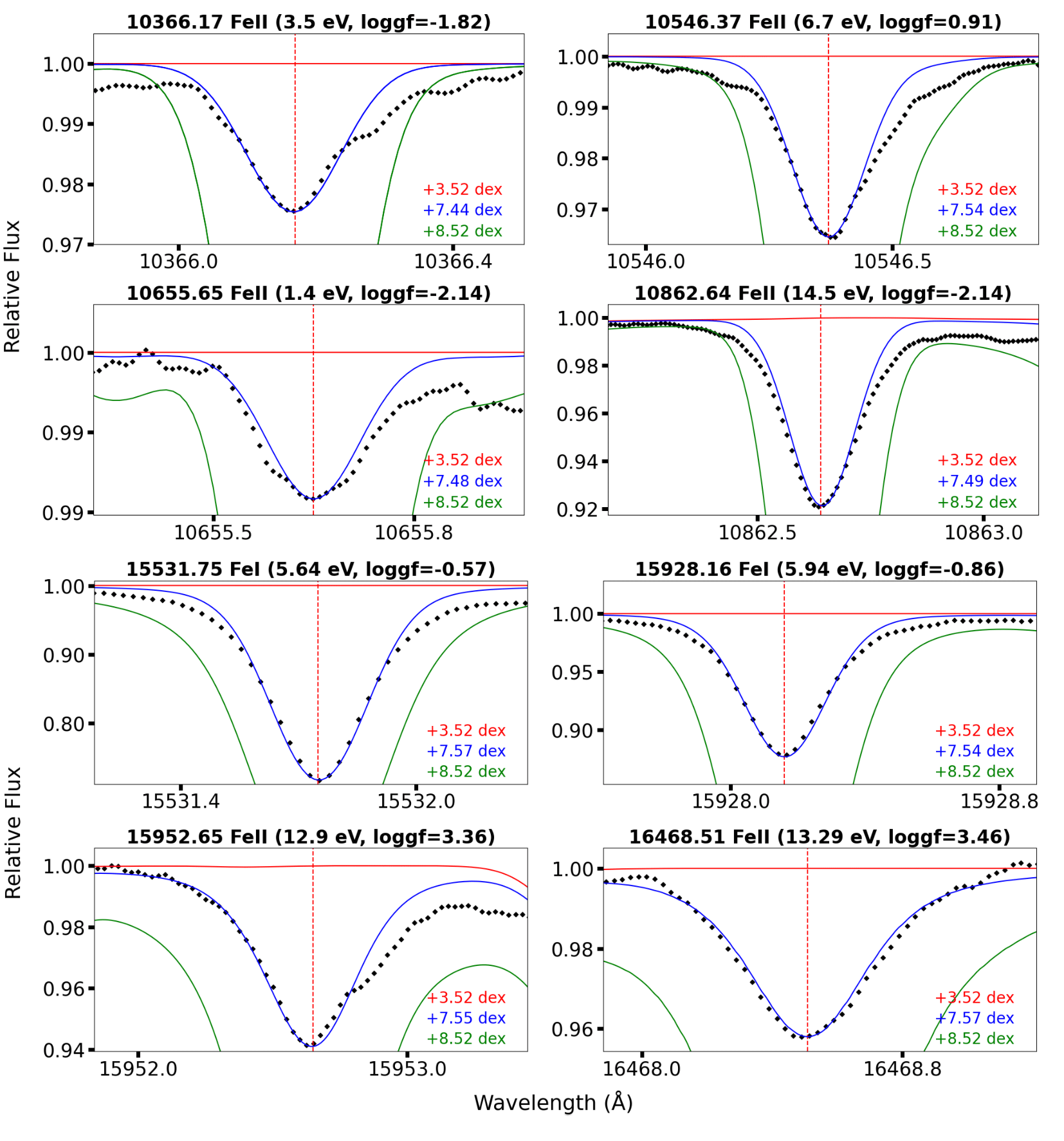}
    \caption{The synthetic spectra generated by the spectrum synthesis method on the KPNO-NIR solar spectrum for the selected neutral and ionized iron lines. The computed profiles illustrate the synthetic spectra for the three varying logarithmic abundances. The red lines are the spectra computed with no contribution from those neutral and ionized iron lines.}
\label{fig:5}
\end{figure}

Parts of the Fe\,{\sc ii} transitions are listed in Table \ref{tab:A7} in bold and asterisk. These lines do not match any of the other line lists examined in this study. The solar abundances for them, obtained from the KPNO-NIR solar spectrum (see Table \ref{tab:A7}), are calculated based on the theoretical $\log gf$ values, yielding NIR solar iron abundances consistent with the NIR mean abundances for both neutral (7.46$\pm$0.07 dex over 204 Fe\,{\sc i} lines) and ionized iron lines (7.50$\pm$0.06 dex over 10 Fe\,{\sc ii} lines). Figure \ref{fig:5} shows the synthetic spectrum generated through the synthesis of the KPNO-NIR solar spectrum spectrum, focusing on the selected Fe\,{\sc i} and Fe\,{\sc ii} lines. Furthermore, as depicted in Figure \ref{fig:5}, an assessment of the line profiles for certain ionized iron lines reveals that they are not suitable for EW analysis (e.g., Fe\,{\sc ii} lines at 10\,366.17 \AA\, and 10\,655.65 \AA). Nonetheless, in this study, these lines were integrated into the analysis by employing the spectrum synthesis method, considering the requirement of Fe\,{\sc ii}, which is extremely scarce in the wavelength ranges spanning the $H$- and $K$-bands. Notably, the abundance values obtained using EW analysis of these lines corresponds to those derived from the spectrum synthesis method. Consequently, the reliability and accuracy of the EW analysis for these lines can be considered satisfactory. The other ionized iron transitions used in the spectral analysis were similarly subjected to spectrum synthesis to verify the reported abundances of these lines using the EW method. The elemental abundances obtained using both methods for these ionized iron transitions were in agreement, with a difference of $\approx$ 0.02 dex.

Some of the iron transitions listed in Tables \ref{tab:A1} and \ref{tab:A7} were classified as both neutral and ionized iron lines in APOGEE DR17 \citep{DR17}. Transitions (neutral/ionized) at 15\,531.75/15\,531.74 \AA, 15\,928.16/15\,928.14 \AA, and 15\,952.63/15\,952.70 \AA\, represent such transitions. With $\log gf$ values from APOGEE DR17, the solar abundances calculated for these lines exhibited notable line-to-line scatter. Because there was no alternative atomic data reference for those with discrepant $\log gf$ values, we were obliged to calculate the astrophysical $\log gf$ values for the Fe\,{\sc ii} lines at 15\,078.21 \AA\, ($\log gf$=2.12 dex), 15\,531.74 \AA\,(4.42 dex), 15\,536.68 \AA\,(2.99 dex), 15\,563.40 \AA\,(2.90 dex), 15\,792.03 \AA\,(2.22 dex), 15\,928.14 \AA\,(3.83 dex), 15\,952.63 \AA\,(3.36 dex), 16\,468.51 \AA\,(3.46 dex), and 17\,000.92 \AA\,(4.25 dex). In the following paragraphs, we provide the details of the tests carried out on these three lines using the EW and spectrum synthesis methods. 

\textbf{Fe\,{\sc i} at 15\,531.74 \AA:}  Among these ionized iron lines, the line at 15\,531 \AA\, appears to be classified as Fe\,{\sc i} in the analyses performed by \citet{Melendez1999} and \citet{Andreasen2016}. Because APOGEE DR17 \citep{DR17} lists the line as both Fe\,{\sc i} and Fe\,{\sc ii}, we queried the multiplet number in the line catalog of \citet{Nave1994}. Subsequently, the presence of other potential lines with the same multiplet number in the spectrum is investigated. The investigation revealed that the 15\,531 \AA\,Fe\,{\sc i} line, listed as a member of multiplet 2061 in Nave's line catalog, belongs to the same multiplet as the line at 15\,906.04 \AA, which is included in our line list. 

Furthermore, we employed spectrum synthesis method to address the identification issue for the line at 15\,531.74 \AA. Assuming this line to be Fe\,{\sc ii}, we used the APOGEE DR17 \citep{DR17} value of $\log gf=-4.84$ dex for abundance analysis. However, it is not possible to create a synthetic profile for this line. Interestingly, EW analysis with the same $\log gf$ value yielded an abundance of 16.70 dex. Because this abundance value was very large, the astrophysical value of $\log gf$ (4.42 dex) was calculated, assuming that it was an Fe\,{\sc ii} line. The spectrum synthesis method suggests an extremely broad line wing for the synthetic line profile of the astrophysical $\log gf$ value. The same $\log gf$ value, along with the EW analysis, implies a logarithmic abundance value of 7.44 dex for the line. 

When assessing the probability of Fe\,{\sc i} in the spectrum synthesis analysis of the line at 15\,531.74\footnote{In this study, we used the $\log gf=$-0.57 dex from \citet{Andreasen2016}.} \AA , the $\log gf$=-0.56 dex from APOGEE DR17 yielded a logarithmic abundance of 7.57 dex. This value is consistent with the average abundance of Fe\,{\sc i} reported in the solar spectrum (Table \ref{tab:3}). Therefore, we conclude that this line is likely an Fe\,{\sc i} transition. 

\textbf{Fe\,{\sc i} at 15\,928.14 \AA:} The line appears to be included as Fe\,{\sc i} in the analyses performed by \citet{Melendez1999} and \citet{Andreasen2016}. It is listed as both Fe\,{\sc i} and Fe\,{\sc ii} in APOGEE DR17 \citep{DR17}. Considering the possibility that this line was Fe\,{\sc i}, the multiplet number listed in the line catalog by \citet{Nave1994} was queried. The presence of other possible lines with the same multiplet number in the spectrum is also investigated. Accordingly, the line at 15\,928 \AA\, which is listed as a member of multiplet 2\,297 in Nave's line catalog, was found to be from the same multiplet as 16\,474.09 \AA\, which is present in our line list. In addition, the identification problem for 15\,928.14 \AA\, was also tested. The line was first assumed to be Fe\,{\sc ii}, and the value $\log gf=-4.53$ dex \citep[reported by][]{DR17} was used for abundance analysis using the spectrum synthesis method. However, it is not possible to create an appropriate synthetic profile for this line. The same $\log gf$ value yielded an abundance of 15.83 dex for the line using EW analysis. The astrophysical $\log gf$ value (3.83) that we calculated for the line, spectrum synthesis and EW analysis methods gave logarithmic abundances of 7.89 dex and 7.47 dex, respectively. When the probability of Fe\,{\sc i} in the spectrum synthesis analysis of the line at 15\,928.14 \AA\, was assessed with $\log gf$ (-0.68 dex) from the APOGEE DR17, the abundance was calculated as 7.34\footnote{It is 7.54 dex when $\log gf=$-0.88 dex from \citet{Melendez1999} was employed.} dex. This value (7.54 dex) is in agreement with the average neutral iron abundance value reported for the solar spectrum in Table \ref{tab:3}. In this context, this line is considered an Fe\,{\sc i} transition.

\textbf{Fe\,{\sc ii} at 15\,952.63 \AA:}  The line appears to be included as Fe\,{\sc i} in the spectroscopic analysis performed by \citet{Melendez1999}. It is listed as both Fe\,{\sc i} and Fe\,{\sc ii} in APOGEE DR17 \citep{DR17}. Considering the possibility that this line is Fe\,{\sc i}, the multiplet number listed by \citet{Nave1994} was queried. The line at 15\,952 \AA\,is listed as a member of multiplet 2\,560 in Nave's line catalog. There were no other atomic transitions listed with the same multiplet number in the Nave's line catalog. In addition, the identification problem for 15\,952.63 \AA\, was also tested using the spectrum synthesis method. The line was first assumed to be Fe\,{\sc ii}, and the $\log gf=$-2.49 dex \citep[reported by][]{DR17} was used for the abundance analysis using the synthesis method. However, it is not possible to compute a synthetic profile for this line. The line is also located on the blue wing of another blend line consisting of Fe\,{\sc i} at 15\,954.09 \AA\, and Mg\,{\sc i} at 15\,954.46 \AA. The EW method with the same $\log gf$ provided a logarithmic abundance of 13.34 dex for the line. The spectrum synthesis and EW methods with astrophysical $\log gf$ (3.36 dex) gave logarithmic abundances of 7.55 dex and 7.49 dex, respectively. When the probability of Fe\,{\sc i} for the line was assessed, spectrum synthesis analysis with the APOGEE DR17 $\log gf$ (-0.61 dex) provided an abundance of 7.33 dex. It is 7.47 dex when $\log gf=$-0.81 dex from \citet{Melendez1999} was employed. This abundance as 7.47 dex agreed with the average neutral iron abundance value reported for the solar spectrum in Table \ref{tab:3}. Although the line was initially identified as an Fe\,{\sc i} transition, our analysis indicates that it can be equally well classified as Fe\,{\sc ii}. When using the APOGEE DR17 $\log gf$ values corresponding to both Fe\,{\sc i} and Fe\,{\sc ii} identifications, the derived abundance remained consistent with the average solar value. Given the relatively low number of Fe\,{\sc ii} transitions in the $H$-band, we chose to classify this line as Fe\,{\sc ii}. However, it is essential to note that this classification is not definitive, and the line can also be considered Fe\,{\sc i}.

The EW analysis, employing the calculated astrophysical $\log gf$ values for Fe\,{\sc ii} lines at 15\,536.68 \AA, 15\,563.40 \AA, and 17\,000.92 \AA, yielded an abundance value consistent with the average solar abundance listed in Table \ref{tab:3}. However, when employing the same $\log gf$ values in spectrum synthesis, a significantly different abundance was obtained, contradicting the EW analysis result (see transitions in red in Tables \ref{tab:A1} and \ref{tab:A7}). Furthermore, the Fe\,{\sc ii} line at 15\,563.40 \AA\, is known to be blended with CN \citep{Ryde2009}, therefore, these lines were excluded from the abundance analysis in this study.

The EW and spectrum synthesis methods for the Fe\,{\sc ii} lines at 15\,078.21 \AA, 15\,792.03 \AA, and 16\,468.51 \AA\, listed in APOGEE DR17 \citep{DR17} provide abundance values consistent with the average solar abundance listed in Table \ref{tab:3}.

\setcounter{table}{0}
\renewcommand{\thetable}{A\arabic{table}}

\begin{table*}
\setlength{\tabcolsep}{0.67pt}
 \caption{Common transitions in APOGEE DR17 wavelength range (15\,000 - 17\,007 \AA) in line list generated in this study. In addition, transitions beyond the 17\,005 \AA\, including the upper wavelength limit of the IGRINS $H$ band, are listed as new. The transitions in red are the lines produce significantly different abundance values via EW and spectrum synthesis methods. They are removed from the analysis.} 
 \label{tab:A1}
\centering
\begin{tabular}{c|c|c|c|c|c||c|c|c|c|c|c||c|c|c|c|c|c}
\hline
\hline
Species& $\lambda$	& LEP & \multicolumn{3}{c||}{$\log gf$ (dex)} &Species&$\lambda$	&LEP & \multicolumn{3}{c||}{$\log gf$ (dex)}  & Species& $\lambda$ &	LEP & \multicolumn{3}{c}{$\log gf$ (dex)} \\
\cline{2-6}
\cline{8-12}
\cline{14-18}
 & (\AA)  &  (eV) &  TS & DR17 & $\Delta$ &  & (\AA)  & (eV) & TS & DR17 & $\Delta$ &  &(\AA) & (eV) & TS & DR17&$\Delta$\\
\hline
\hline
Fe\,{\sc ii}&	15078.21	&	12.13	&	2.12	&	-4.68	&	6.80	&	Fe\,{\sc i}	&	15731.41	&	6.45	&	-0.66	&	-0.47	&	-0.19	&	Fe\,{\sc i}	&	16645.88	&	5.96	&	-0.27	&	-0.20	&	-0.07	\\
Fe\,{\sc i}	&	15080.25	&	6.26	&	-0.84	&	-0.61	&	-0.23	&	Fe\,{\sc i}	&	15733.51	&	6.25	&	-0.76	&	-0.56	&	-0.20	&	Fe\,{\sc i}	&	16648.24	&	6.55	&	-0.50	&	-0.28	&	-0.22	\\
Ca\,{\sc i}	&	15118.20	&	5.02	&	-0.65	&	-2.23	&	1.58	&	Fe\,{\sc ii}&	15792.03	&	12.81	&	2.22	&	-6.20	&	8.42	&	Ca\,{\sc ii}&	16649.88	&	9.24	&	0.57	&	0.57	&	0.00	\\
Fe\,{\sc i}	&	15120.50	&	5.45	&	-1.49	&	-1.02	&	-0.47	&	Fe\,{\sc i}	&	15829.29	&	6.30	&	-1.10	&	-0.83	&	-0.27	&	Fe\,{\sc i}	&	16665.49	&	6.02	&	-0.26	&	-0.26	&	0.00	\\
Fe\,{\sc i}	&	15122.38	&	5.62	&	-0.38	&	-0.38	&	0.00	&	Fe\,{\sc i}	&	15840.19	&	6.36	&	-0.40	&	-0.28	&	-0.12	&	Fe\,{\sc i}	&	16685.58	&	6.34	&	-0.36	&	-0.36	&	0.00	\\
Si\,{\sc i}	&	15127.91	&	7.23	&	-1.90	&	-3.90	&	2.00	&	Fe\,{\sc i}	&	15858.66	&	5.59	&	-1.34	&	-1.22	&	-0.12	&	Fe\,{\sc i}	&	16693.11	&	6.42	&	-0.41	&	-0.26	&	-0.15	\\
Fe\,{\sc i}	&	15176.72	&	5.92	&	-0.95	&	-0.76	&	-0.19	&	Fe\,{\sc i}	&	15868.52	&	5.59	&	-0.17	&	-0.17	&	0.00	&	Fe\,{\sc i}	&	16898.90	&	6.31	&	-0.94	&	-0.78	&	-0.16	\\
Fe\,{\sc i}	&	15179.75	&	5.98	&	-1.24	&	-3.28	&	2.04	&	Ti\,{\sc ii}&	15873.88	&	3.12	&	-1.95	&	-1.95	&	0.00	&	Fe\,{\sc i}	&	16900.23	&	6.30	&	-1.27	&	-1.15	&	-0.12	\\
Fe\,{\sc i}	&	15182.92	&	6.33	&	-0.54	&	-0.54	&	0.00	&	Fe\,{\sc i}	&	15901.53	&	5.92	&	-0.60	&	-0.51	&	-0.09	&	Fe\,{\sc i}	&	16910.69	&	5.87	&	-1.96	&	-1.96	&	0.00	\\
Fe\,{\sc i}	&	15194.50	&	2.22	&	-4.85	&	-4.74	&	-0.11	&	Fe\,{\sc i}	&	15904.35	&	6.36	&	0.45	&	0.25	&	0.20	&	Fe\,{\sc i}	&	16969.91	&	5.95	&	-0.37	&	-0.25	&	-0.12	\\
Fe\,{\sc i}	&	15201.57	&	5.49	&	-1.45	&	\bf new	&	\bf new	&	Fe\,{\sc i}	&	15906.04	&	5.62	&	-0.20	&	-0.25	&	0.05	&	\textcolor{red}{Fe\,{\sc ii}}&	\textcolor{red}{17000.92}	&	\textcolor{red}{13.03}	&	\textcolor{red}{4.25}	&	\textcolor{red}{-0.71}	&	\textcolor{red}{4.96}	\\
Fe\,{\sc i}	&	15207.54	&	5.39	&	0.04	&	0.06	&	-0.02	&	Fe\,{\sc i}	&	15928.16	&	5.95	&	-0.88	&	-0.68	&	-0.20	&	Fe\,{\sc i}	&	17005.45	&	6.07	&	-0.25	&	-0.08	&	-0.17	\\
Fe\,{\sc i}	&	15213.02	&	6.31	&	-0.83	&	-0.63	&	-0.20	&	Fe\,{\sc ii}&	15928.14	&	13.33	&	3.83	&	-4.53	&	8.36	&	Fe\,{\sc i}	&	17009.02	&	6.62	&	-0.37	&	\bf new	&	\bf new	\\
Fe\,{\sc i}	&	15237.77	&	5.79	&	-1.60	&	-1.47	&	-0.13	&	Fe\,{\sc i}	&	15929.48	&	6.31	&	-0.59	&	-0.43	&	-0.16	&	Fe\,{\sc i}	&	17027.62	&	6.62	&	-0.67	&	\bf new	&	\bf new	\\
Fe\,{\sc i}	&	15239.74	&	6.42	&	-0.15	&	-0.05	&	-0.10	&	Fe\,{\sc i}	&	15934.02	&	6.31	&	-0.43	&	-0.21	&	-0.22	&	Fe\,{\sc i}	&	17037.79	&	6.39	&	-0.42	&	\bf new	&	\bf new	\\
Fe\,{\sc i}	&	15267.02	&	5.07	&	-2.49	&	-2.42	&	-0.07	&	Fe\,{\sc i}	&	15940.92	&	5.81	&	-1.42	&	-1.25	&	-0.17	&	Fe\,{\sc i}	&	17052.20	&	6.39	&	-0.60	&	\bf new	&	\bf new	\\
Fe\,{\sc i}	&	15301.56	&	5.92	&	-0.84	&	-0.65	&	-0.19	&	Fe\,{\sc i}	&	15941.85	&	6.31	&	-0.13	&	0.06	&	-0.19	&	Fe\,{\sc i}	&	17067.67	&	6.37	&	-0.17	&	\bf new	&	\bf new	\\
Fe\,{\sc i}	&	15315.66	&	6.28	&	-1.17	&	-2.86	&	1.69	&	Fe\,{\sc i}	&	15952.63	&	6.34	&	-0.81	&	-0.61	&	-0.20	&	Fe\,{\sc i}	&	17072.86	&	5.07	&	-2.41	&	\bf new	&	\bf new	\\
Fe\,{\sc i}	&	15343.81	&	5.65	&	-0.78	&	-0.63	&	-0.15	&	Fe\,{\sc ii}&	15952.70	&	12.90	&	3.36	&	-2.49	&	5.85	&	Fe\,{\sc ii}&	17161.12	&	6.02	&	-0.24	&	\bf new	&	\bf new	\\
Si\,{\sc i}	&	15361.16	&	5.95	&	-2.08	&	-2.03	&	-0.05	&	Fe\,{\sc i}	&	15971.25	&	6.41	&	-0.41	&	-0.32	&	-0.09	&	Fe\,{\sc ii}&	17166.20	&	5.95	&	-0.79	&	\bf new	&	\bf new	\\
Fe\,{\sc i}	&	15381.98	&	3.64	&	-3.03	&	\bf new	&	\bf new	&	Fe\,{\sc i}	&	16019.79	&	6.35	&	-0.87	&	-0.59	&	-0.28	&	Fe\,{\sc i}	&	17200.34	&	6.36	&	-0.91	&	\bf new	&	\bf new	\\
Fe\,{\sc i}	&	15427.61	&	6.45	&	-1.18	&	-0.76	&	-0.42	&	Fe\,{\sc i}	&	16051.74	&	6.26	&	-1.01	&	-0.80	&	-0.21	&	Fe\,{\sc i}	&	17221.43	&	6.43	&	-0.85	&	\bf new	&	\bf new	\\
Fe\,{\sc i}	&	15522.64	&	6.32	&	-1.07	&	-0.92	&	-0.15	&	Fe\,{\sc i}	&	16171.93	&	6.38	&	-0.52	&	-0.43	&	-0.09	&	Fe\,{\sc i}	&	17257.59	&	6.32	&	-0.58	&	\bf new	&	\bf new	\\
Fe\,{\sc i}	&	15524.30	&	5.79	&	-1.51	&	-1.46	&	-0.05	&	Fe\,{\sc i}	&	16225.64	&	6.38	&	-0.03	&	0.11	&	-0.14	&	Fe\,{\sc i}	&	17278.72	&	6.72	&	-0.44	&	\bf new	&	\bf new	\\
Fe\,{\sc i}	&	15527.21	&	6.32	&	-1.11	&	-1.11	&	0.00	&	Fe\,{\sc i}	&	16228.67	&	6.38	&	-1.41	&	-1.13	&	-0.28	&	Fe\,{\sc i}	&	17282.32	&	6.43	&	-0.19	&	\bf new	&	\bf new	\\
Fe\,{\sc i}	&	15531.75	&	5.64	&	-0.57	&	-0.56	&	-0.01	&	Fe\,{\sc i}	&	16231.67	&	6.38	&	0.42	&	0.47	&	-0.05	&	Fe\,{\sc i}	&	17420.83	&	3.88	&	-3.62	&	\bf new	&	\bf new	\\
Fe\,{\sc ii}&	15531.74	&	13.25	&	4.42	&	-4.84	&	9.26	&	Fe\,{\sc i}	&	16235.98	&	5.92	&	-0.39	&	-0.30	&	-0.09	&	Fe\,{\sc i}	&	17433.67	&	6.41	&	-0.48	&	\bf new	&	\bf new	\\
Fe\,{\sc i}	&	15534.26	&	5.64	&	-0.34	&	-0.32	&	-0.02	&	Fe\,{\sc i}	&	16252.57	&	6.32	&	-0.54	&	-0.41	&	-0.13	&	Fe\,{\sc ii}&	17488.61	&	6.41	&	-0.78	&	\bf new	&	\bf new	\\
\textcolor{red}{Fe\,{\sc ii}}&	\textcolor{red}{15536.68}	&	\textcolor{red}{13.51}	&	\textcolor{red}{2.99}	&	\textcolor{red}{-7.77}	&	\textcolor{red}{10.76}	&	Fe\,{\sc i}	&	16258.93	&	6.24	&	-1.03	&	-0.78	&	-0.25	&	Fe\,{\sc i}	&	17500.01	&	5.96	&	-1.10	&	\bf new	&	\bf new	\\
Ti\,{\sc i}	&	15543.78	&	1.88	&	-1.48	&	-1.16	&	-0.32	&	Fe\,{\sc i}	&	16277.50	&	6.32	&	-0.51	&	-0.41	&	-0.10	&	Fe\,{\sc i}	&	17531.20	&	6.64	&	-0.71	&	\bf new	&	\bf new	\\
\textcolor{red}{Fe\,{\sc ii}}&	\textcolor{red}{15563.40}	&	\textcolor{red}{13.31}	&	\textcolor{red}{2.90}	&	\textcolor{red}{-4.08}	&	\textcolor{red}{6.98}	&	Fe\,{\sc ii}&	16468.51	&	13.29	&	3.46	&	-0.47	&	3.93	&	Fe\,{\sc i}	&	17534.80	&	6.64	&	-0.31	&	\bf new	&	\bf new	\\
Fe\,{\sc i}	&	15565.23	&	6.32	&	-0.96	&	-0.77	&	-0.19	&	Fe\,{\sc i}	&	16474.09	&	6.02	&	-0.60	&	-0.62	&	0.02	&	Fe\,{\sc i}	&	17536.93	&	5.91	&	-0.97	&	\bf new	&	\bf new	\\
Fe\,{\sc i}	&	15566.72	&	6.35	&	-0.54	&	-0.35	&	-0.19	&	Fe\,{\sc i}	&	16506.30	&	5.95	&	-0.56	&	-0.41	&	-0.15	&	Fe\,{\sc i}	&	17575.32	&	6.40	&	-1.01	&	\bf new	&	\bf new	\\
Fe\,{\sc i}	&	15579.08	&	6.32	&	-1.09	&	-1.15	&	0.06	&	Fe\,{\sc i}	&	16532.01	&	6.29	&	-0.19	&	-0.08	&	-0.11	&	Fe\,{\sc i}	&	17695.94	&	5.95	&	-0.65	&	\bf new	&	\bf new	\\
Ti\,{\sc i}	&	15602.84	&	2.27	&	-1.64	&	-1.63	&	-0.01	&	Fe\,{\sc i}	&	16544.70	&	6.34	&	-0.42	&	-0.25	&	-0.17	&	Fe\,{\sc i}	&	17714.37	&	6.58	&	-0.76	&	\bf new	&	\bf new	\\
Fe\,{\sc i}	&	15611.15	&	3.41	&	-2.98	&	-3.48	&	0.50	&	Fe\,{\sc i}	&	16552.02	&	6.41	&	-0.08	&	0.12	&	-0.20	&	Fe\,{\sc i}	&	17717.16	&	6.34	&	-0.75	&	\bf new	&	\bf new	\\
Fe\,{\sc i}	&	15648.52	&	5.43	&	-0.70	&	-0.65	&	-0.05	&	Fe\,{\sc i}	&	16559.71	&	6.40	&	-0.35	&	-0.20	&	-0.15	&	Fe\,{\sc i}	&	17747.37	&	5.92	&	-0.76	&	\bf new	&	\bf new	\\
Fe\,{\sc i}	&	15652.87	&	6.25	&	-0.19	&	-0.01	&	-0.18	&	Fe\,{\sc i}	&	16586.06	&	5.62	&	-1.53	&	-1.39	&	-0.14	&	Fe\,{\sc i}	&	17926.40	&	6.74	&	0.05	&	\bf new	&	\bf new	\\
Fe\,{\sc i}	&	15670.13	&	6.20	&	-1.07	&	-0.85	&	-0.22	&	Fe\,{\sc i}	&	16607.65	&	6.34	&	-0.59	&	-0.51	&	-0.08	&	-	        &	-	    &	-	    &	-	    &	-	    &	-	    \\
Fe\,{\sc i}	&	15700.08	&	6.33	&	-1.08	&	-0.98	&	-0.10	&	Fe\,{\sc i}	&	16619.73	&	5.59	&	-1.66	&	-1.47	&	-0.19	&	-	        &	-	    &	-	    &	-	    &	-	    &	-	    \\
\hline
\hline
\end{tabular}
\vspace{2ex}
     \footnotesize{$\Delta \log gf=\log gf_{\rm TS} - \log gf_{\rm DR17}$}\\  
\end{table*}
The discrepancies in the oscillator strengths of the Fe\,{\sc ii} lines within the APOGEE DR17 \citep{DR17} emphasize the necessity for further research in this domain. Our findings also validated the importance of combining EW and spectrum synthesis methods to ensure the reliable determination of abundances. The identification of new Fe lines and the extension of the APOGEE DR17 $H$-band wavelength limit present promising avenues for future investigations using high-resolution NIR spectroscopy. 

This study makes a substantial contribution to the compilation of a comprehensive line list spanning the $Y$-, $J$-, $H$-, and $K$-bands. This resource will be invaluable for NIR spectroscopic analysis of the Sun and solar analog stars, such as HD\,76151, ultimately enabling a more precise and detailed characterization of Fe abundance within their stellar atmospheres. In addition to $\alpha$-element and Fe abundances, the successful application of the line list to HD\,76151 facilitated the derivation of fundamental astrophysical parameters  using NIR spectroscopy.

\section{Convection and non-LTE effects}

In the test conducted to investigate a possible convection effect, the equation reported by \citet{Ludwig1999}, based on a 2D hydrodynamic model, and the equation reported by \citet{Magic2015}, based on a 3D hydrodynamic model, were used to calculate two different mixing length parameters ($\alpha$). The formula of \citet{Magic2015} yields an $\alpha=1.96$ and that of \citet{Ludwig1999} yields an $\alpha=1.58$. In this context, two different {\sc ATLAS9} models \citep{Castelli2003} were created for two different mixing length parameters. The synthetic spectra calculated using these models were compared with the observed HD\,76151 spectra. Although no obvious differences were observed, the synthetic spectrum derived from the mixing length parameter obtained in the study by \citet{Magic2015} was found to be relatively more consistent with the observed spectrum. To calculate element abundances, the model atmospheric parameters had to be refined for the convective model. A correction of +0.1 km s$^{\rm -1}$ was applied to the microturbulence velocity to make the abundances independent of the line EWs. Similarly, a correction of +50 K to the effective temperature value and +0.05 cgs to the gravitational acceleration was required to reconstruct the excitation and ionization equilibrium. The final set of model parameters (5830 K, 4.40 cgs, 0.81 km s$^{\rm -1}$) was found to lead to an average change of 0.01$\pm$0.03 dex for 22 species. For the F-G-K dwarf field stars, the effect of the mixing length parameter on the metallicity determination is expected to be negligible, with an equivalence of less than 0.02 dex for eight dwarf stars \citep{Song2020}. The logarithmic abundance of ionized Fe showed a -0.01 dex decrease for the ATLAS model of $\alpha=1.96$ and the refined model atmospheric parameters from the HARPS spectrum of the star. The largest difference was observed for Na\,{\sc i} with +0.08 dex.

Given that the Fe\,{\sc i} and Fe\,{\sc ii} abundances were used to constrain the model atmospheric parameters in this study, we must consider the non-LTE effects on Fe. These effects were found to be insignificant for Fe\,{\sc ii} lines (\citealp[]{Bergemann2012}; \citealp[]{Lind2012}; \citealp[]{Bensby2014}). According to \citet{Lind2012}, departures from LTE (non-LTE) for Fe\,{\sc ii} lines with low excitation potentials ($<$8 eV) at metallicities [Fe/H] $>$ -3.0 dex were minimal. To account for non-LTE effects on the Fe\,{\sc i} lines, we utilized the 1D non-LTE investigation conducted in several previous studies. The investigation was performed separately for the IGRINS spectra of HD\,76151 and the KPNO-NIR solar spectrum. In the NIR region, the non-LTE corrections\footnote{$\Delta$log$\epsilon$(Fe\,{\sc i}) = log$\epsilon$(Fe\,{\sc i})$_{\rm NLTE}$ - log$\epsilon$(Fe\,{\sc i})$_{\rm LTE}$} calculated for four neutral silicon lines in the Solar spectrum amounted to -0.003 dex (\citealp[]{Si_Optic_IR}{}). Additionally, the non-LTE corrections for the three neutral titanium lines observed in the KPNO-NIR solar spectrum were summed to 0.047 dex (\citealp[]{Ti_Fe_IR}{}). The correction for a single silicon line detected in the IGRINS spectrum of HD\,76151 was -0.001 dex whereas that for a neutral titanium transition at 15\,543.756 \AA\, was 0.08 dex. The IGRINS and KPNO-NIR solar spectra comprised only one ionized titanium line at 15\,873.88 \AA. The non-LTE correction for this line is -0.007 dex. Lastly, the non-LTE correction for the 44 neutral iron lines in both spectra was 0.007 dex (\citealp[]{Ti_Fe_IR}{}).

The non-LTE correction value obtained for the 66 neutral iron lines measured in the KPNO-OPTICAL solar spectrum was 0.006 dex, whereas the 46 neutral Fe lines detected in the HARPS spectrum of HD\,76151 yielded a calculation of 0.008 dex for this value (\citealp[]{Fe_Optic}{}). For Mg\,{\sc i} (one), Si\,{\sc i} (seven), Ca\,{\sc i} (nine), and Mn\,{\sc i} (nine) detected in the HARPS spectrum, the non-LTE correction values were less than 0.03 dex (\citealp[]{Mg_Optic}{}; \citealp[]{Si_Optic_IR}{}; \citealp[]{Ca_Optic}{}; \citealp[]{Mn_Optic}{}), with the largest difference observed for titanium. Accordingly, the non-LTE correction for the 36 neutral titanium lines detected in the HARPS spectrum amounts to 0.06 dex (\citealp[]{Ti_Optic}{}). It was 0.07 dex over the 43 Ti\,{\sc i} lines detected in the KPNO solar spectrum. The non-LTE corrections for the 16 neutral chromium lines in the HARPS and KPNO-OPTICAL solar spectra were 0.05 dex (\citealp[]{Cr_Optic}{}), while the non-LTE correction for Co (\citealp[]{Co_Optic}{}) in the same spectra was approximately 0.1 dex. Thus, the average non-LTE corrections for elemental abundances reported in the optical and NIR solar spectra are consistent with those reported for HD\,76151.



\begin{table*}
\setlength{\tabcolsep}{8.6pt}
\renewcommand{\arraystretch}{1}
\caption{ The $\alpha$-element line list in $Y$-, $J$-, $H$-, and $K$- bands for the analyses of the spectra for HD\,76151 and the Sun. Abundances for individual lines are those obtained for KPNO-NIR spectrum and a model of $T_{\rm eff}$= 5780 K, $\log g =$ 4.40 and $\xi$ = 1.08 km s$^{\rm -1}$.} 
\label{tab:A2}
    \centering    
    \begin{tabular}{c|c|c|c|c|c|c|c|c|c|c}
    \hline
  Species	& $\lambda$ & EW & $\log\epsilon$(X)	&	LEP & $\log \Gamma$ &\multicolumn{5}{c}{$\log gf$} \\
    \cline{2-11} 
 &   (\AA)  &(m\AA)  &(dex) 	&(eV)		&(rad cm$^3$s$^{-1}$)&	VALD & MB99	& DR17 &	BO03	&	NIST\\ 
\cline{1-11}
Ti\,{\sc i}	&	10003.09	&	4.9	    &	4.98	&	2.16	&	$-7.83^{\bf K} $	&	-1.02	&	\bf-1.32	&	-	&	\bf-1.32	&	\bf-1.32	\\
Ti\,{\sc i}	&	10011.74	&	4.6	    &	4.80	&	2.15	&	$-7.83^{\bf K} $	&	\bf-1.18	&	-1.54	&	-	&	-	&	-	\\
Si\,{\sc i}	&	10013.86	&	7.3     &	7.49	&	6.40	&	$-7.05^{\bf V}$	&	-2.32	&	\bf-1.73	&	-	&	-	&	-	\\
Ti\,{\sc i}	&	10034.49	&	4.2	    &	4.98	&	1.46	&	$-7.83^{\bf K} $	&	-1.84	&	\bf-2.09	&	-	&	-2.10	&	-1.77	\\
Si\,{\sc i}	&	10068.33	&	26.5	&	7.54	&	6.10	&	$-7.23^{\bf V}$	&	-1.32	&	\bf-1.40	&	-	&	-	&	-0.80	\\
Si\,{\sc i}	&	10098.55	&	7.1	    &	7.50	&	6.40	&	$-7.06^{\bf V}$	&	-1.83	&	\bf-1.76	&	-	&	-	&	-	\\
Ca\,{\sc i}	&	10249.15	&	5.0	    &	6.47	&	4.53	&	$-7.02^{\bf K} $	&	\bf-1.26	&	-0.96	&	-	&	-	&	-	\\
Si\,{\sc i}	&	10288.94	&	80.7	&	7.50	&	4.92	&	UNS	&	-1.51	&	\bf-1.71	&	-	&	-	&	-1.48	\\
Mg\,{\sc i}	&	10299.24	&	7.3	    &	7.46	&	6.12	&	UNS	&	-2.07	&	-2.06	&	-	&	-	&	\bf-2.08	\\
Si\,{\sc i}	&	10301.41	&	11.6	&	7.52	&	6.10	&	$-7.24^{\bf V}$	&	-2.17	&	\bf-1.83	&	-	&	-	&	-	\\
Mg\,{\sc i}	&	10312.52	&	16.5	&	7.49	&	6.12	&	UNS	&	-1.52	&	-1.71	&	-	&	-	&	\bf-1.73	\\
Si\,{\sc i}	&	10313.20	&	12.5	&	7.55	&	6.40	&	$-7.08^{\bf V}$	&	-0.89	&	\bf-1.56	&	-	&	-	&	-0.96	\\
Ca\,{\sc i}	&	10343.82	&	129.9	&	6.34	&	2.93	&	$-7.57^{\bf K} $	&	\bf-0.30	&	-0.40	&	-	&	-0.49	&	-	\\
Si\,{\sc i}	&	10371.27	&	164.3	&	7.45	&	4.93	&	UNS	&	-0.71	&	-0.80	&	-	&	\bf-0.88	&	-0.71	\\
Ti\,{\sc i}	&	10396.81	&	28.5	&	5.00	&	0.85	&	$-7.84^{\bf K} $	&	-1.54	&	\bf-1.79	&	-	&	-1.83	&	-1.83	\\
Ti\,{\sc i}	&	10496.09	&	23.0	&	4.98	&	0.84	&	$-7.84^{\bf K} $	&	-1.65	&	\bf-1.91	&	-	&	-1.92	&	-1.92	\\
Si\,{\sc i}	&	10627.65	&	121.2	&	7.40	&	5.86	&	$-7.39^{\bf V}$	&	-0.87	&	-0.50	&	-	&	\bf-0.29	&	-	\\
Ti\,{\sc i}	&	10732.87	&	4.0	&	5.01	&	0.83	    &	$-7.84^{\bf K} $	&	-2.52	&	\bf-2.82	&	-	&	-2.87	&	-2.87	\\
Si\,{\sc i}	&	10784.56	&	88.6	&	7.44	&	5.96	&	UNS	&	-0.84	&	\bf-0.72	&	-	&	-0.59	&	-0.64	\\
Si\,{\sc i}	&	10796.11	&	20.7	&	7.50	&	6.18	&	$-7.32^{\bf V}$	&	-1.27	&	\bf-1.49	&	-	&	-	&	-	\\
Ca\,{\sc i}	&	10838.98	&	28.3	&	6.23	&	4.88	&	$-7.63^{\bf K} $	&	0.24	&	\bf0.03	&	-	&	-	&	0.29	\\
Ca\,{\sc i}	&	10861.59	&	9.0	&	6.31	&	4.88	    &	$-7.68^{\bf K} $	&	-0.34	&	-0.49	&	-	&	\bf-0.57	&	-	\\
Ca\,{\sc i}	&	10879.88	&	8.9	&	6.33	&	4.88	    &	$-7.65^{\bf K} $	&	-0.36	&	-0.51	&	-	&	\bf-0.59	&	-	\\
Si\,{\sc i}	&	10885.35	&	132.4	&	7.51	&	6.18	&	$-7.33^{\bf V}$	&	0.22	&	\bf-0.10	&	-	&	-	&	-	\\
Si\,{\sc i}	&	10893.68	&	8.2	&	7.48	&	6.19	    &	$-7.32^{\bf V}$	&	-1.66	&	\bf-1.92	&	-	&	-	&	-	\\
Si\,{\sc i}	&	10894.80	&	14.1	&	7.50	&	6.19	&	$-7.32^{\bf V}$	&	-1.47	&	\bf-1.68	&	-	&	-	&	-	\\
Si\,{\sc i}	&	10984.55	&	88.4	&	7.36	&	6.19	&	$-7.33^{\bf V}$	&	\bf-0.37	&	-0.63	&	-	&	-	&	-	\\
Si\,{\sc i}	&	11627.56	&	19.1	&	7.52	&	5.96	&	$-7.40^{\bf V}$	&	-1.93	&	\bf-1.83	&	-	&	-	&	-	\\
Si\,{\sc i}	&	11640.96	&	95.5	&	7.47	&	6.27	&	$-7.32^{\bf V}$	&	\bf-0.43	&	-0.48	&	-	&	-	&	-0.49	\\
Si\,{\sc i}	&	11863.92	&	37.6	&	7.52	&	5.98	&	$-7.39^{\bf V}$	&	\bf-1.46	&	-1.50	&	-	&	-1.29	&	-	\\
Si\,{\sc i}	&	11900.03	&	20.5	&	7.49	&	5.96	&	$-7.38^{\bf V}$	&	-1.86	&	\bf-1.79	&	-	&	-	&	-	\\
Ca\,{\sc i}	&	11955.95	&	21.5	&	6.30	&	4.13	&	$-7.36^{\bf K} $	&	-0.85	&	\bf-0.91	&	-	&	-	&	-	\\
Si\,{\sc i}	&	12390.17	&	80.0	&	7.46	&	5.08	&	$-7.52^{\bf V}$	&	-1.77	&	-1.93	&	-	&	\bf-1.73	&	-1.97	\\
Si\,{\sc i}	&	12395.84	&	99.7	&	7.50	&	4.95	&	UNS	&	-1.64	&	-1.82	&	-	&	\bf-1.76	&	-1.68	\\
Si\,{\sc i}	&	12583.95	&	64.7	&	7.48	&	6.62	&	$-7.13^{\bf V}$	&	-0.46	&	\bf-0.62	&	-	&	-	&	-	\\
Ca\,{\sc i}	&	12909.08	&	30.4	&	6.25	&	4.43	&	$-7.79^{\bf K} $	&	-0.22	&	\bf-0.50	&	-	&	-	&	-	\\
Si\,{\sc i}	&	13029.54	&	51.8	&	7.58	&	6.08	&	$-7.40^{\bf V}$	&	-0.92	&	\bf-1.37	&	-	&	-	&	-0.71	\\
Si\,{\sc i}	&	13030.97	&	91.2	&	7.35	&	6.08	&	$-7.40^{\bf V}$	&	-0.67	&	-0.99	&	-	&	-	&	\bf-0.71	\\
Ca\,{\sc i}	&	13033.56	&	47.7	&	6.31	&	4.44	&	$-7.79^{\bf K} $	&	-0.06	&	\bf-0.31	&	-	&	-	&	-	\\
Ca\,{\sc i}	&	13134.94	&	59.8	&	6.05	&	4.45	&	$-7.79^{\bf K} $	&	\bf0.09	&	0.14	&	-	&	-	&	-	\\
Si\,{\sc i}	&	13154.56	&	26.3	&	7.48	&	6.62	&	$-7.10^{\bf V}$	&	-1.00	&	\bf-1.20	&	-	&	-	&	-	\\
Ca\,{\sc i}	&	15118.20	&	18.7	&	6.40	&	5.02	&	$-6.98^{\bf K} $	&	-2.22	&	\bf-0.65	&	-2.23	&	-	&	-	\\
Si\,{\sc i}	&	15127.91	&	5.4	&	7.75	&	7.23	    &	UNS	&	\bf-1.90	&	-	&	-3.9	&	-	&	-	\\
Si\,{\sc i}	&	$15361.16^{\bf S} $	&	25.9	&	7.49	    &	5.95	&	UNS	&	\bf-2.08	&	\bf-2.08	&	-2.03	&	-	&	-1.71	\\
Ti\,{\sc i}	&	$15543.78^{\bf S} $	&	18.8	&	5.01	    &	1.88	&	$-7.83^{\bf K} $	&	-1.13	&	\bf-1.48	&	-1.16	&	-	&	-1.08	\\
Ti\,{\sc i}	&	$15602.84^{\bf S} $	&	4.5	&	4.88	&2.27	&	$-7.83^{\bf K} $	&	-1.44	&	-1.81	&	-1.63	&	-	&	-	\\
Ti\,{\sc ii}	&	15873.84	&	21.6	&	4.82	&3.12	&	$-7.86^{\bf K} $ 	&	-1.81	&	-	&	\bf-1.95	&	-	&	-	\\
Ca\,{\sc ii}	&	16649.88	&	32.0	&	6.22	&9.24	&	$-7.39^{\bf K} $  	&	0.64	&	0.59	&	\bf0.57	&	-	&	-	\\
\hline
    \end{tabular}
    \vspace{1ex}
     \footnotesize{($^{S}$) The $\log gf$ values reported by \cite{Smith2013} are -1.93 dex for 15\,361.16 \AA, -1.12 dex for 15\,543.78 \AA, and  {\bf -1.64} dex for 15\,602.84 \AA, respectively. ($^{\textbf{K}}$) Kurucz damping,  ($^{\textbf{V}}$) VALD3 damping, and [UNS] Unsold damping.}
\end{table*}

\begin{table*}
\setlength{\tabcolsep}{9pt}
\caption{The iron and $\alpha$-element line list in $Y$-, $J$-, $H$-, and $K$- bands for the analyses of the spectra for HD\,76151 and the Sun. The transitions in blue are additional lines added to the new line list by RMT analysis applied to the transitions reported by \citet{Andreasen2016}; U/I: Unidentified line by \citet{Nave1994}.} 
\label{tab:A3}
    \centering
    \begin{tabular}{c|c|c|c|c|c|c|c|c|c|c}
    \hline
Species	& $\lambda$ & EW & $\log\epsilon$(X)	&	LEP  &$\log \Gamma$&\multicolumn{4}{c|}{$\log gf$} & RMT \\
    \cline{2-10} 
  &   (\AA)  &  (m\AA)  & (dex)  & (eV) 	&(rad cm$^3$s$^{-1}$)&VALD & MB99	& AN16 &	BO03	 & \\ 
\cline{1-11}
Ca\,{\sc i}	&	19917.19	&	108.3	&	6.18	&	1.90	&	$-7.80^{\bf K} $	&	\bf-2.32	&	-	&	-	&	-	&	-	\\
Si\,{\sc i}	&	20296.36	&	141.3	&	7.22	&	6.72	&	UNS &	\bf-0.16	&	-	&	-	&	-	&	-	 \\
Si\,{\sc i}	&	20343.84	&	111.6	&	7.80	&	6.12	&	UNS &	\bf-1.40	&	-	&	-	&	-	&	-	 \\
Ti\,{\sc i}	&	21782.94	&	34.9	&	4.75	&	1.75	&	$-7.83^{\bf K} $	&	\bf-1.11	&	- 	&	-	&	-	&	-	 \\
Ti\,{\sc i}	&	21897.38	&	18.4	&	4.71	&	1.74	&	$-7.83^{\bf K} $	&	\bf-1.40	&	-	&	-	&	-	&	-	 \\
Fe\,{\sc i}	&	10006.86	&	2.5	    &	7.46	&	5.51	&	$-7.70^{\bf K} $	&	-1.77	&	\bf-1.97	&	-	&	-	&	1998	 \\
Fe\,{\sc i}	&	10019.79	&	9.4	    &	7.51	&	5.48	&	$-7.70^{\bf K} $	&	-1.46	&	\bf-1.44	&	-	&	-	&	1996	 \\
Fe\,{\sc i}	&	10065.05	&	87.8	&	7.36	&	4.84	&	$-7.57^{\bf K} $	&	\bf-0.29	&	-0.57	&	-	&	-	&	1781	 \\
Fe\,{\sc i}	&	10070.52	&	7.0	    &	7.50	&	5.51	&	$-7.70^{\bf K} $	&	-1.93	&	\bf-1.54	&	-1.53	&	-	&	1993	 \\
Fe\,{\sc i}	&	10077.19	&	3.5	    &	7.48	&	2.99	&	$-7.82^{\bf K} $	&	-4.27	&	\bf-4.28	&	-	&	-	&	833	 \\
Fe\,{\sc i}	&	10081.39	&	6.2     &	7.45	&	2.42	&	$-7.83^{\bf K} $	&	-4.54	&	-4.53	&	\bf-4.55	&	-4.42	&	345	 \\
Fe\,{\sc i}	&	10085.07	&	5.3	    &	7.48	&	4.58	&	$-7.76^{\bf K} $	&	-2.13	&	\bf-2.54	&	-	&	-	&	1746	 \\
Fe\,{\sc i}	&	10086.26	&	5.8	    &	7.49	&	2.95	&	$-7.82^{\bf K} $	&	-4.05	&	\bf-4.10	&	-	&	-3.99	&	833	 \\
Fe\,{\sc i}	&	10089.77	&	7.4	    &	7.69	&	5.45	&	$-7.70^{\bf K} $	&	-1.25	&	\bf-1.77	&	-	&	-	&	1993	 \\
Fe\,{\sc i}	&	10137.10	&	9.3	    &	7.52	&	5.09	&	$-7.71^{\bf K} $	&	-1.71	&	\bf-1.83	&	-1.77	&	-	&	1861	 \\
Fe\,{\sc i}	&	10149.08	&	3.5	&7.47&5.10	&	$-7.71^{\bf K} $	&	-2.12	&	\bf-2.23	&	 -	&	-	&	1861	 \\
Fe\,{\sc i}	&	10155.16	&	16.0	&	7.48	&	2.18	&	$-7.76^{\bf K} $	&	-4.23	&	\bf-4.36	&	-4.34	&	-4.25	&	267	 \\
Fe\,{\sc i}	&	10167.47	&	19.5	&	7.50	&	2.20	&	$-7.77^{\bf K} $	&	-4.12	&	\bf-4.26	&	-4.20	&	-4.13	&	267	 \\
Fe\,{\sc i}	&	10195.11	&	22.2	&	7.47	&	2.73	&	$-7.78^{\bf K} $	&	-3.58	&	\bf-3.63 	&	\bf-3.63	&	-	&	620	 \\
Fe\,{\sc i}	&	10216.32	&	120.8	&	7.43	&	4.73	&	$-7.57^{\bf K} $	&	\bf-0.06	&	-0.29	&	-	&	-0.18	&	1781	 \\
Fe\,{\sc i}	&	10218.41 	&41.1	&7.49&	3.07&	$-7.78^{\bf K} $	&	-2.74	&	\bf-2.93	&	-	&	-	&	905	 \\
Fe\,{\sc i}	&	10230.78	&	19.7	&	7.49	&	5.87	&	$-7.58^{\bf K} $	&	-3.32	&	\bf-0.70	&	-0.41	&	~	&	2224	 \\
Fe\,{\sc i}	&	10252.55	&	10.0	&	7.49	&	5.83	&	$-7.58^{\bf K} $	&	-1.03	&	\bf-1.08	&	-	&	-	&	2223	 \\
Fe\,{\sc i}	&	10262.46	&7.7	&7.50&5.48	&	$-7.70^{\bf K} $	&	-1.61	&	     \bf-1.54	&	-	&	-	&	1995	 \\
Fe\,{\sc i}	&	10265.22	&	7.3	&	7.44	&	2.22	    &	$-7.76^{\bf K} $	&	-4.54	&	\bf-4.67	&	\bf-4.67	&	-4.55	&	267	 \\
Fe\,{\sc i}	&	10283.77	&	6.1	&7.54&5.51	&	$-7.70^{\bf K} $	&	\bf-1.66	&	-1.57	&	-	&	-	&	1994	 \\
Fe\,{\sc i}	&	10307.45	&	6.2	&	7.46	&	4.59	    &	$-7.76^{\bf K} $	&	-2.07	&	\bf-2.45	&	-	&	-2.39	&	1746	 \\
Fe\,{\sc i}	&	10332.33	&	10.4	&	7.47	&	3.63	&	$-7.75^{\bf K} $	&	-2.94	&	\bf-3.15 	&	\bf-3.15	&	-3.05	&	1365	 \\
Fe\,{\sc i}	&	10333.18	&	8.6	&	7.46	&	4.59	    &	$-7.76^{\bf K} $	&	-2.59	&	\bf-2.30	&	-	&	-	&	1746	 \\
Fe\,{\sc i}	&	10340.89	&	46.5	&	7.43	&	2.20	&	$-7.77^{\bf K} $	&	-3.58	&	\bf-3.65	&	-3.67	&	-3.59	&	267	 \\
Fe\,{\sc i}	&	10345.20	&	6.4	&	7.41	&	6.16	    &	$-7.39^{\bf K} $	&	-4.44	&	\bf-0.91	&	-	&	-	&	2365	 \\
Fe\,{\sc i}	&	10347.96	&	36.6	&	7.47	&	5.39	&	$-7.70^{\bf K} $	&	-0.55	&	-0.82	&	 \bf-0.75	&	-0.70	&	1995	 \\
Fe\,{\sc i}	&	10353.81	&	23.4	&	7.48	&	5.39	&	$-7.70^{\bf K} $	&	-0.82	&	-1.09	&	 \bf-1.04	&	-0.97	&	1994	 \\
Fe\,{\sc i}	&	10362.70	&	12.1&7.51&5.48	&	$-7.70^{\bf K} $	&	\bf-1.34	&	\bf-1.34	&	-	&	-	&	1993	 \\
Fe\,{\sc i}	&	10364.06	&	17.4	&	7.52	&	5.45	&	$-7.70^{\bf K} $	&	-0.96	&	\bf-1.19	&	-1.13	&	-	&	1995	 \\
Fe\,{\sc i}	&	10379.01	&	18.2	&	7.46	&	2.22	&	$-7.76^{\bf K} $	&	-4.15	&	\bf-4.25	&	\bf-4.25	&	-	&	267	 \\
Fe\,{\sc i}	&	10388.74	&	7.5	&	7.48	&	5.45	    &	$-7.70^{\bf K} $	&	-1.47	&	\bf-1.57	&	-1.53	&	-	&	1994	 \\
Fe\,{\sc i}	&	10395.80	&	61.2	&	7.40	&	2.18	&	$-7.76^{\bf K} $	&	\bf-3.39	&	-3.42	&	-	&	-3.40	&	267	 \\
Fe\,{\sc i}	&	10423.03	&	22.3&7.47&2.69	&	$-7.78^{\bf K} $	&	-3.62	&	\bf-3.68	&	-	&	-3.58	&	620	 \\
Fe\,{\sc i}	&	10423.75	&	29.2	&	7.45	&	3.07	&	$-7.78^{\bf K} $	&	-2.92	&	\bf-3.13	&	-3.12	&	-3.03	&	905	 \\
Fe\,{\sc i}	&	10452.78	&	37.0	&	7.57	&	3.88	&	UNS	&	-	&	\bf-2.30	&	-	&	-	&	U/I 	 \\
Fe\,{\sc i}	&	10469.66	&88.9&	7.32&3.88	&	$-7.72^{\bf K} $	&	\bf-1.18	&	-1.37	&	-	&	-	&	1461	 \\
Fe\,{\sc i}	&	10532.24	&	63.7	&	7.47	&	3.93	&	$-7.73^{\bf K} $	&	-1.48	&	-1.76	&	 \bf-1.70	&	~	&	1461	 \\
Fe\,{\sc i}	&	10535.71	&	26.2	&	7.45	&	6.21	&	$-7.33^{\bf K} $	&	-0.11	&	  -	&	     \bf-0.22	&	-	&	none	 \\
Fe\,{\sc i}	&	10555.65	&	11.8	&	7.50	&	5.45	&	$-7.70^{\bf K} $	&	-1.11	&	\bf-1.39	&	-1.35	&	~	&	1993	 \\
Fe\,{\sc i}	&	10557.52	&	4.7	&	7.4  7&    5.59	        &	$-7.57^{\bf K} $	&	-1.78	&	\bf-1.66	&	-	&	-	&	2077	 \\
Fe\,{\sc i}	&	10577.14	&	16.9	&	7.51	&	3.30	&	$-7.77^{\bf K} $	&	-3.14	&	\bf-3.28	&	-3.24	&	-3.16	&	1060	 \\
\hline
    \end{tabular}
        \vspace{1ex}
     \footnotesize{($^{\textbf{K}}$) Kurucz damping and [UNS] Unsold damping.}
\end{table*}    

\begin{table*}
\setlength{\tabcolsep}{7.4pt}
\caption{The iron lines in the line list in $Y$-, $J$-, $H$-, and $K$- bands for the analyses of the spectra for HD\,76151 and the Sun. The transitions in blue are additional lines added to the new line list by RMT analysis applied to the transitions reported by \citet{Andreasen2016}.} 
\label{tab:A4}
    \centering
    \begin{tabular}{c|c|c|c|c|c|c|c|c|c|c|c}
    \hline
Species	& $\lambda$ & EW & $\log\epsilon$(X)	&	LEP  &$\log \Gamma$&\multicolumn{5}{c|}{$\log gf$} &RMT\\
    \cline{2-11} 
 &   (\AA)  &  (m\AA)  & (dex)  & (eV)	&(rad cm$^3$s$^{-1}$)&VALD & MB99	& AN16 &BO03	&	NIST & \\ 
\cline{1-12}
Fe\,{\sc i}	&	10611.68	&	35.5	&	7.43	&	6.17	&	$-7.33^{\bf K} $	&	0.02	&	0.09	&	\bf-0.07	&	-	&	-	&	none	\\
Fe\,{\sc i}	&	10616.72	&	15.2	&	7.48	&	3.27	&	$-7.77^{\bf K} $	&	-3.13	&	\bf-3.34	&	-3.32	&	-3.22	&	-	&	1060	\\
Fe\,{\sc i}	&	10642.68	&	4.4	&7.52&5.92	&	$-7.58^{\bf K} $	&	\bf-1.44	&	-1.37	&	-	&	-	&	-	&	2224	\\
Fe\,{\sc i}	&	10652.38	&	5.8	&	7.59	&	5.48	    &	$-7.70^{\bf K} $	&	-2.95	&	\bf-1.79	&	-	&	-	&	-	&	none	\\
Fe\,{\sc i}	&	10674.07	&	15.3	&	7.48	&	6.17	&	$-7.33^{\bf K} $	&	-0.47	&	-	&	\bf-0.58	&	-	&	-	&	none	\\
Fe\,{\sc i}	&	10717.81	&	5.4	&	7.49	&	5.54	    &	$-7.52^{\bf K} $	&	-0.44	&	\bf-1.68	&	-	&	-	&	-	&	2082	\\
Fe\,{\sc i}	&	10721.66	&	4.6	&	7.49	&	5.51	    &	$-7.70^{\bf K} $	&	-1.62	&	\bf-1.78	&	-	&	-1.79	&	-	&	2035	\\
Fe\,{\sc i}	&	10725.21	&	16.2	&	7.47	&	3.64	&	$-7.75^{\bf K} $	&	-2.76	&	-2.98	&	\bf-2.95	&	~	&	-	&	1365	\\
Fe\,{\sc i}	&	10753.01	&	39.5	&	7.51	&	3.96	&	$-7.73^{\bf K} $	&	-1.85	&	\bf-2.14	&	-	&	-	&	-	&	1461	\\
Fe\,{\sc i}	&	10780.70	&	9.3	&	7.46	&	3.24	    &	$-7.77^{\bf K} $	&	-3.29	&	-3.59	&	\bf-3.61	&	-3.48	&	-	&	1060	\\
Fe\,{\sc i}	&	10783.05	&	47.4&7.47&3.11	&	$-7.78^{\bf K} $	&	-2.57	&	\bf-2.80	&	-	&	-2.71	&	-	&	905	\\
Fe\,{\sc i}	&	10818.28	&	35.2	&	7.46	&	3.96	&	$-7.73^{\bf K} $	&	-1.95	&	-2.23	&	\bf-2.18	&	-	&	-	&	1461	\\
Fe\,{\sc i}	&	10849.46	&	35.2	&	7.51	&	5.54	&	$-7.58^{\bf K} $	&	-1.44	&	\bf-0.73	&	-	&	-	&	-	&	2080	\\
Fe\,{\sc i}	&	10863.52	&	59.3	&	7.49	&	4.73	&	$-7.58^{\bf K} $	&	-0.90	&	\bf-1.06	&	-	&	-	&	-	&	1780	\\
Fe\,{\sc i}	&	10881.76	&	24.7	&	7.46	&	2.85	&	$-7.77^{\bf K} $	&	-3.60	&	\bf-3.50	&	-	&	-	&	-	&	732	\\
Fe\,{\sc i}	&	10884.26	&	38.4&7.49&3.93	&	$-7.72^{\bf K} $	&	-1.93	&	\bf-2.18	&	-	&	-	&	-	&	1461	\\
Fe\,{\sc i}	&	10896.30	&	42.6	&	7.46	&	3.07	&	$-7.78^{\bf K} $	&	-2.69	&	\bf-2.93	&	-	&	-2.85	&	-	&	905	\\
Fe\,{\sc i}	&	10970.02	&	5.2	&	7.42	&	5.99	    &	$-7.55^{\bf K} $	&	-2.01	&	\bf-1.23	&	-	&	-	&	-	&	2325	\\
Fe\,{\sc i}	&	11026.78	&	15.2	&	7.53	&	3.94	&	$-7.72^{\bf K} $	&	-2.81	&	\bf-2.77	&	-2.49	&	-	&	-	&	1438	\\
Fe\,{\sc i}	&	11045.60	&	21.7	&	7.53	&	5.59	&	$-7.52^{\bf K} $	&	-0.62	&	\bf-1.01	&	-	&	-	&	-	&	2082	\\
Fe\,{\sc i}	&	11071.71	&	12.8	&	7.46	&	3.07	&	$-7.78^{\bf K} $	&	-4.28	&	\bf-3.64	&	-	&	-	&	-	&	none	\\
Fe\,{\sc i}	&	11607.57	&	158.8	&	7.35	&	2.20	&	$-7.77^{\bf K} $	&	\bf-2.01	&	-2.46	&	-	&	-2.27	&	\bf-2.01	&	266	\\
Fe\,{\sc i}	&	11783.26	&	149.0	&	7.28	&	2.83	&	$-7.76^{\bf K} $	&	\bf-1.57	&	-1.86	&	-	&	-	&	\bf-1.57	&	733	\\
Fe\,{\sc i}	&	11882.86	&	193.3	&	7.54	&	2.20	&	$-7.77^{\bf K} $	&	-1.70	&	-2.20	&	-	&	\bf-2.03	&	-1.67	&	266	\\
Fe\,{\sc i}	&	12053.08	&	40.1	&	7.47	&	4.56	&	$-7.58^{\bf K} $	&	-1.49	&	-1.75	&	\bf-1.65	&	-1.56	&	-1.54	&	1712	\\
Fe\,{\sc i}	&	12119.50	&	28.2	&	7.50	&	4.59	&	$-7.76^{\bf K} $	&	-1.64	&	\bf-1.88	&	-1.81	&	-	&	-	&	1745	\\
Fe\,{\sc i}	&	12190.10	&	30.9	&	7.49	&	3.63	&	$-7.77^{\bf K} $	&	-2.56	&	\bf-2.75	&	-	&	-	&	-2.33	&	1364	\\
Fe\,{\sc i}	&	12213.34	&	17.6	&	7.49	&	4.64	&	$-7.59^{\bf K} $	&	-1.85	&	\bf-2.09	&	-2.02	&	-1.93	&	-	&	1712	\\
Fe\,{\sc i}	&	12227.12	&	45.9	&	7.39	&	4.61	&	$-7.58^{\bf K} $	&	-1.37	&	-1.60	&	-	&	\bf-1.45	&	-	&	1712	\\
Fe\,{\sc i}	&	12283.28	&	16.8	&	7.40	&	6.17	&	$-7.32^{\bf K} $	&	-0.54	&	\bf-0.61	&	\bf-0.61	&	-	&	-	&	2385	\\
Fe\,{\sc i}	&	12342.92	&	39.1	&	7.53	&	4.64	&	$-7.59^{\bf K} $	&	-1.46	&	\bf-1.68	&	-1.57	&	-1.57	&	-	&	1712	\\
Fe\,{\sc i}	&	12557.01	&	33.1	&	7.49	&	2.28	&	$-7.84^{\bf K} $	&	-3.63	&	\bf-4.07	&	-4.02	&	-	&	-	&	343	\\
Fe\,{\sc i}	&	12615.93	&	35.9	&	7.53	&	4.64	&	$-7.58^{\bf K} $	&	-1.52	&	\bf-1.77	&	-1.82	&	-	&	-	&	1712	\\
Fe\,{\sc i}	&	12638.72	&	104.2	&	7.42	&	4.56	&	$-7.58^{\bf K} $	&	\bf-0.78	&	-1.00	&	-	&	-	&	-	&	1712	\\
Fe\,{\sc i}	&	12879.78	&	66.1	&	7.49	&	2.28	&	$-7.84^{\bf K} $	&	-3.62	&	\bf-3.61	&	-	&	-	&	-3.46	&	343	\\
Fe\,{\sc i}	&	12896.12	&	22.2	&	7.51	&	4.91	&	$-7.58^{\bf K} $	&	-1.42	&	\bf-1.80	&	-1.79	&	-	&	~	&	1801	\\
Fe\,{\sc i}	&	12934.67	&	25.3	&	7.50	&	5.39	&	$-7.68^{\bf K} $	&	-0.95	&	\bf-1.28	&	-1.14	&	-	&	-	&	1986	\\
Fe\,{\sc i}	&	13006.70	&	31.6	&	7.53	&	2.99	&	$-7.84^{\bf K} $	&	-3.74	&	\bf-3.49	&	-	&	-	&	-	&	832	\\
Fe\,{\sc i}	&	13014.85	&	10.3	&	7.50	&	5.45	&	$-7.66^{\bf K} $	&	-1.69	&	\bf-1.68	&	-1.59	&	-	&	-	&	1987	\\
Fe\,{\sc i}	&	13039.65	&	14.7	&	7.50	&	5.66	&	$-7.70^{\bf K} $	&	-0.73	&	\bf-1.32	&	-	&	-	&	-	&	2116	\\
Fe\,{\sc i}	&	13098.92	&22.3&	7.51&5.01	&	$-7.58^{\bf K} $	&	-1.29	&	\bf-1.73	&	-	&	-	&	-	&	1801	\\
Fe\,{\sc i}	&	13147.93	&	55.4	&	7.47	&	5.39	&	$-7.68^{\bf K} $	&	-0.81	&	-0.93	&	\bf-0.78	&	-	&	-	&	1987	\\
Fe\,{\sc i}	&	13291.78	&	11.1	&	7.43	&	5.48	&	$-7.66^{\bf K} $	&	-1.90	&	\bf-1.58	&	-	&	-	&	-	&	1987	\\
Fe\,{\sc i}	&	13352.18	&	89.0	&	7.48	&	5.31	&	$-7.52^{\bf K} $	&	-0.52	&	-0.55	&	\bf-0.48	&	-	&	-	&	1957	\\
Fe\,{\sc i}	&	13392.11	&	114.7	&	7.43	&	5.35	&	$-7.51^{\bf K} $	&	\bf-0.12	&	-0.25	&	-0.21	&	-	&	-	&	1958	\\
\hline
    \end{tabular}
            \vspace{1ex}
     \footnotesize{($^{\textbf{K}}$) Kurucz damping}
\end{table*}

\begin{table*}
\setlength{\tabcolsep}{7pt}
\caption{The iron lines in the line list in $Y$-, $J$-, $H$-, and $K$- bands for the analyses of the spectra for HD\,76151 and the Sun. The transitions in blue are additional lines added to the new line list by RMT analysis applied to the transitions reported by \citet{Andreasen2016}; U/I: Unidentified line by \citet{Nave1994}.} 
\label{tab:A5}
    \centering
    \begin{tabular}{c|c|c|c|c|c|c|c|c|c|c|c}
    \hline
 Species	& $\lambda$ & EW & $\log\epsilon$(X)	&	LEP  &$\log \Gamma$&\multicolumn{5}{c|}{$\log gf$} &RMT\\
    \cline{2-11} 
 &   (\AA)  &  (m\AA)  & (dex)  & (eV)	&(rad cm$^3$s$^{-1}$)&VALD & MB99	& AN16 &DR17&NIST & \\ 
\cline{1-12}
Fe\,{\sc i}	&	15080.25	&	16.7	&	7.39	&	6.26	&	$-7.71^{\bf K} $	&	\bf-0.84	&	-1.19	&	-	&	-0.61	&	-	&	2465	\\
Fe\,{\sc i}	&	15120.50	&	23.0	&	7.47	&	5.45	&	$-7.71^{\bf K} $	&	\bf-1.49	&	-1.58	&	-	&	-1.02	&	-0.80	&	1983	\\
Fe\,{\sc i}	&	15122.38	&	94.0	&	7.40	&	5.62	&	$-7.55^{\bf K} $	&	-0.21	&	-0.61	&	\bf-0.38	&	\bf-0.38	&	-	&	2066	\\
Fe\,{\sc i}	&	15176.72	&	31.6	&	7.51	&	5.92	&	$-7.58^{\bf K} $	&	-0.50	&	\bf-0.95	&	-0.88	&	-0.76	&	-	&	2217	\\
Fe\,{\sc i}	&	15179.75	&	13.3	&	7.42	&	5.98	&	$-7.58^{\bf K} $	&	\bf-1.24	&	-1.75	&	-	&	-3.28	&	-	&	2217	\\
Fe\,{\sc i}	&	15182.92	&	18.0	&	7.18	&	6.33	&	$-7.70^{\bf K} $	&	-0.53	&	-	&	-	&	\bf-0.54	&	-	&	2426	\\
Fe\,{\sc i}	&	$15194.50^{S}$	&	13.2	&7.47&	2.22	    &	$-7.76^{\bf K} $	&	-4.82	&	\bf-4.85	&	-4.80	&	-4.74	&	-	&	265	\\
Fe\,{\sc i}	&	15201.57	&	28.2	&	7.56	&	5.49	&	UNS	&	-0.16	&	\bf-1.45	&	-1.34	&	-0.41	&	-	&	U/I	\\
Fe\,{\sc i}	&	$15207.54^{S}$	&	193.9	&	7.64&	5.39	&	$-7.52^{\bf K} $	&	0.32	&	-0.10	&	\bf0.04	&	0.06	&	-	&	1956	\\
Fe\,{\sc i}	&	15213.02	&	17.6	&	7.44	&	6.31	&	$-7.70^{\bf K} $	&	-0.47	&	\bf-0.83	&	-0.94	&	-0.63	&	-	&	2453	\\
Fe\,{\sc i}	&	15237.77	&	9.3	&	7.44	&	5.79	    &	$-7.70^{\bf K} $	&	-1.91	&	-1.59	&	\bf-1.60	&	-1.47	&	-	&	2167	\\
Fe\,{\sc i}	&	15239.74	&	49.5	&	7.38	&	6.42	&	$-7.50^{\bf K} $	&	-0.03	&	\bf-0.15	&	-0.03	&	-0.05	&	-	&	2591	\\
Fe\,{\sc i}	&	15267.02	&	6.5	&	7.50	&	5.07	    &	$-7.72^{\bf K} $	&	-2.16	&	\bf-2.49	&	-	&	-2.42	&	-	&	1896	\\
Fe\,{\sc i}	&	15301.56	&	40.5	&	7.52	&	5.92	&	$-7.58^{\bf K} $	&	-0.69	&	\bf-0.84	&	-	&	-0.65	&	-	&	2212	\\
Fe\,{\sc i}	&	15315.66	&	9.0	&	7.43	&	6.28	    &	$-7.45^{\bf K} $	&	-3.40	&	\bf-1.17	&	-	&	-2.86	&	-	&	2395	\\
Fe\,{\sc i}	&	15343.81	&	65.2	&	7.52	&	5.65	&	$-7.56^{\bf K} $	&	-0.58	&	\bf-0.78	&	-	&	-0.63	&	-	&	2066	\\
Fe\,{\sc i}	&	15381.98	&	42.2	&	7.61	&	3.64	&	UNS	&	-	&	\bf-3.03	&	-	&	-	&	-	&	U/I	\\
Fe\,{\sc i}	&	15427.61	&	8.4	&	7.55	&	6.45	    &	$-7.42^{\bf K} $	&	-0.66	&	-0.98	&	\bf-1.18	&	-0.76	&	-	&	2574	\\
Fe\,{\sc i}	&	15522.64	&	12.3	&	7.49	&	6.32	&	$-7.71^{\bf K} $	&	-1.12	&	\bf-1.07	&	-0.88	&	-0.92	&	-	&	2442	\\
Fe\,{\sc i}	&	15524.30	&	13.7	&	7.50	&	5.79	&	$-7.70^{\bf K} $	&	-0.88	&	\bf-1.51	&	-1.45	&	-1.46	&	-	&	2166	\\
Fe\,{\sc i}	&	15527.21	&	10.5	&	7.46	&	6.32	&   UNS	&	-1.81	&	-1.01	&	-0.98	&	\bf-1.11	&	-	&	2422	\\
Fe\,{\sc i}	&	15534.26	&	108.0	&	7.45	&	5.64	&	$-7.55^{\bf K} $	&	-0.38	&	-0.47	&	\bf-0.34	&	-0.32	&	-	&	2066	\\
Fe\,{\sc i}	&	15565.23	&	14.8	&	7.46	&	6.32	&	$-7.70^{\bf K} $	&	-0.56	&	-0.95	&	\bf-0.96	&	-0.77	&	-	&	2424	\\
Fe\,{\sc i}	&	15566.72	&	33.9	&	7.48	&	6.35	&	$-7.71^{\bf K} $	&	-0.68	&	-0.50	&	\bf-0.54	&	-0.35	&	-	&	2469	\\
Fe\,{\sc i}	&	15579.08	&	11.5&7.47&6.32	&   UNS&	-2.12	&	-0.99	&	\bf-1.09	&	-1.15	&	-	&	2420	\\
Fe\,{\sc i}	&	15611.15	&	50.2	&	7.43	&	3.41	&	$-7.78^{\bf K} $	&	-3.77	&	-3.20	&	\bf-2.98	&	-3.48	&	-	&	1224	\\
Fe\,{\sc i}	&	$15648.52^{S}$	&	90.6	&	7.46&	5.43	&	$-7.52^{\bf K} $	&	-0.60	&	-0.80	&	-0.73	&	-0.65	&	-	&	1956	\\
Fe\,{\sc i}	&	15652.87	&	74.1&7.49&6.25	&	$-7.70^{\bf K} $	&	-0.16	&	\bf-0.19	&	-0.17	&	-0.01	&	-	&	2421	\\
Fe\,{\sc i}	&	15670.13	&	14.5	&	7.45	&	6.20	&	$-7.43^{\bf K} $	&	-0.99	&	-1.04	&	\bf-1.07	&	-0.85	&	-	&	2378	\\
Fe\,{\sc i}	&	15700.08	&	8.8	&	7.34	&	6.33	    &	$-7.30^{\bf K} $	&	-2.15	&	\bf-1.08	&	-	&	-0.98	&	-	&	2420	\\
Fe\,{\sc i}	&	15731.41	&	17.9	&	7.35	&	6.45	&	$-7.49^{\bf K} $	&	-0.34	&	\bf-0.66	&	-0.87	&	-0.47	&	-	&	2578	\\
Fe\,{\sc i}	&	15733.51	&	26.1	&	7.46	&	6.25	&	$-7.71^{\bf K} $	&	-0.98	&	\bf-0.76	&	-0.90	&	-0.56	&	-	&	2439	\\
Fe\,{\sc i}	&	15829.29	&	11.4	&	7.44	&	6.30	&	$-7.70^{\bf K} $	&	-1.20	&	\bf-1.10	&	-	&	-0.83	&	-	&	2410	\\
Fe\,{\sc i}	&	15840.19	&	39.3	&	7.40	&	6.36	&	$-7.70^{\bf K} $	&	-1.16	&	\bf-0.40	&	-	&	-0.28	&	-	&	2412	\\
Fe\,{\sc i}	&	15858.66	&	26.5	&	7.45	&	5.59	&	$-7.58^{\bf K} $	&	-0.49	&	\bf-1.34	&	-1.08	&	-1.22	&	-	&	2064	\\
Fe\,{\sc i}	&	15868.52	&	147.4	&	7.51	&	5.59	&	$-7.48^{\bf K} $	&	0.09	&	-0.26	&	-0.24	&	\bf-0.17	&	-	&	2063	\\
Fe\,{\sc i}	&	15901.53	&	62.3&7.49&5.92	&	$-7.58^{\bf K} $	&	-0.88	&	\bf-0.60	&	-0.57	&	-0.51	&	-	&	2209	\\
Fe\,{\sc i}	&	15904.35	&	137.1	&	7.36	&	6.36	&	$-7.69^{\bf K} $	&	\bf0.45	&	0.25	&	-	&	0.25	&	-	&	2483	\\
Fe\,{\sc i}	&	15906.04	&	137.9	&	7.49	&	5.62	&	$-7.48^{\bf K} $	&	-0.34	&	-0.34	&	\bf-0.20	&	-0.25	&	-	&	2061	\\
Fe\,{\sc i}	&	15929.48	&	33.0&7.45&6.31	&	$-7.71^{\bf K} $	&	-0.38	&	\bf-0.59	&	-0.61	&	-0.43	&	-	&	2441	\\
Fe\,{\sc i}	&	15934.02	&	42.8&7.43&6.31	&	$-7.71^{\bf K} $	&	-0.29	&	\bf-0.43	&	-0.44	&	-0.21	&	-	&	2441	\\
Fe\,{\sc i}	&	15940.92	&	17.1	&	7.51	&	5.81	&	$-7.70^{\bf K} $	&	-1.59	&	\bf-1.42	&	-1.23	&	-1.25	&	-	&	2186	\\
Fe\,{\sc i}	&	15941.85	&	70.7	&	7.46	&	6.36	&	$-7.53^{\bf K} $	&	0.27	&	-0.04	&	\bf-0.13	&	0.06	&	-	&	none	\\
Fe\,{\sc i}	&	15971.25	&	35.0	&	7.38	&	6.41	&	$-7.49^{\bf K} $	&	-0.14	&	\bf-0.41	&	-	&	-0.32	&	-	&	2576	\\
Fe\,{\sc i}	&	16019.79	&	19.9	&	7.50	&	6.35	&	$-7.70^{\bf K} $	&	-0.59	&	-0.81	&	\bf-0.87	&	-0.59	&	-	&	2411	\\
Fe\,{\sc i}	&	16051.74	&	16.8	&	7.48	&	6.26	&	$-7.71^{\bf K} $	&	-0.94	&	-0.95	&	\bf-1.01	&	-0.80	&	-	&	2463	\\
Fe\,{\sc i}	&	16171.93	&	29.7	&	7.37	&	6.38	&	$-7.70^{\bf K} $	&	-0.45	&	\bf-0.52	&	-0.47	&	-0.43	&	-	&	2483	\\
\hline
    \end{tabular}
        \vspace{1ex}
     \footnotesize{($^{S}$) The $\log gf$ (values reported by \cite{Smith2013} are -4.78 dex for 15\,194.50 \AA, 0.08 dex for 15\,207.54 \AA, and {\bf -0.70} dex for 15\,648.52 \AA, respectively. ($^{\textbf{K}}$) Kurucz damping, and [UNS] Unsold damping.}
\end{table*}  
\begin{table*}
\setlength{\tabcolsep}{7.5pt}
\caption{The iron lines in the line list in $Y$-, $J$-, $H$-, and $K$- bands for the analyses of the spectra for HD\,76151 and the Sun. The transitions in blue are additional lines added to the new line list by RMT analysis applied to the transitions reported by \citet{Andreasen2016}.} 
\label{tab:A6}
    \centering
    \begin{tabular}{c|c|c|c|c|c|c|c|c|c|c|c}
    \hline
 Species	& $\lambda$ & EW & $\log\epsilon$(X)	&	LEP &$\log \Gamma$&\multicolumn{5}{c|}{$\log gf$} &RMT\\
    \cline{2-11} 
 &   (\AA)  &  (m\AA)  & (dex)  & (eV)&(rad cm$^3$s$^{-1}$)&	VALD & MB99	& AN16 &DR17&NIST & \\ 
\cline{1-12}
Fe\,{\sc i}	&	16225.64	&	80.6	&	7.44	&	6.38	&	$-7.70^{\bf K} $	&	0.30	&	\bf-0.03	&	-	 &	0.11	&	-	 &	2483	\\
Fe\,{\sc i}	&	16228.67	&	6.3	    &	7.53	&	6.38	&	$-7.67^{\bf K} $	&	-0.89	&	-1.18	&	-	 &	-1.13	&	\bf-1.41	 &	none	\\
Fe\,{\sc i}	&	16231.67	&	142.2	&	7.41	&	6.38	&	$-7.67^{\bf K} $	&	0.60	&	\bf0.42	&	0.31	 &	0.47	&	-	&	2551	\\
Fe\,{\sc i}	&	16235.98	&	86.0	&	7.48	&	5.92	&	$-7.58^{\bf K} $	&	-0.02	&	-0.45	&	\bf-0.39	 &	-0.30	&	-	&	2218	\\
Fe\,{\sc i}	&	16252.57	&	38.0	&	7.46	&	6.32	&	$-7.55^{\bf K} $	&	-0.31	&	-0.50	&	\bf-0.54	 &	-0.41	&	-	&	2423	\\
Fe\,{\sc i}	&	16258.93	&	19.2	&	7.54	&	6.24	&	$-7.43^{\bf K} $	&	\bf-1.03	&	-0.90	&	-	 &	-0.78	&	-	&	2378	\\  
Fe\,{\sc i}	&	16277.50	&	34.4	&	7.38	&	6.32	&	$-7.43^{\bf K} $	&	-3.07	&	\bf-0.51	&	-	 &	-0.41	&	-	&	2417	\\
Fe\,{\sc i}	&	16474.09	&	58.0	&	7.50	&	6.02	&	$-7.57^{\bf K} $	&	-0.50	&	\bf-0.60	&	-0.57	 &	-0.62	&	-0.96	&	2297	\\
Fe\,{\sc i}	&	16506.30	&	71.0	&	7.53	&	5.95	&	$-7.57^{\bf K} $	&	-0.46	&	\bf-0.56	&	-	 &	-0.41	&	-	&	2284	\\
Fe\,{\sc i}	&	16532.01	&	74.7	&	7.47	&	6.29	&	$-7.70^{\bf K} $	&	-0.82	&	\bf-0.19	&	\bf-0.19	 &	-0.08	&	-	&	2463	\\
Fe\,{\sc i}	&	16544.70	&	46.4	&	7.46	&	6.34	&   UNS &	-0.03	&	\bf-0.42	&	-0.3	 &	-0.25	&	-	&	2549	\\
Fe\,{\sc i}	&	16552.02	&	75.9	&	7.47	&	6.41	&	$-7.66^{\bf K} $	&	0.34	&	-0.01	&	\bf-0.08	 &	0.12	&	-	&	2552	\\
Fe\,{\sc i}	&	16559.71	&	44.5	&	7.42	&	6.40	&	$-7.66^{\bf K} $	&	0.21	&	\bf-0.35	&	-0.31	 &	-0.20	&	-	&	2551	\\
Fe\,{\sc i}	&	16586.06	&	19.7	&	7.49	&	5.62	&	$-7.48^{\bf K} $	&	-0.75	&	\bf-1.53	&	-1.52	 &	-1.39	&	-	&	2063	\\
Fe\,{\sc i}	&	16607.65	&	31.9	&	7.43	&	6.34	&	$-7.41^{\bf K} $	&	-1.60	&	\bf-0.59	&	-0.57	 &	-0.51	&	-	&	2407	\\
Fe\,{\sc i}	&	16619.73	&	15.5	&	7.48	&	5.59	&	$-7.48^{\bf K} $	&	-0.86	&	\bf-1.66	&	-	 &	-1.47	&	-	&	2063	\\
Fe\,{\sc i}	&	16645.88	&	97.1	&	7.47	&	5.96	&	$-7.58^{\bf K} $	&	-0.03	&	-0.34	&	\bf-0.27	 &	-0.20	&	-	&	2209	\\
Fe\,{\sc i}	&	16648.24	&	24.0	&	7.38	&	6.55	&	$-7.49^{\bf K} $	&	\bf-0.50	&	-0.45	&	-	 &	-0.28	&	-	&	2644	\\
Fe\,{\sc i}	&	16665.49	&	97.8	&	7.52	&	6.02	&	$-7.57^{\bf K} $	&	-0.04	&	-0.30	&	\bf-0.26	 &	\bf-0.26	&	-	&	2291	\\
Fe\,{\sc i}	&	16685.58	&	54.2	&	7.49	&	6.34	&	$-7.27^{\bf K} $	&	-2.95	&	-0.56	&	-	 &	\bf-0.36	&	-	&	2404	\\
Fe\,{\sc i}	&	16693.11	&	37.1	&	7.40	&	6.42	&	$-7.66^{\bf K} $	&	-0.14	&	\bf-0.41	&	-	 &	-0.26	&	-	&	2552	\\
Fe\,{\sc i}	&	16898.90	&18.1&7.48	&6.31	&	$-7.70^{\bf K} $	&	-0.91	&	\bf-0.94	&	-	 &	-0.78	&	-	&	2451	\\
Fe\,{\sc i}	&	16900.23	&	7.1	&	7.37	&	6.30	    &	$-7.70^{\bf K} $	&	-0.81	&	\bf-1.27	&	-	 &	-1.15	&	-	&	2408	\\
Fe\,{\sc i}	&	16910.69	&	4.0	&	7.41	&	5.87	    &	$-7.70^{\bf K} $	&	-1.61	&	-1.94	&	-	 &	\bf-1.96	&	-	&	2246	\\
Fe\,{\sc i}	&	16969.91	&	91.2	&	7.52	&	5.95	&	$-7.57^{\bf K} $	&	-0.07	&	\bf-0.37	&	-	 &	-0.25	&	-	&	2287	\\
Fe\,{\sc i}	&	17005.45	&	94.2	&	7.52	&	6.07	&	$-7.57^{\bf K} $	&	0.01	&	\bf-0.25	&	-	 &	-0.08	&	-	&	2295	\\
Fe\,{\sc i}	&	17009.02	&	33.4	&	7.48	&	6.62	&	$-7.70^{\bf K} $	&	-0.30	&	-0.20	&	\bf-0.37	 &	-	&	-	&	2657	\\
Fe\,{\sc i}	&	17027.62	&	20.5	&	7.54	&	6.62	&	$-7.70^{\bf K} $	&	-	&	\bf-0.67	&	-	 &	-	&	-	&	2657	\\
Fe\,{\sc i}	&	17037.79	&	41.6	&	7.45	&	6.39	&	$-7.68^{\bf K} $	&	-0.85	&	\bf-0.42	&	-0.52	 &	-	&	-	&	2599	\\
Fe\,{\sc i}	&	17052.20	&	27.8	&	7.42	&	6.39	&	$-7.68^{\bf K} $	&	-0.74	&	\bf-0.60	&	-	 &	-	&	-	&	2599	\\
Fe\,{\sc i}	&	17067.67	&	64.2	&	7.43	&	6.37	&	$-7.46^{\bf K} $	&	-2.02	&	\bf-0.17	&	-	 &	-	&	-	&	none	\\
Fe\,{\sc i}	&	17072.86	&	8.0	&	7.44	&	5.07        &	$-7.72^{\bf K} $	&	-1.91	&	\bf-2.41	&	-	 &	-	&	-	&	1893	\\
Fe\,{\sc i}	&	17161.12	&	92.2	&	7.46	&	6.02	&	$-7.57^{\bf K} $	&	-0.07	&	-0.32	&	\bf-0.24	 &	-	&	-	&	2284	\\
Fe\,{\sc i}	&	17166.20	&	48.6&7.53&5.95	&	$-7.57^{\bf K} $	&	-0.81	&	\bf-0.79	&	-	 &	-	&	-	&	2291	\\
Fe\,{\sc i}	&	17200.34	&	14.2	&	7.38	&	6.36	&	$-7.69^{\bf K} $	&	-0.64	&	\bf-0.91	&	-	 &	-	&	-	&	2481	\\
Fe\,{\sc i}	&	17221.43	&	13.4	&	7.36	&	6.43	&	$-7.64^{\bf K} $	&	-0.71	&	\bf-0.85	&	-0.74	 &	-	&	-	&	2629	\\
Fe\,{\sc i}	&	17257.59	&	38.0	&	7.50	&	6.32	&	$-7.70^{\bf K} $	&	-0.38	&	\bf-0.58	&	-	 &	-	&	-	&	2421	\\
Fe\,{\sc i}	&	17278.72	&	15.4	&	7.27	&	6.72	&	$-7.70^{\bf K} $	&	\bf-0.44	&	-0.39	&	-	 &	-	&	-	&	2702	\\
Fe\,{\sc i}	&	17282.32	&	60.0	&	7.46	&	6.43	&	$-7.64^{\bf K} $	&	0.10	&	-0.16	&	\bf-0.19	 &	-	&	-	&	2629	\\
Fe\,{\sc i}	&	17420.83	&	6.3	&	7.42	&	3.88	    &	$-7.76^{\bf K} $	&	-2.88	&	\bf-3.62	&	-	 &	-	&	-	&	1437	\\
Fe\,{\sc i}	&	17433.67	&	37.0	&	7.47	&	6.41	&	$-7.66^{\bf K} $	&	0.03	&	-0.43	&	\bf-0.48	 &	-	&	-	&	2550	\\
Fe\,{\sc i}	&	17488.61	&	25.1	&	7.57	&	6.41	&	$-7.66^{\bf K} $	&	-0.13	&	\bf-0.78	&	-	 &	-	&	-	&	2550	\\
Fe\,{\sc i}	&	17500.01	&	24.7	&	7.48	&	5.96	&	$-7.58^{\bf K} $	&	-0.76	&	\bf-1.10	&	-1.28	 &	-	&	-	&	2207	\\
Fe\,{\sc i}	&	17531.20	&	9.3	&	7.24	&	6.64	    &	$-7.70^{\bf K} $	&	\bf-0.71	&	-0.68	&	-	 &	-	&	-	&	2696	\\
Fe\,{\sc i}	&	17534.80	&	23.1	&	7.26	&	6.64	&	$-7.70^{\bf K} $	&	-0.13	&	\bf-0.31	&	-	 &	-	&	-	&	2696	\\
Fe\,{\sc i}	&	17536.93	&	38.6	&	7.54	&	5.91	&	$-7.70^{\bf K} $	&	-	&	\bf-0.97	&	-	 &	-	&	-	&	none	\\
\hline
    \end{tabular}
                \vspace{1ex}
     \footnotesize{($^{\textbf{K}}$) Kurucz damping and [UNS] Unsold damping.}
\end{table*} 
\begin{table*}
\setlength{\tabcolsep}{6.9pt}
\caption{The iron lines in the line list in $Y$-, $J$-, $H$-, and $K$- bands for the analyses of the spectra for HD\,76151 and the Sun. Wavelengths for iron transitions listed in the APOGEE DR17 as both Fe\,{\sc i} and Fe\,{\sc ii} are shown in bold.}
\label{tab:A7}
    \centering
    \begin{tabular}{c|c|c|c|c|c|c|c|c|c|c|c}
    \hline
 Species	& $\lambda$ & EW & $\log\epsilon$(X)	&	LEP &$\log \Gamma$ &\multicolumn{5}{c|}{$\log gf$} &RMT\\
    \cline{2-11} 
 &   (\AA)  &  (m\AA)  & (dex)  & (eV)&(rad cm$^3$s$^{-1}$)&	VALD & MB99	& AN16 &DR17&NIST & \\ 
\cline{1-12}
Fe\,{\sc i}	&	17575.32	&	10.1	&	7.36	&	6.40	&	$-7.43^{\bf K} $	&	-0.82	&	\bf-1.01	&	-	&	-	&	-	&	2501	\\
Fe\,{\sc i}	&	17695.94	&	60.9	&	7.52	&	5.95	&	$-7.56^{\bf K} $	&	-0.52	&	\bf-0.65	&	-	&	-	&	-	&	2291	\\
Fe\,{\sc i}	&	17714.37	&	12.1	&	7.36	&	6.58	&	$-7.55^{\bf K} $	&	\bf-0.76	&	-0.64	&	-0.40	&	-	&	-	&	2656	\\
Fe\,{\sc i}	&	17717.16	&	24.0	&	7.46	&	6.34	&	$-7.70^{\bf K} $	&	-0.36	&	\bf-0.75	&	-	&	-	&	-	&	2561	\\
Fe\,{\sc i}	&	17747.37	&	50.9	&	7.50	&	5.92	&	$-7.58^{\bf K} $	&	-1.91	&	\bf-0.76	&	-	&	-	&	-	&	2209	\\
Fe\,{\sc i}	&	17926.40	&	61.6	&	7.48	&	6.74	&	$-7.70^{\bf K} $	&	\bf0.05	&	0.27	&	-	&	-	&	-	&	2702	\\
Fe\,{\sc i}	&	19923.34	&	46.4	&	7.43	&	5.02	&	$-7.59^{\bf K} $	&	\bf-1.56	&	-	&	-1.55	&	-	&	-	&	1800	\\
Fe\,{\sc i}	&	20363.27	&	46.1	&	7.46	&	6.07	&	$-7.57^{\bf K} $	&	-1.13	&	-	&	\bf-0.63	&	-	&	-	&	2295	\\
Fe\,{\sc i}	&	20716.94	&	87.3	&	7.21	&	6.02	&	$-7.57^{\bf K} $	&	\bf-0.01	&	-	&	-	&	-	&	-	&	2286	\\
Fe\,{\sc i}	&	21178.16	&	8.5	&	7.33	&	3.02	    &	$-7.78^{\bf K} $	&	-4.01	&	-	&	\bf-4.24  	&	-	&	-	&	902	\\
Fe\,{\sc i}	&	21238.47	&	73.5	&	7.49	&	4.96	&	$-7.58^{\bf K} $	&	-1.42	&	-	&	\bf-1.37 	&	-	&	-	&	1800	\\
Fe\,{\sc i}	&	21284.35	&	4.7	&	7.39	&	3.07	    &	$-7.71^{\bf K} $	&	-4.26	&	-	&	\bf-4.51 	&	-	&	-	&	902	\\
Fe\,{\sc i}	&	21515.15	&	11.6	&	7.64	&	5.34	&	$-7.58^{\bf K} $	&	\bf-2.17	&	-	&	-1.88	&	-	&	-	&	1971	\\
Fe\,{\sc i}	&	21553.30	&	11.1	&	7.51	&	6.70	&	$-7.55^{\bf K} $	&	-0.50	&	-	&	\bf-0.81	&	-	&	-	&	2735	\\
Fe\,{\sc i}	&	21735.46	&	27.3	&	7.45	&	6.17	&	$-7.38^{\bf K} $	&	-0.72	&	-	&	\bf-0.80  	&	-	&	-	&	2344	\\
Fe\,{\sc i}	&	21813.98	&	8.2	&	7.48	&	5.85	    &	$-7.66^{\bf K} $	&	-1.39	&	-	&	\bf-1.69	&	-	&	-	&	2162	\\
Fe\,{\sc i}	&	21851.38	&	10.0	&	7.38	&	3.64	&	$-7.84^{\bf K} $	&	\bf-3.61	&	-	&	-3.59	&	-	&	-	&	1362	\\
Fe\,{\sc i}	&	22257.11	&	138.6	&	7.49	&	5.06	&	$-7.58^{\bf K} $	&	\bf-0.71	&	-	&	-0.87	&	-	&	-	&	1856	\\
Fe\,{\sc i}	&	22385.10	&	23.3	&	7.21	&	5.32	&	$-7.58^{\bf K} $	&	-1.36	&	-	&	\bf-1.42	&	-	&	-	&	1941	\\
Fe\,{\sc i}	&	22392.88	&	61.9	&	7.48	&	5.10	&	$-7.59^{\bf K} $	&	-1.25	&	-	&	\bf-1.35	&	-	&	-	&	1856	\\
Fe\,{\sc i}	&	22419.98	&	58.3	&	7.42	&	6.22	&	$-7.38^{\bf K} $	&	-0.16	&	-	&	\bf-0.30	&	-	&	-	&	2344	\\
Fe\,{\sc i}	&	23308.48	&	27.7	&	7.35	&	4.08	&	$-7.73^{\bf K} $	&	-2.31	&	-	&	\bf-2.66  	&	-	&	-	&	1484	\\
\cline{2-11}
Fe\,{\sc i}	&	\bf 15531.75	&	87.1	&	7.49	&5.64	&	$-7.48^{\bf K} $  	&	-0.24	&	-0.73	  &	\bf-0.57	&	-0.56	&	-	&	2061	\\
Fe\,{\sc ii}	&	\bf15531.74*	&	87.1	&	7.44&13.25	&	UNS	&	-3.27	&	-	  &	-	&	-4.84	&	-	&	-	\\
Fe\,{\sc i}	&	\bf 15928.16	&	38.1	&	7.50	&5.95   &	$-7.57^{\bf K} $	&	-0.68	&	\bf-0.88	  &	-0.86	&	-0.68	&	-	&	2297	\\
Fe\,{\sc ii}	&	\bf15928.14*	&	38.1	&7.47&	13.33	&	UNS	&	-4.53	&	-	  &	-	&	-4.53	&	-	&	-	\\
Fe\,{\sc i}	&	\bf 15952.63 	&	27.9	&	7.61&6.34	    &	$-7.30^{\bf K} $ 	&	-3.25	&	\bf-0.81	  &	-	&	-0.61	&	-	&	2560	\\
Fe\,{\sc ii}	&	\bf15952.70*	&	27.9	&7.49&	12.9	&	UNS	&	-2.49	&	-	  &	-	&	-2.49	&	-	&	-	\\
\cline{2-11}
Fe\,{\sc ii}	&	10366.17	&	3.5	&	7.50	&	6.72	&	$-7.90^{\bf K} $	&	\bf-1.82	&	-1.76	  &	-	&	-	&	-1.83	&	-	\\
Fe\,{\sc ii}	&	10490.90	&	2.6	&	7.51	&	5.55	&	$-7.89^{\bf K} $	&	-2.91	&	-2.95	  &	-	&	-	&	\bf-3.00	&	-	\\
Fe\,{\sc ii}	&	10501.50	&	18.1	&	7.50	&5.55	&	$-7.89^{\bf K} $	&	\bf-2.01 &	-2.17	  &	-1.93	&	-	&	-2.09	&	-	\\
Fe\,{\sc ii}	&	10546.37	&	6.7	&	7.55	&	9.65	&	$-7.69^{\bf K} $	&	-	&	\bf0.91	  &	-	&	-	&	-0.58	&	-	\\
Fe\,{\sc ii}	&	10655.65	&	1.4	&	7.45	&	6.81	&	$-7.90^{\bf K} $	&	\bf-2.14	&	-2.12	  &	-	&	-	&	-	&	-	\\
Fe\,{\sc ii}	&	10862.64	&	14.5	&	7.50	&5.59	&	$-7.89^{\bf K} $	&	\bf-2.14	&	-2.11	  &	-2.04	&	-	&	-2.2	&	-	\\
Fe\,{\sc ii}	&	15078.21*	&	6.4	&	7.51	&	12.13	& UNS &	-4.68	&	-	  &	-	&	-4.68	&	-	&	-	\\
\textcolor{red}{Fe\,{\sc ii}}&	\textcolor{red}{15536.68*}	&	\textcolor{red}{5.8}	    &	\textcolor{red}{7.51}	&	\textcolor{red}{13.51}	&UNS&\textcolor{red}{-7.77}	&\textcolor{red}{-}	&	\textcolor{red}{-}	&	\textcolor{red}{-7.77}	&	\textcolor{red}{-}	&	\textcolor{red}{-}	\\
\textcolor{red}{Fe\,{\sc ii}}&	\textcolor{red}{15563.40*}	&	\textcolor{red}{6.5}	    &	\textcolor{red}{7.51}	&	\textcolor{red}{13.31}	& UNS &\textcolor{red}{-4.08}	&	\textcolor{red}{-}	&	\textcolor{red}{-}	&	\textcolor{red}{-4.08}	&	\textcolor{red}{-}	&	\textcolor{red}{-}	\\
Fe\,{\sc ii}&	15792.03*	&	3.3	    &	7.51	&	12.81	& UNS &	-6.20	&	-	&	-	&	-6.20	&	-	&	-	\\
Fe\,{\sc ii}&	16468.51*	&	22.1    &	7.50	&	13.29	&$-7.34^{\bf K} $ &-0.47	&	-	&	-	&	-0.47	&	-	&	-	\\
\textcolor{red}{Fe\,{\sc ii}}&	\textcolor{red}{17000.92*}	&	\textcolor{red}{92.2}	&	\textcolor{red}{7.50}	&	\textcolor{red}{13.03}	&$-7.62^{\bf K} $ &\textcolor{red}{-0.69}	&	\textcolor{red}{-}	&	\textcolor{red}{-}	&	\textcolor{red}{-0.71}	&	\textcolor{red}{-}	&	\textcolor{red}{-}	\\
\hline
    \end{tabular}
        \vspace{1ex}
     \footnotesize{($^{*}$) The $\log gf$ values were astrophysical and were calculated in this study. ($^{\textbf{K}}$) Kurucz damping, and [UNS] Unsold damping.}
\end{table*}

\begin{table*}
\scriptsize
\setlength{\tabcolsep}{1.9pt}
\renewcommand{\arraystretch}{0.95}
\caption{The model atmosphere parameters, [$\alpha$/Fe] abundances, and the fundamental parameters for HD\,76151 were compiled from literature. H (HARPS), I (IGRINS), and H+I denote spectroscopic analyses utilizing combined optical and NIR line lists.}
\label{tab:A8}
\centering
\begin{tabular}{c|c|c|c|c|c|c|c|c|c|c}
\hline
$T_{\rm eff}$ & $\log g$  &  [Fe/H]	&  [Mg/Fe]   & [Si/Fe] & [Ca/Fe]  & [Ti/Fe] &  $R$	&  $M$   & $t$ & Reference\\
\cline{1-11}
(K)           & (cgs)     & \multicolumn{5}{c|}{(dex)}  &  ($R_{\odot}$)   & ($M_{\odot}$) & (Gyr) &\\
\hline
\hline
5780$\pm$88   & 4.35$\pm$0.16 & 0.14$\pm$0.08 & -0.15$\pm$0.11 & 0.03$\pm$0.13 & 0.00$\pm$0.14 & -0.04$\pm$0.16 & 1.125$\pm$0.008 &	1.015$\pm$0.037	&	6.8$\pm$1.6	& TS(H)	\\
5780$\pm$178  & 4.31$\pm$0.25 & 0.21$\pm$0.09 & --             & 0.01$\pm$0.17 & --            & 0.10$\pm$0.18  & 1.190$\pm$0.012 & 1.046$\pm$0.069 & 6.8$\pm$2.6&  TS(I) \\
5790$\pm$170  & 4.35$\pm$0.18 & 0.24$\pm$0.12 & -0.09$\pm$0.14 &-0.03$\pm$0.16 & 0.02$\pm$0.16 & -0.02$\pm$0.18 & 1.125$\pm$0.012 & 1.053$\pm$0.062& 5.5$\pm$2.3 & TS(H+I) \\
5600          & 4.44$\pm$0.20 &-0.09$\pm$0.15 & 0.09           & --	           & 0.16          & 0.05           & -- &--&--& 1 \\
5763          & 4.37          & 0.01          & 0.11           & -0.02         & -0.06 & 0.03  &--              & --&--&  2\\
5902          & 4.11          & 0.14      	  & --             & --	           &  --  	       & --             & -- &--&--& 3 \\
5763          & 4.37          & 0.01          & --             & --            &  --           & --             & -- &--&--& 4 \\
5687          & 4.37          & 0.02          & --             & --	           &  --  	       & --             & -- &--&--& 5 \\
5727          & 4.50          & 0.10          & 0.10           & 0.02	       & 0.02	       & -0.10          & -- &--&--&  6 \\
5825$\pm$50   & 4.62$\pm$0.15 & 0.15$\pm$0.06 & --             & --	           & --  	       & --             & --		&	1.07	&	--&7 \\
5825          & 4.62          & 0.15          & --             & -0.01         &  -0.04        & -0.01          & -- &--&--& 8  \\
5750$\pm$50   & 4.40$\pm$0.05 & 0.08$\pm$0.04 & --             & --	           & --  	       & --             & -- &--&--& 9  \\
5759$\pm$70   & 4.45$\pm$0.10 & 0.06$\pm$0.07 & --             & --	           & --  	       & --             & 1.02$\pm$0.04	&	1.05	&	--	&	10\\
5776$\pm$20   & 4.40$\pm$0.20 & 0.05$\pm$0.10 & 0.01           & 0.04          & --            & --             & -- &--&--& 11 \\
5803$\pm$29   & 4.50$\pm$0.08 & 0.14$\pm$0.04 & --             & --	           & --  	       & --             & -- &1.07&--&  12  \\
5803$\pm$50   & 4.50$\pm$0.15 & 0.14$\pm$0.05 & 0.06$\pm$0.07  &  --           &  --           & --             & -- &--&--& 13 \\
5770          & 4.39          & 0.03          & 0.08           & 0.03          & -0.06         & 0.03           & -- &--&--&14 \\
5773          & 4.49          & 0.02          & --             & --            & --            & --             & --	&	1.02	&	-- & 15 \\
5790$\pm$44   & 4.55$\pm$0.06 & 0.11$\pm$0.03 & --             & -0.01         & --            & --             &	0.979$\pm$0.017&1.24$\pm$0.12	&	3.60	& 16  \\
5790$\pm$44   & 4.55$\pm$0.06 & 0.11$\pm$0.03 & --             & -0.01         & --            & --             &	0.979$\pm$0.017	&	1.05$\pm$1.03	&	3.60& 16  \\
5765$\pm$100  & 4.44$\pm$0.10 & 0.08$\pm$0.03 & 0.23           & 0.01          & -0.01         & 0.01           & -- &--&--&  17  \\
5692          & 4.28          & 0.08          & --             & --            & -             & --             & -- &--&--&  18 \\
5678          & 4.49          & 0.22          & --             & --            & --            & --             & --&-- &--&  19 \\
5769          & 4.475         & 0.125         & --             & -0.026        & --            & -0.067         & --	&	1.02	&	3.71$\pm$1.23	& 20   \\
5768          & 4.45          & 0.06          & --             & --	           & --  	       & --             & -- &--&--&  21 \\
5788$\pm$23   & 4.48$\pm$0.02 & 0.12$\pm$0.02 & --	           & --  	       & --            & --             & -- &1.037&--&22  \\
5760          & 4.40          & 0.01          & --             & --            & --            & --             & -- &--&--& 23 \\
5788          & 4.48          & 0.12          & -0.02          & -0.01         & --            & --             & -- &--&--&  24 \\
5770$\pm$25   & 4.40$\pm$0.07 & 0.11$\pm$0.02 & --             & --	           & --  	       & --             & 0.98 &0.88&2.00$\pm$2.50&  25  \\
5787$\pm$80   & 4.43          & 0.031         & --             & --	           & --  	       & --             & -- &--&--& 26  \\
5787$\pm$80   & 4.43 & 0.031  & --            & --	           & --  	       & --            & --             & --&--& 26  \\
5773$\pm$59	  & 4.42$\pm$0.17 & 0.11$\pm$0.04 & --	           & --  	       & --            & --             & -- &--&--& 27  \\
5758          & 4.42          & 0.07          & --             & --            & --            & --             & -- &--&--&  28 \\
5748$\pm$38   & 4.42$\pm$0.07 & 0.15$\pm$0.03 & --             & --	           & --  	       & --             & -- &--&--&29  \\
5793$\pm$26   & 4.47$\pm$0.05 & 0.13$\pm$0.04 & --             & --            & --            & --             & -- &--&--&  30 \\
5788          & 4.48          & 0.12          & 0.003          & -0.003        & 0.008 & 0.003 & --             & -- &--& 31  \\
5859$\pm$24   & 4.77$\pm$0.07 & 0.23$\pm$0.03 & --	           & --  	       & --            & --             & -- &--&--&32  \\
5756$\pm$50   & 4.42$\pm$0.03 & 0.09$\pm$0.04 & --             & --	           & --  	       & --             & -- &	1.02$\pm$0.03	&	1.40$\pm$0.30	&	33 \\
5786$\pm$12	  & 4.51$\pm$0.04 &0.101$\pm$0.002&-0.051$\pm$0.010&-0.131$\pm$0.030&-0.111$\pm$0.010&-0.111$\pm$0.020& -- &--&--&34 \\
5687          & 4.37          & -0.08         & --             & --            & --            & --             & --&--&--& 35 \\
5776          & 4.40          & 0.05          & 0.06           & 0.09 & -0.01  & --            & --             & -- &--&36 \\
5781$\pm$17   & 4.44$\pm$0.05 & 0.12$\pm$0.01 & --  	       & --            & --            & --	            & --&	1.054$\pm$0.010	&	0.295$\pm$0.211	&37 \\
5752          & 4.51          & 0.12          & --             & --            &  --           & --             & -- &--&--& 38  \\
5786$\pm$34	  & 4.40$\pm$0.14 & 0.12$\pm$0.04 & -0.01$\pm$0.04 &-0.04$\pm$0.04 & -0.02$\pm$0.06& -0.03$\pm$0.05 & --	&1.05$\pm$0.03&	2.50$\pm$1.40	&	39\\
5831          & 4.64 & 0.13   & --            & --             &  --           & --            & --             & --&--& 40 \\
5702          & 4.39 & 0.14   & --            & --             &  --           & --            & --             & -- &--& 41 \\
5776$\pm$10	  & 4.54$\pm$0.02 &0.119$\pm$0.028& --             &-0.001$\pm$0.037&-0.003$\pm$0.036&-0.033$\pm$0.049& -- &--&--&42 \\
5780          & 4.43          & 0.10$\pm$0.05 & 0.08$\pm$0.09  & 0.04$\pm$0.08 & 0.05$\pm$0.11 &  0.04$\pm$0.09 & --	&	1.01$\pm$0.10	&	4.56	&	43 \\
5787          & 4.43          & 0.10          & --             & --            & --            & --             & -- &--&--&44 \\
5794$\pm$36   & 4.69$\pm$0.08 & 0.02$\pm$0.04 & --             & --	           & --  	       & --             & -- &--&--& 45  \\
5781          & 4.44          & 0.12          & --             & --            & --            & --             & -- &  -- &--&46 \\
5761          & 4.43          & 0.10$\pm$0.04 & -0.05$\pm$0.16 & 0.00$\pm$0.05 &-0.01$\pm$0.07 & -0.02$\pm$0.07 & --&--&--& 47 \\
5748$\pm$38   & 4.42$\pm$0.07 & 0.15$\pm$0.03 &--              & --            & --            & --             & --&--&--&48\\
5816$\pm$30   & 4.51$\pm$0.25 & 0.13$\pm$0.06 & -0.06$\pm$0.11 & 0.15$\pm$0.06 & 0.08$\pm$0.06 & 0.07$\pm$0.06 & 1.018$\pm$0.013 &	1.067$\pm$0.030 &	2.068$\pm$1.276 &	49\\ 
5776          & 4.49          & 0.10          & --             & --            & --            & --             &--&--&--& 50  \\
5780          & 4.40          & 0.15          & --             & --            & --            & --             & 1.01$\pm$0.02	&	1.04$\pm$0.04	&	--	&	51\\
5777$\pm$139  & 4.47$\pm$0.08 & 0.10$\pm$0.01 & --             & --	           & --  	       & --             & 0.984$\pm$0.053	&	1.04$\pm$0.13	&	--	&	52 \\
5785$\pm$3    & 4.44$\pm$0.01 & 0.11$\pm$0.01 & --             & --	           & --  	       & --             & --	&	1.06$\pm$0.02	&	1.40$\pm$0.50	&	53 \\
\hline
\end{tabular}
\begin{minipage}{18cm}
\vspace{1ex}
 [01] \cite{Cayrel1981}, [02] \cite{Edvardsson1993}, [03] \cite{Marsakov1995}, [04] \cite{Tomkin1995}, [05] \cite{Gratton1996}, [06] \cite{Thevenin1998}, [07] \cite{Santos2001},[08] \cite{Bodaghee2003},  [09] \cite{Heiter2003}, [10] \cite{Fuhrmann2004}, [11] \cite{Mishenina2004} , [12] \cite{Santos2004},  [13] \cite{Beirao2005}, [14] \cite{Soubiran2005}, [15] \cite{Takeda2005}, [16] \cite{Valenti2005}, [17] \cite{Luck2006}, [18] \cite{Cenarro2007}, [19] \cite{Schiavon2007}, [20] \cite{Takeda2007},[21] \cite{Soubiran2008}, [22] \cite{Sousa2008}, [23] \cite{Onehag2009}, [24] \cite{DelgadoMena2010}, [25] \cite{Ghezzi2010}, [26] \cite{Casagrande2011}, [27] \cite{DaSilva2011}, [28] \cite{Katz2011}, [29] \cite{Prugniel2011}, [30] \cite{Wu2011}, [31] \cite{Adibekyan2012}, [32] \cite{Maldonado2012}, [33] \cite{Ramirez2012}, [34] \cite{Tabernero2012}, [35] \cite{Carretta2013},  [36] \cite{Mishenina2013}, [37] \cite{Tsantaki2013}, [38] \cite{Datson2015}, [39] \cite{DaSilva2015}, [40] \cite{Maldonado2015}, [41] \cite{Paletou2015}, [42] \cite{Mahdi2016}, [43] \cite{Luck2017}, [44] \cite{Netopil2017}, [45] \cite{Rich2017}, [46] \cite{Suarez2017}, [47] \cite{Luck2018}, [48] \cite{Park2018},  [49] \cite{Soto2018}, [50] \cite{casali2020}, [51] \cite{Hirsch2021}, [52] \cite{Paegert2021}, [53] \cite{Martos2023}, TS: This Study.
\end{minipage}
\end{table*}

\end{appendix}
\end{document}